\documentstyle[12pt,dina4p,german,twoside,epsfig]{report}
\originalTeX

\newlength{\bundsteg}
\newlength{\breite}

\def\thefigure{\arabic{chapter}-\arabic{figure}}

\reversemarginpar

\newcommand{\dx}[1]{\!\!{d#1}\;}
\newcommand{\li}{\!\!\left}
\newcommand{\re}{\right}
\newcommand{\eqlabel}[1]{\label{#1}}
\newcommand{\plabel}[1]{\label{#1}}

\newcommand{\stackrelss}[2]
           {\stackrel{\scriptscriptstyle #1}{\scriptscriptstyle #2}}
\newcommand{\DK}{{P\!K}}
\newcommand{\sgn}[1]{\sigma_{\!\underline{#1}}}
\newcommand{\qt}[1]{\bar{\tau}_{#1}}

\newfont{\comfont}{cmss10}

{\comfont\par\noindent{\rm.}\dotfill\ Kommentar\ \dotfill{\rm.}\\}%
{\\\noindent{\rm.}\dotfill\ Ende~des~Kommentars\ \dotfill{\rm.}\par}

\begin{document}
%
%
\begin{titlepage}
\hspace*{\fill}{\large DESY 96-115}
\vspace*{6cm}
\begin{center}

{\LARGE\bf
An Investigation of Stochastic Cooling\\
in the Framework of Control Theory\\}
\vspace{2cm}
\Large
Olaf Meincke\\
\large\sl 
Deutsches Elektronen-Synchrotron DESY, Hamburg 
\end{center}
\end{titlepage}

\thispagestyle{empty}
\vspace*{10cm}
\newpage

\begin{titlepage}
\vspace*{4cm}
\begin{center}

{\LARGE\bf
Systemtheoretische Untersuchung\\
der stochastischen K\"uhlung\\}
\vspace{4cm}
{\large\bf
Dissertation\\
zur Erlangung des Doktorgrades\\
des Fachbereichs Physik\\
der Universit\"at Hamburg\\
\vspace{2cm}
vorgelegt von\\
Olaf Meincke\\
aus Hamburg\\
\vspace{2cm}
Hamburg\\
1995}
\end{center}
\end{titlepage}

\thispagestyle{empty}
\vspace*{15.5cm}
\begin{tabular}{l@{\hspace{5mm}}l}
Gutachter der Dissertation: & Prof. Dr. R.-D. Kohaupt \\
                            & Prof. Dr. P. Schm\"user  \\[2ex]
Gutachter der Disputation:  & Prof. Dr. R.-D. Kohaupt \\
                            & Prof. Dr. Dr. h. c. G.-A. Voss \\[2ex]
Datum der Disputation:      & 31.8.1995 \\[2ex]
\parbox[b]{5cm}{Sprecher des \\ Fachbereichs Physik und \\ Vorsitzender des \\ 
Promotionsausschusses:}     & Prof. Dr. B. Kramer
\end{tabular}
\newpage

\pagenumbering{roman}
\germanTeX
\section*{Kurzfassung}

Diese Arbeit betrachtet das stochastische K\"uhlen ungebunchter Strahlen unter 
dem Aspekt der Systemtheorie. Der Schwerpunkt liegt dabei auf der Untersuchung 
der kollektiven Strahl\-bewegung, mit dem Ziel, Stabilit\"atsaussagen f\"ur 
einen Strahl in einem aktiven K\"uhlsystem zu erhalten. Denn ein 
stochastisches K\"uhlsystem bildet einen R\"uckkopplungskreis und ist daher 
vergleichbar mit den Feedback-Systemen, die zur D\"ampfung kollektiver 
Instabilit\"aten eingesetzt werden. Da jedes System, das auf sich selbst 
zur\"uckwirkt, potentiell instabil ist, erfordern derartige 
R\"uckkopplungskreise eine sorgf\"altige Analyse ihrer Stabilit\"at.

Ausgehend von einer linearen K\"uhlwechselwirkung wird f\"ur das transversale 
K\"uhlen eine selbstkonsistente L\"osung der Strahlbewegung hergeleitet. Dazu 
wird die kollektive Bewegung des Strahls in seine koh\"arenten Moden zerlegt. 
Die Rechnung ber\"ucksichtigt die Lokalit\"at von Detektor und Kicker und die 
daraus resultierende zeitdiskrete Struktur in der Teil\-chendynamik. Aus der 
selbstkonsistenten L\"osung wird dann ein Stabilit\"atskriterium f\"ur jede 
Mode des Strahls abgeleitet. Die erhaltenen Ausdr\"ucke erlauben auch eine 
\"Uberlappung der Frequenzb\"ander im Spektrum des Strahls und liefern 
demzufolge \"uber den gesamten Frequenzbereich g\"ultige Aussagen.

Nachdem so die Grenzen der Stabilit\"at festgelegt worden sind, erfolgt eine 
Beschreibung der K\"uhlung durch die Fokker-Planck-Gleichung. Die Berechnung 
ihrer Drift- und Diffusi\-onskoeffizienten wird im Frequenzbereich 
durchgef\"uhrt. Auch sie betrachtet den Detektor und Kicker des K\"uhlsystems 
als lokale Objekte und beinhaltet somit die Taktung in der 
K\"uhlwechselwirkung. Die Fokker-Planck-Gleichung liefert eine statistische 
Beschreibung, die kollektive Effekte nicht einbezieht und daher implizit die 
Stabilit\"at des Strahls voraussetzt. Die hieraus folgenden Vorhersagen 
\"uber die K\"uhlung sind folglich nur innerhalb der herge\-leiteten 
Stabilit\"atsgrenzen physikalisch sinnvoll. Daher wird am Ende gepr\"uft, ob 
die ermit\-telten Parameter, mit denen das K\"uhlsystem am effizientesten 
arbeitet, vertr\"aglich sind mit der Stabilit\"at des Strahls.
\newpage
\originalTeX
\section*{Abstract}

This thesis provides a description of unbunched beam stochastic cooling in the
framework of control theory. The main interest in the investigation is 
concentrated on the beam stability in an active cooling system. A stochastic 
cooling system must be considered as a closed-loop, similar to the feedback 
systems used to damp collective instabilities. These systems, which are able 
to act upon themselves, are potentially unstable and therefore their stability 
must be carefully analysed.

Assuming a linear transverse cooling interaction, the self-consistent solution
for the beam motion is derived by means of a mode analysis of the collective 
beam motion. Furthermore the calculation treats the pick-up and kicker of the 
cooling system as localized objects which impose a discrete time structure on 
the dynamics of the beam particles. This solution then yields a criterion for 
the stability of each collective mode. The expressions which have been obtained
also allow for overlapping frequency bands in the beam spectrum and thus are 
valid over the entire frequency range.

Having established the boundaries of stability in this way, the Fokker-Planck 
equation is used to describe the cooling process. The drift and diffusion
coefficients are derived in the frequency domain taking into account the 
localization of  pick-up and kicker and the sampled nature of the cooling 
interaction. The Fokker-Planck equation provides a purely statistical 
description, which does not include collective effects and thus a stable beam 
must be assumed. Hence the predictions about the cooling process following 
from the Fokker-Planck equation only make physical sense within the boundaries 
of beam stability. Finally it is verified that the parameters of the cooling 
system which give the best cooling results are compatible with the stability 
of the beam.

\tableofcontents
\chapter*{Introduction}
\pagenumbering{arabic}
\addcontentsline{toc}{chapter}{Introduction}
\plabel{prolpart1}

\section*{Motivation}
\addcontentsline{toc}{section}{Motivation}
\plabel{prolpart2}

Storage rings aim to supply intense particle beams with long lifetimes and to 
bring them into collision with high luminosity. Hence small beam dimensions 
are desirable because they can improve both the lifetime and the luminosity. 
In this regard electron beams distinguish themselves from proton beams%
\footnote{The same is true for the corresponding anti-particle beams.} because 
they have an inherent damping mechanism. The quantum-like energy loss, due to 
their synchrotron radiation, in combination with the acceleration in the 
rf-cavities hold the beam dimensions in an equilibrium state and thus make 
electron beams insensitive to small perturbations \cite{san}. For proton beams 
this radiation damping is negligible for the beam energies which currently can 
be attained, so that already small excitations of the beam lead to a 
continuous increase  of the beam dimensions. In the HERA proton ring, for 
example, typical emittance growth rates of $1\ \pi\,\mbox{mm}\,\mbox{mrad/h}$ 
have been observed during beam collisions \cite{wil}. Because this emittance 
growth degrades the luminosity, various theories have been studied to explain 
this effect \cite{bru, zim}. At present, the dimensions of proton beams can 
only be reduced by means of an external system which provides an artificial 
damping mechanism. This active process of emittance reduction is called {\sl 
beam cooling}.

The cooling of a particle beam can be understood as an increase of its 
phase-space density. This process concentrates the particles in the center of 
their distribution and thus reduces the phase-space volume which they occupy. 
This requires a local interaction in phase space which acts individually on 
the particles. In other words, the system has to resolve small fractions of 
the phase space of the beam in order to manipulate the internal phase-space 
structure.

In stochastic cooling systems this interaction happens by means of an external 
feedback loop, similar to the feedback systems used to damp collective 
instabilities \cite{sac, moe, bi2, ch2}. The basic idea is that a pickup 
detects from each particle the quantity to be reduced and a kicker feeds the 
amplified signals back to the particles with an appropriate phase shift to 
reduce the measured offsets. In order to act on the particles with their 
proper corrections the cooling system must be arranged in such a way that the 
time delay of the signals in the electronic components matches the transit 
time of the particles between pickup and kicker. However, due to their finite 
bandwidth, stochastic cooling systems are not able to resolve single particles 
so that the correction signals always contain the information of many 
particles. Besides the coherent self-interaction which provides damping a 
particle receives the signals from other particles being processed at the same 
time. In the case of random particle motions the latter produces an incoherent 
contribution in the cooling interaction which causes a diffusion. This 
diffusion counteracts the damping, and thus degrades the cooling. In 
particular, the cooling system has to operate in such a way that a net cooling 
effect remains, which limits the system gain.

The description so far assumes random particle motion which strictly speaking 
only holds true until the first correction has been applied. The feedback of 
the cooling system introduces correlations among the particles which are 
manifested as a coherent beam motion causing a loss of the statistical 
independence of the particles. Different frequencies of the particles, 
however, lead to a decoherence of the collective motion and thus can destroy
these correlations. For a sufficiently large frequency spread the phases will 
randomize within one revolution removing the correlations completely between 
successive cooling steps. This perfect phase mixing results in the shortest 
cooling time. Decays of correlations which take more than one revolution 
deteriorate the cooling because remaining correlations increase the diffusion 
effect. On the other hand, if the buildup of correlations occurs faster 
than their decay by the phase mixing, the correlations will continuously grow. 
Because then all beam particles participate in a collective motion, the 
resulting coherent interaction is many times stronger than the single particle 
self-interaction and therefore dominates the dynamics. In that case the 
collective particle motion arising from the correlations leads to instability 
of the beam.

In this picture stochastic cooling is divided into two competing processes, 
the cooling through the self-interaction and the diffusion. For a rigorous 
mathematical description which includes both effects one studies the particle 
density distribution in phase space. The time evolution of the phase space 
density reflects the dynamics of the cooling process and is usually derived 
from a Fokker-Planck equation. This approach, which is also used in this work, 
provides analytical expressions for the parameters which characterize the 
performance of the cooling system and allows quantitative predictions of the 
maximum attainable cooling rates. 

The Fokker-Planck equation describes the cooling process at the microscopic 
particle level. Its derivation presumes the statistical independence of the 
individual particles and thus completely neglects the collective effects of 
the beam motion. The following remarks emphasize this point:
\begin{itemize}
\item%
Initially the Fokker-Planck equation determines the time evolution of the 
probability density of a single particle in phase space. Applying this result 
to all $N$ particles of the beam and thus identifying the probability density 
with the phase space density of the beam requires the statistical independence 
of the particles.
\item%
The calculations of its drift and diffusion coefficients use a perturbation 
expansion in a small parameter $\epsilon$ which measures the strength of the 
feedback force. This perturbation series converges only if the feedback force 
remains bounded which implies a stable beam motion.
\end{itemize}
Hence one obtains physically reasonable results from the Fokker-Planck 
description only within the stability boundaries of the beam. Since stochastic 
cooling systems close a feedback loop in which the beam acts upon itself, they 
are potentially unstable like any feedback system. Therefore a thorough 
stability analysis of the collective beam motion in cooling systems is a 
prerequisite for the applicability of the Fokker-Planck equation. 

The existing reports about stochastic cooling either omit the verification of 
this requirement and implicitly assume a stable beam or consider simplified 
cases and thus obtain results which are valid only within certain limits 
\cite{wei}. For that reason the main interest in the investigation of this 
thesis is concentrated on the beam stability in an active cooling system.
In the following chapters transverse stochastic cooling of unbunched beams is 
studied for the case of a linear cooling interaction.

The analysis benefits from the distinct time scales underlying instabilities 
and stochastic cooling. Instabilities typically develop within milliseconds 
whereas the cooling times span a range from a few seconds to many hours. On 
the time scale of instabilities the internal phase space configuration of the 
beam only changes immaterially by means of the cooling interaction and 
therefore can be regarded as constant. Hence a separate treatment of the two 
processes becomes possible so that the beam stability is studied decoupled 
from the phase space cooling. The short time scale of instabilities also 
suggests an investigation in frequency domain. The collective motion of the 
beam is decomposed into the coherent beam modes and stability criteria 
are derived for each mode which allow predictions about the beam stability 
over the entire frequency range. 

The method which is used to obtain these results differs from the usual 
treatment of instabilities by the Vlasov-theory. It is based on the theory 
of multi-bunch feedback systems which was derived from the control theory of 
discrete time signals \cite{ko1}. This theory already includes the discrete 
time structure of the interaction originating in the localized pickup and 
kicker which is essential for a careful stability analysis. 

For bunched beams, however, this method does not succeed in the same way due 
to the basically different longitudinal motions of particles in bunched and 
unbunched beams. Although first results for bunched beams have been obtained, 
they still necessitate further investigations and hence will not be presented 
in this work. 

\section*{Outline}
\addcontentsline{toc}{section}{Outline}
\plabel{prolpart3}

First, Chapter \ref{cofopart1} reviews the theoretical and practical aspects 
of stochastic cooling. Chapter \ref{basepart1} introduces the basic terms and 
concepts of the control theory which become important in the following 
investigations. This formalism is used in Chapter \ref{stabpart1} to derive 
the stability criteria of the coherent beam modes. These results determine the 
range within which the parameters of the cooling system preserve beam 
stability. Chapter \ref{coolpart1} gives a mathematical description of the 
cooling process, including the calculations of the cooling parameters. 
Finally, the results of Chapter \ref{coolpart1} are compared with the limits 
derived in Chapter \ref{stabpart1} in order to verify their compatibility with 
the stability of the beam. 

\chapter{Theory and Applications of Stochastic Cooling} 
\plabel{cofopart1}

\section{Applications of Stochastic Cooling} \plabel{cofopart2}

Stochastic cooling was invented by S. van der Meer in 1968\footnote{He 
published his idea for the first time in 1972 \cite{mee}.} but was only 
experimentally demonstrated seven years later. Since that time large 
improvements have been achieved in the technical realization of cooling
systems, which has opened more and more new application fields for stochastic 
cooling. This section gives an overview of these practical uses.

\subsubsection{Production of Intense Antiproton Beams}

An important application of stochastic cooling is the accumulation of 
antiprotons which render an efficient operation of $p\bar{p}$ storage rings 
possible. The existing $p\bar{p}$ accelerators -- the Super-Proton-Synchrotron 
(SPS) at CERN and the TEVATRON at Fermilab -- have made considerable 
contributions to high-energy physics, in particular the discoveries of the 
$Z^0$ and $W^\pm$ bosons as well as of the top quark. A prerequisite for these 
successes had been intense antiproton beams without which the necessary 
luminosity could never had been delivered.

The production of antiprotons uses a high-energy proton beam which is directed
at a metal target. The production rate of the antiprotons is however small 
and the delivered beam has a broad momentum spread and large transverse 
emittances. To obtain an intense antiproton beam the antiprotons are collected 
over a long time. This process is called accumulation and takes place in 
storage rings specially designed for that purpose.

Accumulation becomes possible by virtue of longitudinal stochastic cooling. 
The principle is based on the fact that the mean energy of the antiprotons 
differs from their storage energy in the accumulator ring. The energy 
difference is chosen such that a newly-injected antiproton beam does not 
affect the stored beam. Longitudinal stochastic cooling then adjusts the 
energies of the incoming antiprotons to the storage energy, and thus provides 
the longitudinal phase-space required for the following antiprotons. Since the 
antiprotons stay in these accumulator rings for a long time (up to 24 hours), 
they are, in addition, cooled in the longitudinal and transverse directions in 
order to preserve the increased phase-space densities over the period of 
accumulation. \pagebreak

Examples of accumulator rings are the Antiproton-Accumulator (AA) and the 
Antiproton-Collector (ACOL) at CERN and the Debuncher/Accumulator-complex at 
Fermilab. These rings enhance the longitudinal phase-space density by a factor 
of $\stackrelss{>}{\sim} 10^4$, and the transverse phase-space densities by 
factors $10$ to $100$ \cite{bi2}.

\subsubsection{Improvement of the Beam Properties}

Stochastic cooling makes it feasible to generate narrowly-collimated and 
almost mono-energetic beams without any loss of particles. Longitudinal 
stochastic cooling systems allow a reduction of the energy spread of the beam 
and thus improve the energy resolution at the experiments. Transverse cooling 
decreases the horizontal and vertical beam emittances and therefore raises the 
luminosity. Hence stochastic cooling can substantially contribute to better 
experimental conditions. 

In the Low-Energy-Antiproton-Ring (LEAR) at CERN, for example, transverse 
emittances of $\epsilon_{x,z} \stackrelss{<}{\sim} 3\pi\,\mbox{mm}\, 
\mbox{mrad}$ and a momentum spread of $\Delta p/p < 0.2\%$ were attained 
in a beam with $\sim 5 \cdot 10^{10}$ particles \cite{cas}. Owing to its 
stochastic cooling systems, LEAR can deliver high-quality antiproton beams for 
precision measurements.

\subsubsection{Preservation of the Beam Quality}

In stored beams various effects, e.g.\ intra-beam scattering, residual gas 
scattering or beam-beam interaction, can cause a growth of the transverse 
emittances and of the energy spread which in general results in particle loss. 
In this case stochastic cooling can be used to compensate the undesired 
increase, and thus preserves the beam quality during long storage times. 

Especially ion storage rings profit from this process, e.g.\ the 
Experimental-Storage-Ring (ESR) at the GSI in Darmstadt, the 
Cooler-Synchrotron (COSY) at the KfK J\"ulich, the Test-Storage-Ring (TSR) at 
the  MPI in Heidelberg, CELSIUS in Uppsala and ASTRID in Aarhus, to mention 
just a few.

\subsubsection{Bunched Beam Cooling}

The application fields considered so far all refer to the cooling of unbunched
beam. Indeed concrete efforts exist to apply stochastic cooling also to 
bunched beams. Fermilab aims at cooling the bunched proton and antiproton beam 
in the TEVATRON, both horizontally and vertically. The planned systems are to 
counteract the emittance growth of the beams which is mainly caused by 
power-supply ripples and electronic noise \cite{jac}. The idea is to raise 
the luminosity lifetime so that the beams can be stored over longer periods. 
Since in that case more time becomes available for the antiproton 
accumulation, one ends up with more intense antiproton beams. On the other 
hand, the beams have to be replaced less frequently so that 
altogether the useful time for beam collisions increases. In that way one 
hopes to double the integrated luminosity \cite{jac}.

First tests have already been carried out with a vertical cooling system for 
the proton beam, but measurable changes in the emittance growth rate could 
not be observed so far \cite{jac}. Before stochastic cooling can efficiently 
be applied to bunched beams, a lot more research and development will be 
necessary.

\section{Theoretical Description of Stochastic Cooling}
\plabel{cofopart3}

The theoretical formulation of stochastic cooling can follow various ways.
One possibility is to look at the process purely in the time domain. This 
provides a very intuitive picture of the cooling but does not allow precise 
quantitative predictions. On the other hand, one can analyse the beam dynamics 
in the frequency domain by means of the spectrum generated by the particles. 
This more rigorous mathematical treatment yields reliable predictions about 
the cooling process. Both formulations represent statistical descriptions 
which rely on the signal fluctuations due to the discreteness of the beam 
particles. In this microscopic view, the collective beam motion is completely 
disregarded. 

Another approach to stochastic cooling is given by the kinetic theory. It 
investigates the time evolution of the $1-, 2-, \ldots, N-$particle 
distribution functions and hence takes into account the correlations among 
the particles. Thus predictions about collective effects become possible. On 
the other hand, this method requires a substantial mathematical effort and one 
obtains analytical solutions only for simple systems. Here, we will not 
further pursue this path. More detailed information can be found in 
\cite{bi2, ch2}.

In the next two sections we elaborate on the sample picture in the time domain 
and the description by a Fokker-Planck equation which in its relevant parts
is performed in the frequency domain.

\subsubsection{The Sample Picture}

We now discuss the formulation of stochastic cooling in the {\sl sample 
picture}, as it has been developed in \cite{moe}. For this purpose we consider 
a transverse cooling system for unbunched beams. However the major concern 
does not aim at detailed mathematical derivations but much more at a 
discussion of the assumptions underlying this description. For the most 
part the argumentation follows \cite{moe, bi2} where further details can also 
be found. 

The formulation is based on the idea of dividing the beam into samples. 
Because of its finite bandwidth, $W$, a cooling system cannot resolve the 
individual particles in a dense beam and thus always processes many particles 
simultaneously which in each case define a sample. The size of the sample, 
i.e.\ its number of particles, is given by $N_{\!S} = N/2WT_0$. Here, $N$ is 
the total number of particles in the beam and $T_0$ denotes the nominal 
revolution time. The larger the bandwidth $W$, the smaller the samples 
processed by the system, and the more distinctly the individual signal 
contribution of each sample particle will emerge. In order to come closer to 
the ideal case in which each particle is cooled separately, we have to make 
the system bandwidth as large as possible.

Since all sample particles contribute to the correction of a particular sample 
particle $i$, we can write its displacement $\check{x}_i$ after the correction 
as 
\begin{equation} \eqlabel{cofo1}
\check{x}_i = x_i - \lambda \sum_{j=1}^{N_{\!S}} x_j,
\end{equation}
where $\lambda$ is the strength with which the measured particle displacements
are fed back. Hence it follows, for the difference of $x_i^2$ before and after 
the correction
\begin{equation} \eqlabel{cofo2}
\Delta x_i^2 = - 2\lambda \sum_{j=1}^{N_{\!S}} x_i x_j + \lambda^2 
\sum_{j=1}^{N_{\!S}} \sum_{j'=1}^{N_{\!S}} x_j x_{j'}.
\end{equation}
Before the first correction the sample particles can be considered as 
statistically independent so that averaging over their displacements yields 
$\langle x_j x_{j'} \rangle = \delta_{jj'}$. So we obtain for Eq.\ 
(\ref{cofo2}) 
\begin{equation} \eqlabel{cofo3}
\langle \Delta x^2 \rangle = - 2\lambda \langle x^2 \rangle + 
\lambda^2 N_{\!S} \langle x^2 \rangle.
\end{equation}
The first term arises from the self-interaction of the particles. It is the 
{\sl coherent} contribution which effects the intrinsic cooling. The second 
term describes the impact of the other sample particles and represents the
{\sl incoherent} contribution. It results in a diffusion which increases the 
amplitudes of the particle motions, and thus can be interpreted as a heating 
of the beam. 

The change (\ref{cofo3}) is valid only for the first correction since after 
this correction the particles are correlated and the assumption $\langle x_j 
x_{j'} \rangle = \delta_{jj'}$ is no longer justified. According to Eq.\ 
(\ref{cofo1}), we can write after the correction
\[ \check{x}_i \check{x}_j = \left( x_i - \lambda \sum_{k=1}^{N_{\!S}} x_k 
\right) \left( x_j - \lambda \sum_{k'=1}^{N_{\!S}} x_{k'} \right). \]
Even if $\langle x_i x_j \rangle = 0$ for $i \not= j$ is satisfied before the 
correction, the corresponding expression $\langle \check{x}_i \check{x}_j 
\rangle$ after the correction contains non-zero terms of the form 
$- \lambda\langle x_i^2 \rangle, - \lambda \langle x_j^2 \rangle, \ldots$ 
expressing the correlations among the particles. 

On the other hand, different revolution frequencies of the particles can 
destroy these correlations. In the sample picture this process is called {\sl 
mixing} and illustrated by a change of the sample population rendering 
the sample particles again statistically independent. Hence mixing requires 
that the decay of correlations occurs faster than their build-up by the 
cooling interaction, otherwise the correlations would continuously increase 
and the collective motion of the particles would dominate. In that case the
coherent particle motions can cause an instability of the beam.

The sample picture describes the correlations by a constant, time-independent
mixing factor which serves as a measure of how fast the samples are 
rearranged. The time evolution of the displacements $x_i$, and thus the 
dynamics of the particles, is disregarded in this picture. The initial 
decrement (\ref{cofo3}) is transferred to all successive cooling steps, 
thereby neglecting the correlations which are introduced by the cooling 
process. This can be seen clearly in the calculation of the cooling rate where 
the time evolution of the mean-squared amplitude is deduced from the initial 
change (\ref{cofo3}) obtained from statistically independent particles,
\[ \frac{d}{dt} \langle x^2 \rangle \longrightarrow
\frac{\langle \Delta x^2 \rangle}{T_0}. \]

The stochastic cooling formulation in the sample picture entirely relies on 
the mixing assumption, i.e.\ the fact that the reorganization of the samples
is guaranteed. Based on this assumption, the description disregards both 
the individual particle motions and the collective motion of the beam. 
Consequently, beam stability is an indispensable prerequisite for the validity 
of the mixing assumption. 

\noindent
Finally, we quote the expression which the sample picture yields for the 
cooling rate \cite{moe}
\begin{equation} \eqlabel{cofo4}
\frac{1}{\tau} = \frac{W}{N} \left[ 2g (1 - \tilde{M}^{-2}) - g^2 (M + U) 
\right].
\end{equation}

Since the derivation does not take into account the particle motions, the 
dynamic effects of the cooling process must explicitly be introduced by 
means of empirical arguments. \newpage \noindent
In detail, we find in Eq.\ (\ref{cofo4}): 
\begin{itemize}
\item%
The factor $(1 - \tilde{M}^{-2})$ describes the undesirable migration of 
particles into other samples on their way from the pickup to the kicker. This 
{\sl unwanted mixing} only effects the self-interaction of the particles and 
degrades the coherent contribution.
\item%
The desired mixing of the sample between the kicker and pickup is represented 
by the mixing factor $M$ and interpreted as a weighting of the incoherent 
contribution. The worse the sample reorganization between successive cooling 
steps, the larger the mixing factor $M$, and the more pronounced the 
incoherent contribution to the cooling rate will be.
\item%
The quantity $U$ models the electronic noise in the cooling system and leads 
to an additional enhancement of the incoherent contribution.
\item%
Once more the importance of the bandwidth $W$ is stressed. We recognize that 
an increase of bandwidth results in a faster cooling rate and thus improves 
the cooling. For that reason the bandwidth is of fundamental concern in any
cooling system.
\end{itemize}

Newer cooling systems have bandwidths of up to 4~GHz and typically cover one 
of the frequency bands from 1-2~GHz, 2-4~GHz or 4-8~GHz. The number of 
particles in the beams varies in these systems between $10^8$ and $10^{12}$.
LEAR, for example, stores antiproton beams of $\sim 10^{10}$ particles. The 
cooling times attained in these systems reach from a few seconds up to some 
hours.

\subsubsection{The Fokker-Planck Equation}

A more rigorous mathematical representation of stochastic cooling which 
provides exact quantitative predictions about the cooling process is given by 
the Fokker-Planck equation. In this section the basic ideas of this 
formulation are compiled, intending to disclose the approximations which enter 
into the description. More details and the results following from the 
Fokker-Planck treatment of stochastic cooling can be found in Chapter 
\ref{coolpart1}. Information going beyond that can be found in \cite{bi2, ch2, 
kam, we2}.

The Fokker-Planck equation determines, for each particle, the time development 
of its probability density in the quantity being cooled. In principle these 
densities allow the calculation of all statistical moments of the relevant 
quantity and therefore provide a complete description of the cooling process. 
For statistically independent particles the probability densities are all the 
same and can be identified with the corresponding distribution function of 
the beam \cite{ch2, kam}. 

To illustrate this we once more consider transverse cooling of the betatron 
motions. In this case an appropriate variable to describe the cooling process 
is the action $I$ of each particle. In connection with this variable we 
define a density $\rho(I, t)$ such that $\rho(I, t)dI$ gives the number of 
particles with actions between $I$ and $I + dI$ at a time $t$. The density 
$\rho(I, t)$ is connected with the probability density $\psi(I, t)$ of the 
independent particles by the relation $\rho(I, t) = N \psi(I, t)$ and its time 
development is determined by the following Fokker-Planck equation \cite{we2}
\[ 
\frac{\partial}{\partial t} \rho(I, t) = 
- \frac{\partial}{\partial I} \left\{ F(I) \rho(I, t) - \frac{1}{2} 
D(I) \frac{\partial}{\partial I} \rho(I, t) \right\}.
\]
The drift and diffusion coefficients $F(I)$ and $D(I)$ are obtained from the 
expressions
\[
F(I) = \Big\langle \frac{\Delta I}{\Delta t} \Big\rangle 
\qquad\mbox{and}\qquad
D(I) = \Big\langle \frac{(\Delta I)^2}{\Delta t} \Big\rangle,
\]
where $\Delta I$ denotes the change of action in the time interval $\Delta t$, 
and the square brackets indicate an average over the initial conditions. The 
calculation of $\Delta I$ appears difficult because the instantaneous change 
$\dot{I}$ of the action depends on the actions of all $N$ particles in the 
beam. Representing the cooling interaction by an appropriate function 
$G(I_1, \ldots, I_N, t)$, we can write 
\[ \dot{I} = G(I_1, \ldots, I_N, t). \]
Integration of this expression over the time interval $\Delta t$ does not 
yield the desired value $\Delta I$ because the integrand itself depends on 
the yet unknown actions, 
\[ \Delta I = \int\limits_0^{\Delta t} \dx{t} G(I_1, \ldots, I_N, t). \]
To overcome this difficulty we are forced to use a perturbation expansion and 
hence substitute in the integrand the unperturbed actions $I^0$ which are 
derived from the known zero-order particle motions, 
\[ \Delta I = \int\limits_0^{\Delta t} \dx{t} G(I_1^0, \ldots, I_N^0, t). \]
In general this step allows the evaluation of the integral and with it the 
determination of the drift and diffusion coefficients necessary to solve the 
Fokker-Planck equation. This proceeding corresponds to Picard-Lindel\"of's 
iteration method which converges only if the function $G(I_1, \ldots, I_N, t)$ 
satisfies a Lipschitz condition \cite{bro}. In the present case this implies 
that the feedback force of the cooling interaction must remain bounded which 
can only be ensured by a stable beam motion.

It should be mentioned that the same approximation is made in the sample 
picture by applying the first correction (\ref{cofo3}) valid only for 
uncorrelated particles to all successive cooling steps. By that the 
modification of the particle motions is neglected, thus assuming that the 
particles still move along their initial, unperturbed trajectories. Therefore 
the above considerations hold true for the sample picture as well. 

Since Chapter \ref{coolpart1} will give full details of the cooling 
description by a Fokker-Planck equation, we restrict ourselves here to some 
general remarks. The information about cooling interaction and the structure 
of the cooling system are contained in the drift and diffusion coefficients so 
that their derivations include
\begin{itemize}
\itemsep0ex
\item the individual particle motions,
\item the positions of pickup and kicker and
\item the signal transfer through the electronic components.
\end{itemize}
The calculations benefit from a treatment in the frequency domain, as in 
\cite{col}. There it has been shown that the particles can be discriminated by 
their frequencies, and over long times only interact via common frequencies in 
their spectra (see also Sect.\ \ref{basepart6}). The representation in the 
frequency domain allows a clearer interpretation of the inter-particle 
correlations than a time domain description, and furthermore quantitative 
predictions about the mixing become possible. 

A particle with tune $Q$ and revolution frequency $\omega$ generates, at a 
pickup, a spectrum of lines at the frequencies $(m + Q)\omega$ with $m = 0, 
\pm 1, \pm 2, \ldots$ and can be coherently excited only at these frequencies 
(see Sect.\ \ref{basepart5}). The number of particles producing, in their 
spectra, the same frequencies $(m + Q)\omega$, determines the strength of the 
diffusion and is a measure how well mixing occurs. More generally, all dynamic 
properties of the cooling process which had to be introduced empirically into 
the sample picture emerge automatically from the Fokker-Planck description. 
This follows from the consideration of the particle motions and the 
localization of the pickup and kicker in the calculations of the drift and 
diffusion coefficients, and will become apparent in the discussion of the 
results in the Sections \ref{coolpart9} and \ref{coolpart10}. 

The cooling descriptions presented above have in common that they are 
restricted to the microscopic interactions between individual particles 
without involving the coherent beam motion. They only consider the long-term 
behaviour of the beam which is governed by the cooling, and presume beam 
stability in the active cooling system. The major deficiency of these 
descriptions is the absence of the necessary stability analysis so that they 
rely on unfounded assumptions. This thesis could remove these shortcomings. 
Here, the existing descriptions have been extended by the stability 
investigation omitted so far, providing them with a solid physical and 
mathematical foundation. 

Especially when stochastic cooling is applied to complex accelerators, such as 
HERA, the TEVATRON or the LHC, the collective beam dynamics must be thoroughly 
understood. An over-simplified description cannot rule out that problems will 
later arise from coherent beam signals, preventing the operation of the 
system. This method has already proven true in the realization of the feedback 
systems at DESY whose reliability is largely due to the fact that their 
conceptional designs are based on detailed theoretical investigations.

\chapter{Stability in Feedback Loops} 
\plabel{basepart1}

\section{Stochastic Cooling Systems as Closed Loops}
\plabel{basepart2}

\begin{figure}[htb] 
\centerline{\epsfig{figure=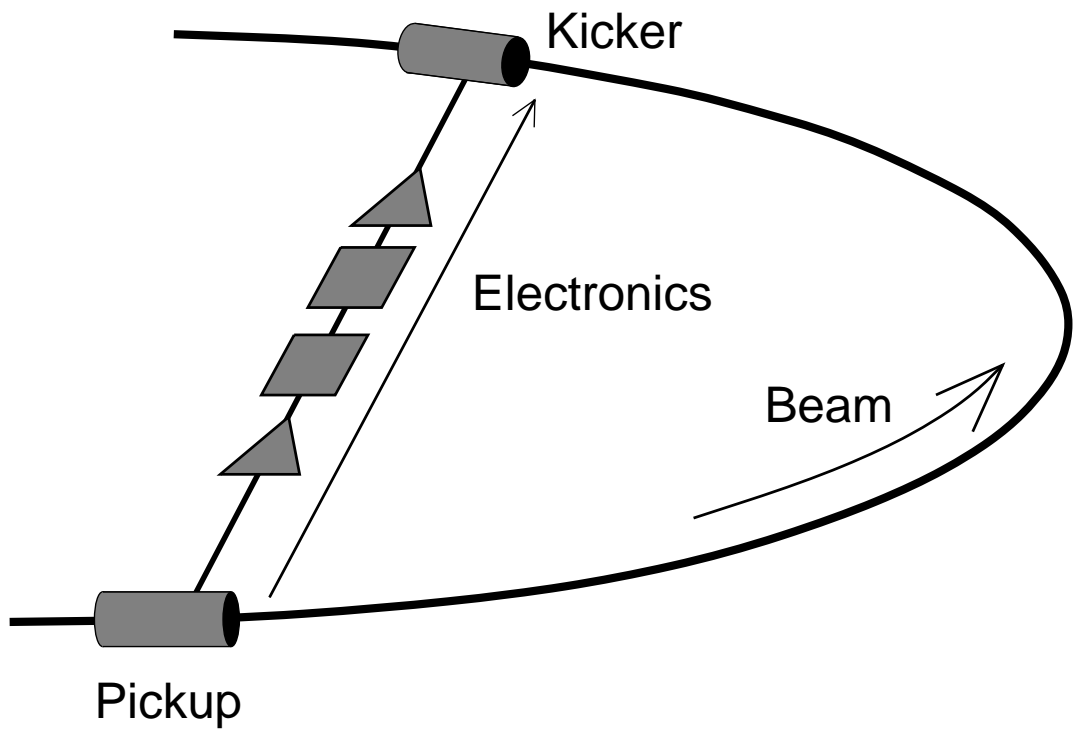,height=9.5cm}}
\caption{\label{cool}\sl Schematic view of a cooling system.}
\end{figure}

The mathematical description of stochastic cooling involves two different 
aspects. One is mainly interested in predictions about the cooling performance 
which are usually derived within a model which assumes statistically 
independent particles and thus neglects any correlations among them. Since 
stochastic cooling systems close a loop in which the beam can act upon itself 
(see Fig.\ \ref{cool}), they are potentially unstable and therefor their 
stability must be analysed. In this chapter we develop the methods which will 
later be used to investigate the collective behaviour of the particles.

\noindent
In principle stochastic cooling systems operate in the following way:
\begin{itemize}
\itemsep0ex
\item%
The particles generate signals at the pickup which contain the information 
necessary for their corrections.
\item%
The signals are processed in the electronics of the system which in general 
changes their amplitude and phase.
\item%
The kicker feeds the modified signals back to the beam in order to reduce the 
measured offsets.
\end{itemize}
Later in this chapter we will see that a thorough investigation of the 
stability has to consider the pickup and kicker as {\sl localized} objects, 
leading to a discrete time structure of the signals. 

In this thesis we study the stability within the framework of multi-bunch 
feedback theory which has been developed from the control theory of discrete 
time signals \cite{ko1, ko2}. In the following sections we introduce the basic 
concepts of the theory by means of simple examples.

\section{The Transfer Function}
\begin{figure}[htb] 
\centerline{\epsfig{figure=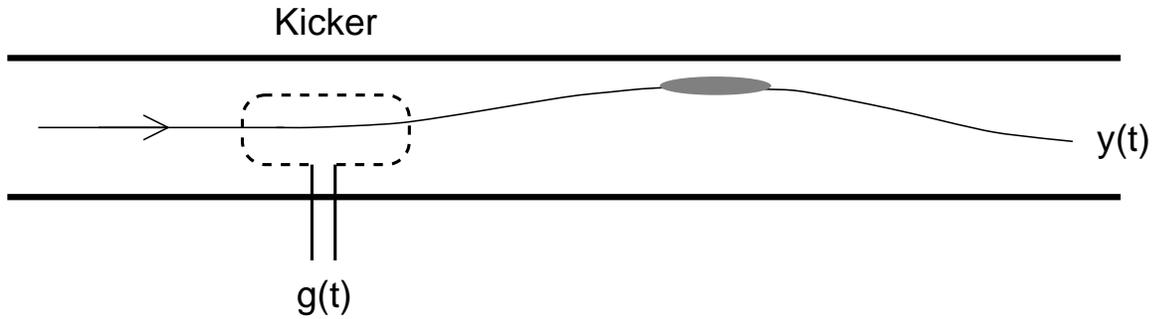,height=4cm}}
\caption{\label{kick}\sl Beam excitation by an external force.}
\end{figure}

\plabel{basepart3}
The transfer function describes the relation between cause and consequence 
of an interaction.   In Fig.\ \ref{kick}, for example, an external kicker 
excites betatron oscillations in a beam. The cause in this case is the kick 
$g(t)$ applied to the beam (at the kicker) and the consequence is the 
displacement $y(t)$ of the beam due to the resulting oscillation.

\noindent
For a linear relation between cause $g(t)$ and consequence $y(t)$ we write 
\begin{equation} \eqlabel{sys1} 
y(t) = \int\limits_{-\infty}^{+\infty}\!\dx{t'} G(t, t') g(t')
\end{equation}
which defines the impulse response $G(t,t')$. Having changed variables
according to $\{t, t'\} \longrightarrow \{t, t-t'\}$, we stipulate the 
additional properties of the impulse response $G$:
\[
\begin{array}{lcl}
\mbox{(1) $G$ is stationary} & \Longleftrightarrow & G(t, t-t') = G(t-t') \\
\mbox{(2) $G$ is causal} & \Longleftrightarrow & G(t-t') \equiv 0 
\quad\mbox{for}\quad t-t' \le 0 \\
\mbox{(3) $G$ is real}
\end{array} \]
Using Eq.\ (\ref{FTE5}), the general Fourier transformation (\ref{FTt2w}) of 
Eq.\ (\ref{sys1}) yields the product 
\[\widetilde{y}(w) = \widetilde{G}(w) \widetilde{g}(w).\]
The Fourier transform $\widetilde{G}(w)$ of the impulse response is called 
{\sl transfer function}. 
Since the time functions $y(t)$, $G(t)$ and $g(t)$ are real, it follows that 

\[
\widetilde{y}^*(w) = \widetilde{y}(-w^*) \mbox{ ,}\quad 
\widetilde{G}^*(w) = \widetilde{G}(-w^*) \quad\mbox{and}\quad 
\widetilde{g}^*(w) = \widetilde{g}(-w^*).
\]

\section{The General Fourier Transformation} 
\plabel{vFT}

\begin{figure}[htb] 
\centerline{\epsfig{figure=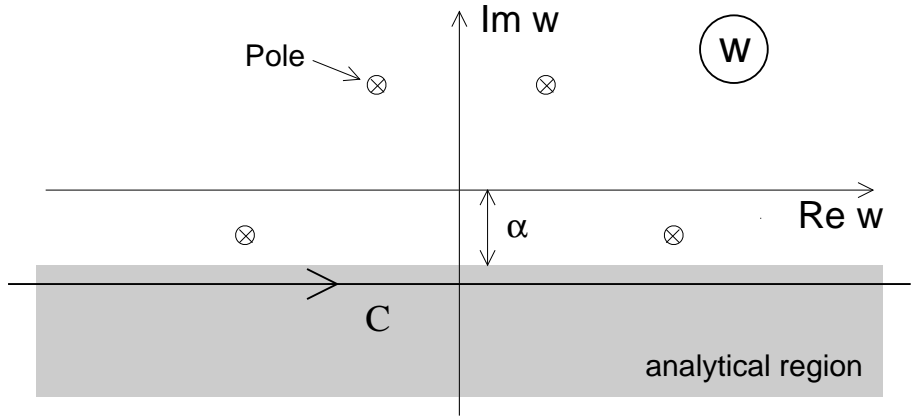,height=5.5cm}}
\caption{\label{plane}\sl The integral contour $C$ of the inverse 
transformation.}
\end{figure}

In accelerator physics it is reasonable to assume that the relevant functions 
do not grow faster in time than  
\begin{equation} \eqlabel{FTBed}
|f(t)| \le Me^{\alpha t} \quad\mbox{for}\quad t \ge 0 
\end{equation}
with real constants $M, \alpha > 0$. Allowing for complex frequencies $w$, 
the general Fourier transform of $f(t)$ is defined by \cite{mys} 
\begin{equation}\eqlabel{FTt2w}
\widetilde{f}(w) = \frac{1}{2\pi} \int\limits^\infty_0\dx{t} f(t)e^{-iwt}.
\end{equation}
$\widetilde{f}(w)$ is an analytical function at least in the lower $w$-plane 
for ${\sf Im}\,w < -\alpha$. The inverse transformation is given by
\begin{equation}\eqlabel{FTw2t}
f(t) = \int\limits_C\dx{w} \widetilde{f}(w)e^{iwt}.
\end{equation}
The path $C$ has to lie in the analytical region of $\widetilde{f}(w)$. It can 
be chosen as a straight line parallel to the real axis with ${\sf Im}\,w < 
-\alpha$ and thus is below the singularities of $\widetilde{f}(w)$. For our 
purposes we need to consider only {\sl simple poles} (see Fig.\ \ref{plane}).

The time evolution of $f(t)$ is completely determined by the poles of 
$\widetilde{f}(w)$. Evaluating the integral (\ref{FTw2t}) with the residue 
theorem yields \cite{mys}
\begin{equation} \eqlabel{sys2}
f(t) = 2\pi i \sum_k {\rm res}_{w_k} \li\{ \widetilde{f}(w) \re\} e^{iw_kt}
\end{equation}
where the sum extends over all poles $w_k$ of $\widetilde{f}(w)$. The poles 
result in oscillating terms with exponentially growing or decreasing 
amplitudes depending on the sign of the imaginary part ${\sf Im}\,w_k$.
In the upper $w$-plane the imaginary parts are positive and give damped 
solutions, since with $w_k = \omega_k + i\alpha_k$ 
\[ f(t) \sim e^{iw_kt} \sim e^{i\omega_kt}e^{-\alpha_kt} \longrightarrow 0
\qquad\mbox{for}\qquad t \longrightarrow \infty \quad\mbox{and}\quad
\alpha_k > 0. \]
Correspondingly poles in the lower $w$-plane which have ${\sf Im}\,w_k < 0$ 
describe growing, unstable solutions.

Another advantage of the general Fourier transformation appears in the 
solution of linear differential equations: the time derivatives of a function 
transform into simple algebraic expressions in the frequency domain which 
moreover include the initial conditions, e.g.\
\begin{eqnarray*}
\widetilde{\left[\dot{f}(t)\right]} & = & iw\widetilde{f}(w) - 
\frac{1}{2\pi} f(0)                                                        \\
\widetilde{\left[\ddot{f}(t)\right]} & = & -w^2\widetilde{f}(w) - 
\frac{iw}{2\pi} f(0) - \frac{1}{2\pi} \dot{f}(0)
\end{eqnarray*}
where $f(0)$ and $\dot{f}(0)$ denote the values of $f(t)$ and $\dot{f}(t)$ at
the time $t = 0$. The general expression together with some other
useful properties of the general Fourier transformation can be found in 
Appendix \ref{FTE}, and in example \ref{hoxspart2} the formalism is applied to 
a free harmonic oscillator.

\section{The Feedback Mechanism}
\plabel{basepart4}

The concept of transfer functions introduced in Section \ref{basepart3} also 
applies if the cause which modifies the beam motion originates in the beam 
itself, i.e.\ if the beam indirectly acts upon itself. In the mechanism of 
such a feedback the concept of impedance plays an important part. The 
impedance describes the relation between an excitation and the resulting 
response in frequency domain, analogous to the transfer function in time 
domain. Impedance and transfer function are connected through the general 
Fourier transformation. The general feedback mechanism splits into two basic 
steps:
\begin{itemize}
\item[(i)]%
The collective motion of the beam generates electromagnetic fields through 
the \\impedances of the storage ring.
\item[(ii)]%
These electromagnetic fields act upon the beam and thus modify its motion.
\end{itemize}
Given the right phase relation between beam motion and reacting fields 
the oscillation amplitude of the beam will continuously grow, and the beam 
motion becomes unstable. In order to predict the stability behaviour of a beam 
in a feedback loop, we need a self-consistent description of the beam motion. 
The following simple example will clarify this further. 

For that purpose we consider a harmonic oscillator with frequency $\Omega_0$ 
which acts upon itself. Being stationary for times $t \le 0$, the oscillator 
is excited by a $\delta$-pulse $g(t) = A\delta(t)$. The equation of motion can 
be written as
\[ \ddot{x}(t) + \Omega_0^2\, x(t) = F[x](t) + g(t). \]
Assuming a linear response, the reacting force $F[x](t)$ reads (see Eq.\ 
(\ref{sys1}))
\[ F[x](t) = \int\limits_0^\infty\dx{t'} G(t-t') x(t') \]
where $G(t-t')$ denotes the transfer function. Hence follows
\begin{equation} \eqlabel{sys3}
\ddot{x}(t) + \Omega_0^2\, x(t) = \int\limits_0^\infty\dx{t'}G(t-t')x(t') 
 + g(t).
\end{equation}
The general Fourier transformation of Eq.\ (\ref{sys3}) yields
\[
(-w^2 + \Omega_0^2)\, \widetilde{x}(w) = \widetilde{G}(w)\widetilde{x}(w)
 + \widetilde{g}(w).
\]
The problem of the self-consistent description of the motion reduces to the 
solution of an algebraic equation in the frequency domain. For the present 
case we easily obtain 
\begin{equation} \eqlabel{sys4}
\widetilde{x}(w) = 
\frac{\widetilde{g}(w)}{-w^2 + \Omega_0^2 - \widetilde{G}(w)}.
\end{equation}
Since the time behaviour of $x(t)$ is completely determined by the poles of 
$\widetilde{x}(w)$ (see Sect.\ \ref{vFT}), we must find the zeros in the
denominator of Eq.\ (\ref{sys4})
\[ -w^2 + \Omega_0^2 = \widetilde{G}(w). \]
This is demonstrated in the example \ref{hoxspart3} for a given impedance 
$\widetilde{G}(w)$. 

\section{Discrete Time Signals}
\plabel{basepart5}

The formalism developed so far describes continuous self-interactions, such as 
through the broad-band impedance of the storage ring. In feedback systems, 
however, the interaction takes place via pickup and kicker which are localized 
objects in the storage ring. Hence the particles generate signal pulses at the 
pickup and they sample the forces at the kicker with the revolution time which 
results in a {\sl discrete} time structure of the interaction.

\begin{figure}[htb]
\centerline{\epsfig{figure=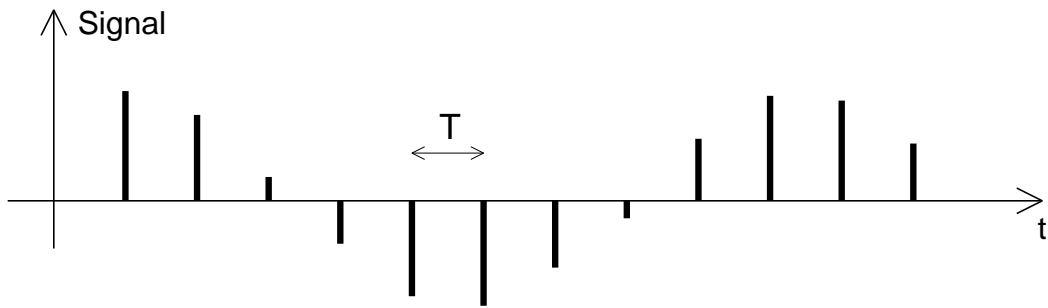,height=4cm}}
\caption{\label{pulse}\sl The discrete time signal generated by a 
transversally oscillating particle in a localized pickup.}
\end{figure}

We will illustrate this for the transverse signal at a pickup produced by a 
particle which executes betatron oscillations. Because the particle passes the 
pickup only once per turn, it does not produce a continuous signal but a 
series of amplitude-modulated $\delta$-pulses separated by its revolution time 
$T$ (see Fig.\ \ref{pulse}).
 \\
Assuming that for $t=0$ the particle is at the pickup, we write its signal 
$S(t)$ as
\begin{equation} \eqlabel{sys5}
S(t) = x(t) \sum_{(k)} T \delta(t-kT) = \sum_{(k)} x(kT) \,T \delta(t-kT)
\quad\mbox{where}\quad \sum_{(k)} \equiv \sum_{k=-\infty}^{+\infty}.
\end{equation}
$x(t)$ describes the betatron oscillation of the particle and $T$ denotes
its revolution time. The general Fourier transformation of $S(t)$ yields
\begin{equation} \eqlabel{sys6}
\widetilde{S}(w) = \frac{1}{2\pi} \int\limits_0^\infty\dx{t} S(t) e^{-iwt} =
\frac{1}{2\pi} \int\limits_0^\infty\dx{t} \sum_{(k)} x(kT) T \delta(t-kT)
e^{-iwt} = \frac{T}{2\pi} \sum_{k=0}^\infty x(kT) e^{-iwkT}.
\end{equation}
We express the displacement $x(kT)$ by its Fourier transformation 
\begin{equation} \eqlabel{sys7}
x(kT) = \int\limits_C\dx{w'} \widetilde{x}(w') e^{iw'kT} 
\end{equation}
and thus obtain
\begin{equation} \eqlabel{sys8}
\widetilde{S}(w) = \frac{1}{\omega} \sum_{k=0}^\infty \;\int\limits_C\dx{w'} 
                   \widetilde{x}(w') e^{iw'kT} e^{-iwkT} 
                 = \int\limits_C\dx{w'} \widetilde{x}(w')\, 
                   \frac{1}{\omega} \sum_{(k)} e^{i(w'-w)kT} 
\end{equation} 
where $\omega = 2\pi/T$ is the revolution frequency of the particle. 
Since $\widetilde{x}(w')$ is analytic in the lower $w$-plane, the integral
in Eq.\ (\ref{sys7}) does not contribute if $k < 0$, and thus the summation in 
Eq.\ (\ref{sys8}) can be extended over all values $k$. Using Poission's 
formula \cite{mys}
\[ \sum_{(k)} e^{i(w'-w)kT} = \omega \sum_{(m)} \delta(w'-w-m\omega) \]
we find 
\[ \widetilde{S}(w) = \int\limits_C\dx{w'} \widetilde{x}(w') \sum_{(m)} 
\delta(w'-w-m\omega) = \sum_{(m)} \widetilde{x}(w+m\omega). \]
Here we define the periodic function $\hat{x}(w)$ by
\[ \hat{x}(w) = \sum_{(m)} \widetilde{x}(w+m\omega). \]
which has the period $\omega$, since for any integer $l$ 
\[ \hat{x}(w+l\omega) = \sum_{m=-\infty}^\infty \widetilde{x}(w+[m+l]\omega)
  = \sum_{k=-\infty}^\infty \widetilde{x}(w+k\omega) = \hat{x}(w) \]
where $k=m+l$ has been substituted.

The spectrum of a particle which generates a pulsed signal at a pickup in time 
with its revolution time $T$ consists of an infinite series of lines separated 
by its revolution frequency $\omega$. An equivalent result follows for a 
particle with revolution frequency $\omega$ which samples the force at a 
localized kicker. Given that the particle oscillates with the betatron 
frequency $\Omega = {\sf Re}\,w$, it can be excited coherently at any 
frequency $\Omega + m\omega$ with $m = 0, \pm 1, \pm 2, \ldots$.

The single particle results can easily be extended to $N$ particles equally 
distributed around the storage ring. Since in this case the signals at the 
pickup are separated by the time difference $\Delta t = T/N$, the signal 
spectrum is periodic with the frequency $2\pi/\Delta t = 2\pi/(T/N) = 
N\omega$. For the same reasons a kicker can excite a coherent oscillation of 
the $N$ particles only at frequencies $\Omega + lN\omega$ with $l = 0, \pm 1, 
\pm 2, \ldots$.

\section{Overlapping Frequency Bands}
\plabel{basepart6}

In the previous section we have assumed equal revolution frequencies which is 
unlikely for the particles in an unbunched beam. Owing to the momentum spread 
of the particles which is always present in an unbunched beam, the revolution 
frequencies are spread over an interval. The signal spectrum generated by the 
beam particles at a pickup now consists of a series of {\sl frequency bands} 
which mirror the frequency distribution of the particles. The width of the 
bands increases to higher frequency so that beyond a certain frequency they 
overlap.
\begin{figure}[ht]
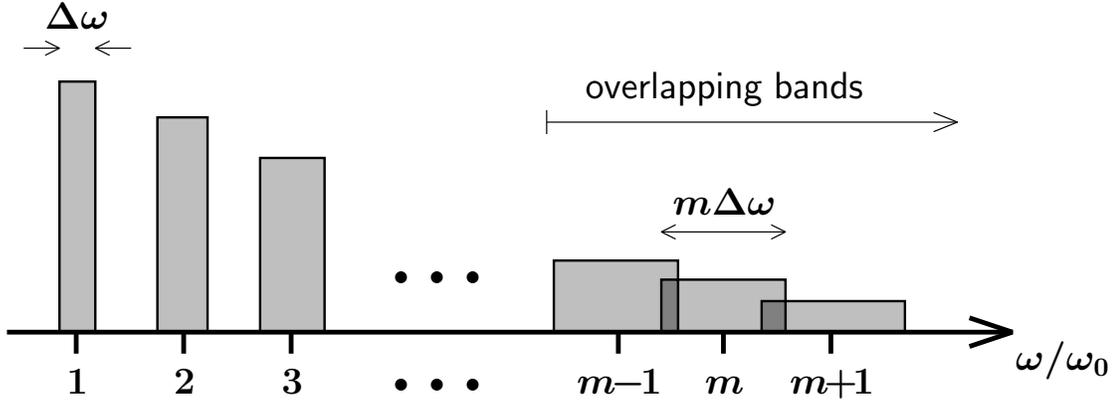

\include{bands}
\caption{\label{bands}\sl Band spectrum for a rectangular frequency 
distribution.}
\end{figure}

As an example we will discuss the spectrum of the longitudinal beam signal. 
For the sake of simplicity we assume a rectangular distribution of the 
revolution frequencies with a center frequency $\omega_0$. So the frequencies 
are equally spread over the interval $\Delta\omega = \omega_{max} - 
\omega_{min}$ where $\omega_{max}$ and $\omega_{min}$ denote the upper and 
lower cut-off frequency of the distribution. 

A particle $j$ of the distribution with a revolution frequency $\omega_j$ 
generates a spectrum of lines at the revolution harmonics $m\omega_j$, $m = 0, 
\pm1, \pm2, \ldots$ (see Sect.\ \ref{basepart5}). Consequently the lines for 
the lower and upper cut-off frequency of the distribution appear at the 
frequencies $m\omega_{min}$ and $m\omega_{max}$ respectively, so that the 
width of the revolution band $m$ is given by $m\omega_{max} - m\omega_{min} = 
m(\omega_{max} - \omega_{min}) = m\Delta\omega$. Thus the bands become broader 
for increasing $|m|$. Since each band contains the same number of particles, 
the amplitude of the bands decreases as their width increases. This behaviour 
is shown in Fig.\ \ref{bands}.

\begin{figure}[h]
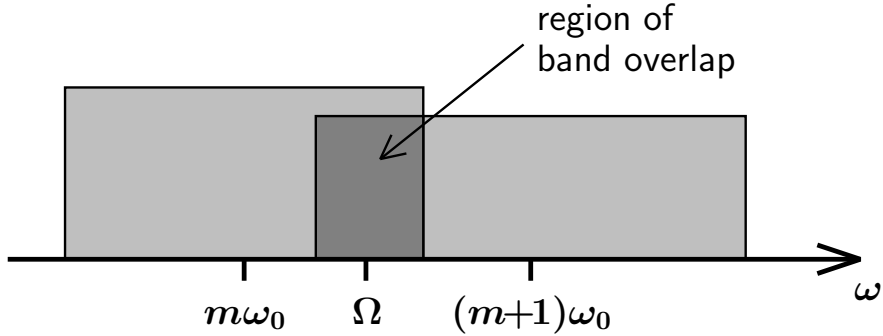

\include{overlap}
\caption{\label{overlap}\sl Two overlapping bands with harmonic numbers $m$ 
and $m\!+\!1$. In the overlap region revolution frequencies $\omega_j$ and 
$\omega_{j'}$ exist which satisfy $\Omega = m\omega_j = (m\!+\!1)\omega_{j'}$.}
\end{figure}

For sufficiently high harmonic numbers $m$ the width of the bands will satisfy 
$m\Delta\omega \stackrelss{>}{\sim} \omega_0$, and hence adjacent frequency 
bands will overlap (see Fig.\ \ref{bands}). In this region of band overlap, 
particles with different revolution frequencies $\omega_j, \omega_{j'}, 
\omega_{j''}, \ldots$ from different revolution bands $m, m', m'', \ldots$ 
will contribute to the spectrum at the same frequency $\Omega$ if their 
harmonics fulfil the resonance condition $\Omega = m\omega_j = m'\omega_{j'} = 
m''\omega_{j''} = \ldots$. In this case the particles can no longer be 
distinguished by their revolution frequencies $\omega_j$. Fig.\ \ref{overlap} 
shows the overlap of two adjacent bands.

An analogous result follows for a localized force which the beam particles
sample with their revolution times $T_j$. A periodically changing force with a 
frequency $\Omega$ falling into the overlap region of the bands will 
coherently excite particle with different revolution frequencies $\omega_j,
\omega_{j'}, \omega_{j''}, \ldots$ in different revolution bands $m, m', m'', 
\ldots$ if they meet the resonance condition $\Omega = m\omega_j = 
m'\omega_{j'} = m''\omega_{j''} = \ldots$.

Thus, in feedback systems the motion of particles with different revolution 
frequencies can couple through overlapping frequency bands. A particle with 
frequency $\omega_j$ generating signals at frequencies $\Omega = m\omega_j$ 
can act coherently on a particle with frequency $\omega_{j'}$, provided that 
the frequency bands $m$ and $m'$ (which satisfy the condition $m\omega_j = 
\Omega = m'\omega_{j'}$) overlap within the bandwidth of the system.

\chapter{Stability of the Unbunched Beam}
\plabel{stabpart1}

\section{The Existing Treatment of the Problem}
\plabel{stabpart2}

The formalism developed in the previous chapter is now used to investigate 
the stability of an unbunched beam subjected to transverse stochastic cooling.
To this end we must find a self-consistent description of the beam motion 
which treats the pickup and kicker as localized objects. In Section 
\ref{basepart6}, we saw that this localization together with a frequency 
distribution in the beam leads to a band spectrum which extends over the 
entire frequency range. The width of the bands increases to higher 
frequencies, and beyond a certain frequency the bands overlap. The feedback 
interaction in the region of overlapping frequency bands couples the motions 
of almost all particles, complicating the derivation of the self-consistent 
solution of the beam motion. Because of this, most of the work on stochastic 
cooling disregards the localization of pickup and kicker, and thus obtains 
predictions about the beam stability which are only valid in the frequency 
range of non-overlapping bands \cite{moe, wei, der}. Since the stochastic 
cooling operation profits to a certain degree from these overlapping bands, 
the results have only limited physical relevance. Although self-consistent 
solutions have been derived which thoroughly take into account the 
localization \cite{bi1, me2}, the beam stability in such a cooling system has 
not yet been investigated.

In this chapter, we present a careful analysis of the beam motion in order to 
predict the stability of the beam. First, we derive the equation of motion of 
a single particle undergoing stochastic cooling: the driving term in this 
equation contains the feedback force at the kicker which depends on the 
displacements of all particles at the pickup. For the stability analysis the 
decisive factor is not the individual particle motion, but the {\sl 
collective} motion of all particles which is determined by averaged 
macroscopic quantities of the beam. Summing the single particle equations over 
all beam particles and ignoring the discreteness of the particles, we obtain 
equations describing the motion of a continuous beam. These equations are then 
solved by means of the methods discussed in the previous section, resulting 
in the self-consistent solution finally used to derive the stability criteria 
of the beam. 

\section{Particle Motion in an Unbunched Beam}
\plabel{stabpart3}

\begin{figure}[ht]
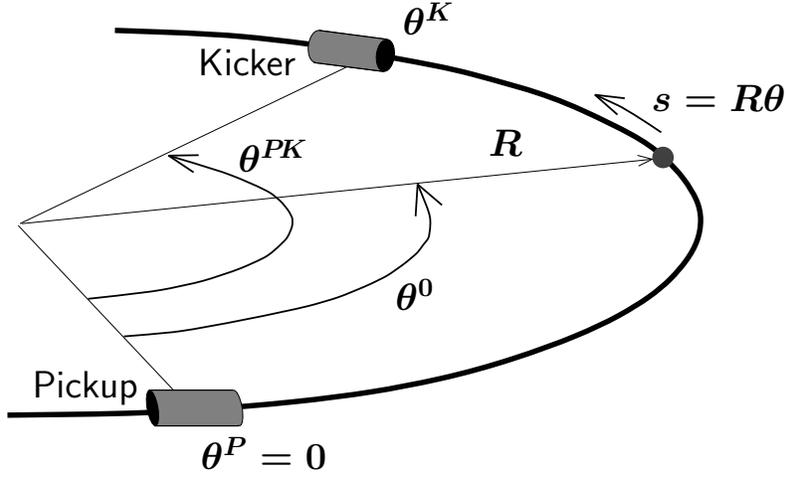

\include{coords}
\caption{\label{stabfig1}\sl The longitudinal coordinates of a particle.}
\end{figure}

We consider a single particle which is freely circulating in the longitudinal 
direction with a revolution frequency $\omega$. If $\theta^0$ denotes the 
initial azimuth at the time $t = 0$ (see Fig.\ \ref{stabfig1}), the azimuthal
coordinate $\theta(t)$ of the particle can be written as $\theta(t) = \omega t 
+ \theta^0.$

Assuming uncoupled horizontal and vertical motions of the particle, we need 
not distinguish between the planes and we will refer to both as simply 
transverse motion. 
To describe the transverse motion we define the quasi-time $\tau$ by \cite{ko1}
\begin{equation} \eqlabel{stab1}
\tau(t) = \frac{\varphi(s[t])}{\Omega}
\end{equation}
where $\Omega$ denotes the betatron frequency. $\varphi(s[t])$ describes the 
betatron phase advance of the particle as a function of its orbit coordinate 
$s[t]$. The reference point is chosen such that $\tau(0) = 0$. 
Since the betatron phase advance $\varphi$ of a particle is the same for each 
revolution, it follows 
\begin{equation} \eqlabel{stab2} 
\tau(t + T) = \tau(t) + T
\end{equation}
with the revolution time $T$. It can also be shown that
\begin{equation} \eqlabel{stab3}
\frac{d\tau}{dt} = \frac{\omega}{\Omega} \frac{R}{\beta(s)}
\end{equation}
where $\beta(s)$ is the beta function and $R$ the radius of the accelerator.

It is convenient to express the transverse displacement $X(\tau)$ of the 
particle in Courant-Snyder variables which are defined by
\[ x(\tau[t]) = \frac{X(\tau[t])}{\beta(s[t])} \]
since then the transverse motion of the particle obeys the equation of motion 
of a harmonic oscillator.

\section{The Signal at a Pickup}
\plabel{stabpart4}

We now derive the signal which the beam particles generate in a pickup at the 
position $\theta^P = 0$. The particle $j$ passes the pickup at times $t_k = 
t_j^0 + kT_j$ with $k = 0, \pm1, \pm2, \ldots$ where $t_j^0$ is the time it 
takes to cover its initial azimuth $\theta_j^0$, i.e.\ $\theta_j^0 = 
\omega_j t_j^0$. Hence the signal $S(t)$ at the pickup can be written as 
\begin{equation} \eqlabel{stab4}
S(t) = \sqrt{\beta_P} \sum_{j=1}^N x_j\li( \tau_j(t - t_j^0) \re) 
       \sum_{(k)} T_j\,\delta(t - t_j^0 - kT_j).
\end{equation}
Here, $\beta_P$ refers to the value of the beta function at the pickup, and 
the notation $(k)$ implies a summation over all integers,
\[\sum_{(k)} \equiv \sum_{k=-\infty}^{+\infty}.\]
Writing the time dependence of $x_j$ in this way, we explicitly take into
account the betatron phase advance which corresponds to the initial azimuthal 
distance $\theta_j^0$ of the particles from the pickup, and thus ensure that 
the transverse displacements of the particles at the pickup are summed with 
the proper phase. Using the periodicity (\ref{stab2}), we obtain for 
Eq.\ (\ref{stab4})
\begin{eqnarray} \eqlabel{stab5}
S(t) & = & \sqrt{\beta_P} \sum_{j=1}^N \sum_{(k)} 
           x_j\!\left( \tau_j(kT_j) \right) 
           T_j\,\delta(t - t_j^0 - kT_j)                   \nonumber\\
     & = & \sqrt{\beta_P} \sum_{j=1}^N \sum_{(k)} 
           x_j\!\left( \tau_j(0) + kT_j \right)
           T_j\,\delta(t - t_j^0 - kT_j)                   \nonumber\\
     & = & \sqrt{\beta_P} \sum_{j=1}^N \sum_{(k)} x_j(kT_j)\, 
           T_j\,\delta(t - t_j^0 - kT_j)                   \nonumber\\
     & = & \sqrt{\beta_P} \sum_{j=1}^N \sum_{(k)} x_j(t - t_j^0)\, 
           T_j\,\delta(t - t_j^0 - kT_j).
\end{eqnarray}
Inserting for $x_j(t - t_j^0)$ the general Fourier transform 
\[ 
x_j(t - t_j^0) = \int\limits_C\dx{w} \widetilde{x}_j(w) e^{iw(t - t_j^0)}, 
\]
and expanding the periodic $\delta$-function in a Fourier series,
\[ \sum_{(k)} T_j\, \delta(t - t_j^0 - kT_j) =  
   \sum_{(m)} e^{im\omega_j(t - t_j^0)}, \]
yields 
\begin{equation} \eqlabel{stab6}
S(t) = \sqrt{\beta_P} \sum_{j=1}^N \sum_{(m)} \int\limits_C\dx{w}
       \widetilde{x}_j(w) e^{-iwt_j^0} e^{-im\theta_j^0} e^{i(w + m\omega_j)t}.
\end{equation}
At this point it is necessary to explain the model of the unbunched beam on 
which the further investigations of the collective beam motion is based.

\section{The Model of the Unbunched Beam}
\plabel{stabpart5}

\subsection{Decomposition of the Motion into Coherent Modes}
\plabel{stabpart6}

In order to describe the {\sl collective} motion of the beam, we assume that 
the initial azimuths $\theta_j^0$ of the $N$ particles are equally distributed 
around the ring. Of course any other distribution is just as possible, but 
with a view to analyse the macroscopic beam motion the equally spaced 
distribution is the simplest choice which also seems physically reasonable. 
Within this assumption, the betatron oscillations of the particles can be 
decomposed with respect to the initial azimuths, 
\begin{equation} \eqlabel{modes}
\widetilde{x}_j(w) e^{-iwt_j^0} = \frac{1}{N} \sum_{l=-N/2}^{N/2-1}
\widetilde{C}_l(w) e^{il\theta_j^0}. 
\end{equation}
Because the expansion will always be used in the limit $N \to \infty$ (see 
Sect.\ \ref{stabpart7}), an even $N$ can be assumed without loss of 
generality. Performing this limit, the {\sl symmetric} summation ensures 
that the index range equally extends to $\pm\infty$ whilst its center remains 
fixed at the origin at $l = 0$, allowing a clear identification of the 
conjugate expansion coefficients. 

The exponential functions $e^{il\theta_j^0}$ in Eq.\ (\ref{modes}) form a 
complete mutually orthogonal set, and hence the behaviour of 
$\widetilde{x}_j(w)$ is completely determined by the coefficients 
$\widetilde{C}_l(w)$. Given that all beam particles have the same revolution 
frequency $\omega_0$, the expansion leads to the normal modes of the beam and 
the coefficients $\widetilde{C}_l(w)$ yield the mode amplitudes \cite{ko1}. 
In order to predict the stability of the collective beam motion it is 
sufficient to show that the expansion coefficients are bounded: we thus 
restrict further investigations to the stability of these coefficients. 

\subsection{The Limit of the Continuous Beam}
\plabel{stabpart7}

If the initial azimuths of the $N$ particles are equally distributed around 
the ring, adjacent particles will be separated by $\Delta\theta^0 = 2\pi/N$ so 
that the initial azimuth of particle $j$ can be written as $\theta_j^0 = 
j\cdot\Delta\theta^0$. For a large number of particles, i.e.\ in the limit 
$N \to \infty$, the distance $\Delta\theta^0$ becomes infinitesimal small, 
suggesting the definition of a continuous azimuthal distribution function. The 
summations over the initial azimuths $\theta_j^0 = j \cdot \Delta \theta^0$ 
can now be replaced by integrations over a continuous variable $\theta^0$.

By the same argument the summations with respect to the revolution frequencies 
$\omega_j$ can be substituted by corresponding integrations. When the 
frequencies of the particles are spread over an interval $\Delta\omega$, the
mean frequency difference is given by $\delta\omega = \Delta\omega/N$. In 
the limit of large particle numbers this difference approaches zero and we 
can assume a continuous frequency distribution $f_0(\omega)$. 

Since the particle index $j$ marks both the initial azimuth $\theta_j^0$ 
and the revolution frequency $\omega_j$, the substitution of a summation over 
particles leads to a double integral over the corresponding variables 
$\theta^0$ and $\omega$. Defining a distribution function $\bar{f}(\theta^0, 
\omega)$ normalized to unity, we can write for any function $F(\theta^0, 
\omega)$
\[ 
\sum_{j=1}^N F(\theta_j^0, \omega_j) \longrightarrow 
N \int\limits_0^{2\pi}\dx{\theta^0} \hspace{-0.8em} 
\int\limits_0^\infty \dx{\omega} \bar{f}(\theta^0, \omega) F(\theta^0, \omega).
\]
Assuming that the revolution frequencies and initial azimuths are 
uncorrelated, the distribution function $\bar{f}(\theta^0, \omega)$ 
factorizes, and hence reads for equally distributed initial azimuths
\[ \bar{f}(\theta^0, \omega) = f_\theta(\theta^0) f_0(\omega)
 = \frac{1}{2\pi} f_0(\omega) \qquad\mbox{with}\qquad 
\int\limits_0^\infty \dx{\omega} f_0(\omega) = 1. \]
Under these conditions, the substitution can be written as
\begin{equation} \eqlabel{sum2int} 
\sum_j F(\theta_j^0, \omega_j) \longrightarrow N \int\!\frac{d\theta^0}{2\pi}\!
\int\dx{\omega} f_0(\omega) F(\theta^0, \omega).
\end{equation}

We will also give a physical argument for describing the beam by continuous 
variables. 
Electronic systems have only a limited bandwidth, and thus are unable to 
resolve individual particles of a dense beam. By averaging over many 
particles, these systems smooth out the inherent discrete time structure of 
the signals which hence change continuously in time. Thus the measured signals 
in an azimuthal equally distributed beam do not reveal the discreteness of the 
particles, and it seems reasonable to assume a continuous beam. 

The change to a continuous frequency distribution is justified by the fact 
that the typical frequency difference $\delta\omega$ can be resolved only 
after a time $t \sim 1/\,\delta\omega$ which lies far beyond the time scales 
considered in the following investigations. Moreover, the frequencies have to 
be constant over this time interval, requiring a stability of the accelerator 
components which technically cannot be realized. Fluctuations in the 
electronics (e.g.\ power-supply ripples or electronical noise) make it 
impossible to attach fixed frequencies to the particles.

\section{The Signal Transfer from Pickup to Kicker}
\plabel{stabpart8}

We now apply the continuous beam model to the signal $S(t)$ at the pickup. 
Starting at Eq.\ (\ref{stab6}), we insert the expansion (\ref{modes}) and 
obtain
\[
S(t) = \sqrt{\beta_P} \sum_j \sum_{(m)} \int\dx{w} \frac{1}{N} \sum_l 
\widetilde{C}_l(w) e^{il\theta_j^0} e^{-im\theta_j^0} e^{i(w + m\omega_j)t}.
\]
For a continuous beam the particle sum can be replaced according to Eq.\ 
(\ref{sum2int}), yielding 
\[
S(t) = \sqrt{\beta_P} \sum_{(m)} \sum_{(l)} \int\!\! dw\! 
\int\!\frac{d\theta^0}{2\pi}\! \int\dx{\omega} f_0(\omega)
\widetilde{C}_l(w) e^{i(l-m)\theta^0} e^{i(w + m\omega)t}.
\]
Since
\[ \int\!\frac{d\theta^0}{2\pi}\; e^{i(l-m)\theta^0} = \delta_{ml}, \]
it follows that 
\[
S(t) = \sqrt{\beta_P} \sum_{(m)} \int\!\!dw\! \int\dx{\omega} f_0(\omega) 
\widetilde{C}_m(w) e^{i(w + m\omega)t}.
\]
The general Fourier transform of $S(t)$ reads
\begin{equation} \eqlabel{stab7}
\widetilde{S}(w) = \frac{\sqrt{\beta_P}}{2\pi} \sum_{(m)} \int\dx{\omega}
                   f_0(\omega) \widetilde{C}_m(w - m\omega).
\end{equation}
Assuming a linear impulse response $G(t)$ which models the complete cooling 
system including the pickup and kicker, we write the signal transfer from the 
pickup to kicker as (see Sect.\ \ref{basepart3}) 
\[ G\left[S\right](t) = \int\limits_0^\infty\dx{t'} G(t - t') S(t'). \]
Using Eq.\ (\ref{FTE5}), the Fourier transformation yields
\[ \widetilde{G\left[S\right]}(w) = \widetilde{G}(w)\widetilde{S}(w), \]
so that after the inverse transformation the transferred signal is given by 
\begin{equation} \eqlabel{stab8}
G\left[S\right](t) = \int\dx{w} e^{iwt} \widetilde{G}(w)\widetilde{S}(w).
\end{equation}
Eqs.\ (\ref{stab7}) and (\ref{stab8}) determine the force acting on the 
beam at the kicker.

\section{The Self-Consistent Solution to the Beam Motion}
\plabel{stabpart9}

\subsection{The Motion of a Single Particle}
\plabel{stabpart10}

\subsubsection{The Equation Of Motion}

In the notation of Section \ref{stabpart3} the equation of motion of the 
particle $j$ reads
\begin{equation} \eqlabel{stab9}
\frac{d^2}{d\tau_j^2}\, x_j\li( \tau_j(t - t_j^0) \re) + 
\Omega_j^2\, x_j\li( \tau_j(t - t_j^0) \re) =
\Omega_j^2 \beta_K^{3/2} \Big( F_j(t) + g_j(t) \Big)
\end{equation}
where $\beta_K$ is the value of the beta function at the kicker. $F_j(t)$ and 
$g_j(t)$ denote the feedback force and an external excitation respectively. 
Because the particle passes the localized kicker only at discrete times, the 
forces $F_j(t)$ and $g_j(t)$ are sampled quantities.

\subsubsection{Sampled Forces at the Kicker}

Let $\theta^K$ denote the position of the kicker, and $\theta^\DK = \theta^K - 
\theta^P$ its azimuthal distance to the pickup (see Fig.\ \ref{stabfig1}). 
The particle samples the forces at the kicker at times $t_k = t_j^0 + t_j^\DK 
+ kT_j$ with $k = 0, \pm 1, \pm 2, \ldots$ where $t_j^\DK$ is the transit time 
of the particle between pickup and kicker, i.e.\ $\theta^\DK = \omega_j 
t_j^\DK$. Hence the sampled forces $F_j(t)$ and $g_j(t)$ can be written as 
\begin{eqnarray} \eqlabel{stab10}
F_j(t) & = & \sum_{(k)} T_j\,\delta(t - t_j^0 - t_j^\DK - kT_j)\,
             G\left[S\right](t) \nonumber\\
       & = & \sum_{(k)} T_j\,\delta(t - t_j^0 - t_j^\DK - kT_j)\,
             G\left[S\right](t_j^0 + t_j^\DK + kT_j)
\end{eqnarray}
and
\begin{eqnarray} \eqlabel{stab11}
g_j(t) & = & \sum_{(k)} T_j\,\delta(t - t_j^0 - t_j^\DK - kT_j)\,
             g(t) \nonumber\\
       & = & \sum_{(k)} T_j\,\delta(t - t_j^0 - t_j^\DK - kT_j)\,
             g(t_j^0 + t_j^\DK + kT_j).
\end{eqnarray}
$G\left[S\right](t)$ is given by Eq.\ (\ref{stab8}) and $g(t)$ describes the 
continuous external excitation.

\subsubsection{Transformation of the Equation of Motion}

By substituting $\bar{\tau}_j = \tau_j(t - t_j^0)$ in Eq.\ (\ref{stab9}) and 
observing that $d\bar{\tau}_j = d\tau_j$, we obtain 
\begin{equation} \eqlabel{stab12}
\frac{d^2}{d\bar{\tau}_j^2}\, x_j(\bar{\tau}_j) + \Omega_j^2\, 
x_j(\bar{\tau}_j) = \Omega_j^2 \beta_K^{3/2} \Big( F_j(t) + g_j(t) \Big).
\end{equation}
The general Fourier transform with respect to the quasi-time $\bar{\tau}_j$ is 
defined by 
\[
\widetilde{f}(w) = \frac{1}{2\pi} \int\limits_0^\infty \dx{\bar{\tau}_j}
f(\bar{\tau}_j) e^{-iw\bar{\tau}_j}
\]
and thus the transformed equation of motion (\ref{stab12}) reads 
\begin{equation} \eqlabel{stab13}
(-w^2 + \Omega_j^2)\,\widetilde{x}_j(w) =
\Omega_j^2 \beta_K^{3/2} \left(\widetilde{F}_j(w) + \widetilde{g}_j(w)\right), 
\end{equation}
where
\begin{equation} \eqlabel{stab14}
\widetilde{F}_j(w) = \int \dx{\bar{\tau}_j} F_j(t) e^{-iw\bar{\tau}_j} 
\end{equation}
and
\begin{equation} \eqlabel{stab15}
\widetilde{g}_j(w) = \int \dx{\bar{\tau}_j} g_j(t) e^{-iw\bar{\tau}_j}.
\end{equation}

\subsubsection{Transformation of the Feedback Force}

Now the Fourier transform of the feedback force $\widetilde{F}_j(w)$ is 
further evaluated. Starting with Eq.\ (\ref{stab14}), 
\begin{eqnarray*}
\widetilde{F}_j(w) 
& = & \int \dx{\bar{\tau}_j} F_j(t) e^{-iw\bar{\tau}_j}     \\
& = & \int \dx{\tau_j} F_j(t) e^{-iw\tau_j(t - t_j^0)},
\end{eqnarray*}
we replace the integration over $\tau_j$ by an integration over $t$ by using 
the relation (\ref{stab3}) and obtain by Eq.\ (\ref{stab10}) 
\[
\widetilde{F}_j(w) = \frac{\omega_j}{\Omega_j}\frac{R}{\beta_K} \int \dx{t}
                     \sum_{(k)} T_j\,\delta(t - t_j^0 - t_j^\DK - kT_j)\,
                     G\left[ S \right](t_j^0 + t_j^\DK + kT_j)
                     e^{-iw\tau_j(t - t_j^0)}.
\]
Using of the periodicity (\ref{stab2}), we find for the quasi-time in the 
exponent after integration 
\[ \tau_j(t_j^0 + t_j^\DK + kT_j - t_j^0) = \tau_j(t_j^\DK + kT_j) =
\tau_j^\DK + kT_j \]
where $\tau_j^\DK = \tau_j(t_j^\DK)$. Together with Eq.\ (\ref{stab8}), it 
follows 
\begin{eqnarray} \eqlabel{stab16}
\widetilde{F}_j(w) & = & 
      \frac{1}{\Omega_j} \frac{2\pi R}{\beta_K}\, \sum_{(k)}
      e^{-iwkT_j} e^{-iw\tau_j^\DK} 
      \int \dx{w'} \widetilde{G}(w') \widetilde{S}(w')
      e^{iw'(t_j^0 + t_j^\DK + kT_j)}                     \nonumber\\
& = & e^{-iw\tau_j^\DK} \frac{1}{\Omega_j} \frac{2\pi R}{\beta_K}
      \int \dx{w'} \widetilde{G}(w') \widetilde{S}(w') e^{iw' t_j^\DK}
      e^{iw' t_j^0}\, \sum_{(k)} e^{i(w' - w)kT_j}.
\end{eqnarray}
Since
\[ 
\sum_{(k)} e^{i(w' - w)kT_j} = \omega_j \sum_{(m)} \delta(w'- w - m\omega_j),
\]
this reduces to 
\begin{equation} \eqlabel{stab17}
\widetilde{F}_j(w) = 
     e^{-iw\tau_j^\DK} \;\frac{\omega_j}{\Omega_j} \frac{2\pi R}{\beta_K}\,
     \sum_{(m)} \widetilde{G}(w + m\omega_j) \widetilde{S}(w + m\omega_j)
     e^{i(w + m\omega_j)t_j^\DK} e^{i(w + m\omega_j)t_j^0}.
\end{equation}
Finally, we define a periodic function $\widehat{F}_j(w)$ by 
\begin{equation} \eqlabel{stab18}
\widehat{F}_j(w) =  e^{iw\tau_j^\DK} \widetilde{F}_j(w),
\end{equation}
which has the period $\omega_j$ because $\widehat{F}_j(w + l\omega_j) = 
\widehat{F}_j(w)$ for any integer $l$ (see Sect.\ \ref{basepart5}).

\subsubsection{Transformation of the External Excitation}

Proceeding with Eqs.\ (\ref{stab11}) and (\ref{stab15}) in the same way, we 
obtain an expression similar to the result (\ref{stab16})
\begin{eqnarray*}
\widetilde{g}_j(w) & = & 
      \frac{1}{\Omega_j} \frac{2\pi R}{\beta_K}\, \sum_{(k)}
      e^{-iwkT_j} e^{-iw\tau_j^\DK} \int \dx{w'} 
      \widetilde{g}(w') e^{iw'(t_j^0 + t_j^\DK + kT_j)}              \\
& = & e^{-iw\tau_j^\DK} \frac{1}{\Omega_j} \frac{2\pi R}{\beta_K}
      \int \dx{w'} \widetilde{g}(w') e^{iw' t_j^\DK} e^{iw' t_j^0}\,
      \sum_{(k)} e^{i(w' - w)kT_j}                         
\end{eqnarray*}
so that 
\begin{equation} \eqlabel{stab19}
\widetilde{g}_j(w) =
     e^{-iw\tau_j^\DK} \;\frac{\omega_j}{\Omega_j} \frac{2\pi R}{\beta_K}\, 
     \sum_{(m)} \widetilde{g}(w + m\omega_j) 
     e^{i(w + m\omega_j)t_j^\DK} e^{i(w + m\omega_j)t_j^0}.
\end{equation}
Analogously, we define the periodic function $\widehat{g}_j(w)$ by
\begin{equation} \eqlabel{stab20}
\widehat{g}_j(w) =  e^{iw\tau_j^\DK} \widetilde{g}_j(w)
\end{equation}
with the property $\widehat{g}_j(w + l\omega_j) = \widehat{g}_j(w)$.

\subsection{The Transition to the Continuous Beam}
\plabel{stabpart11}

The single particle results which have been obtained thus far are now employed 
to derive the self-consistent solution to the collective beam motion in the 
framework of the continuous beam model established in Section \ref{stabpart5}. 
At this time, however, we will only present the important results and refer 
the detailed calculations to Appendix \ref{somopart1}.

Before proceeding, it is necessary to make some approximations which are of 
use in the derivations.
\begin{itemize} \plabel{Qtaudef}
\item[(i)] %
We assume zero chromaticity, $\xi = 0$. Then all particles have the 
same tune $Q$ and their betatron frequencies are given by $\Omega_j = 
Q\omega_j$.\footnote{This means no essential restriction because a 
frequency-dependent tune could easily be taken into account in the further
formalism by the substitution $Q \to Q(\omega)$. However, this would produce 
unnecessary complex equations which basically do not provide any new physical 
insight.}
\item[(ii)] %
If $\varphi^\DK$ denotes the betatron phase advance of the particles between 
pickup and kicker, we can write $\tau_j^\DK = \varphi^\DK/Q\omega_j$.
\end{itemize}
The derivation starts at the transformed equation of motion (\ref{stab13})
which by Eqs.\ (\ref{stab18}) and (\ref{stab20}) can be written as
\begin{equation} \eqlabel{stab20b}
(-w^2 + \Omega_j^2)\,\widetilde{x}_j(w) = \Omega_j^2 \beta_K^{3/2} 
e^{-iw\tau_j^\DK} \left( \widehat{F}_j(w) + \widehat{g}_j(w) \right).
\end{equation}
From this equation we can derive the self-consistent solution to the beam
signal $S(t)$ at the pickup. According to Eq.\ (\ref{stab26a}), the general 
Fourier transform $\widetilde{S}(w)$ of $S(t)$ is  
\begin{equation} \eqlabel{stab26}
\widetilde{S}(w) = \frac{\widetilde{R}(w) \widetilde{g}(w)}
{1 - \widetilde{R}(w) \widetilde{G}(w)}
\end{equation}
with the definition
\begin{equation} \eqlabel{stab25a}
\widetilde{R}(w) = \kappa N \int\limits_0^\infty \dx{\omega} f_0(\omega)\,
\omega^2 \sum_{(m)} \frac{e^{-i(w + m\omega)\varphi^\DK/Q\omega}}
{(Q\omega)^2 - (w + m\omega)^2}\, e^{iw\theta^\DK/\omega}
\end{equation}
where $\kappa = RQ\sqrt{\beta_{\!P}\,\beta_{\!K}}$.

Using Eq.\ (\ref{stab26}), we can immediately derive the force acting on the 
beam, and a determination of the coefficients $\widetilde{C}_l(w)$ becomes 
possible. This step is also carried out in Appendix \ref{somopart1}, yielding 
the result (\ref{stab31b}), 
\begin{equation} \eqlabel{stab31} 
\widetilde{C}_l(w) = \bar{\kappa} N \int\limits_0^\infty \dx{\omega} 
f_0(\omega)\, \omega^2\, \frac{e^{-iw\varphi^\DK/Q\omega}}{(Q\omega)^2 - w^2}\,
\frac{\widetilde{g}(w + l\omega) e^{i(w + l\omega)\theta^\DK/\omega}}
{1 - \widetilde{G}(w + l\omega) \widetilde{R}(w + l\omega)}
\end{equation}
with $\bar{\kappa} = 2\pi RQ\sqrt{\beta_K}$. Now the stage is set for an 
investigation of the stability behaviour of the beam.

\section{Stability of the Beam}
\plabel{stabpart12}

The results of the preceding section provide the basis for a stability 
analysis of the coefficient $C_l(t)$. The time behaviour and so the stability 
of each $C_l(t)$ is completely determined by the singularities of the Fourier 
transform $\widetilde{C}_l(w)$ (see Sect.\ \ref{vFT}), and hence we must find 
the zeros in the denominator of Eq.\ (\ref{stab31}). At first glance, we would 
expect that the singularities are determined by the two conditions
\begin{eqnarray} 
\eqlabel{stab34} (1) & & (Q\omega')^2 - w_0^2 = 0              \\
\eqlabel{stab35} (2) & & \widetilde{G}(w_l + l\omega')\,
                        \widetilde{R}(w_l + l\omega') = 1.
\end{eqnarray}
However within the bandwidth of the system, i.e.\ for $\widetilde{G}(w_0 + 
l\omega') \not= 0$, the zeros $w_0$ of Eq.\ (\ref{stab34}) do not result in
singularities, since for each term $(Q\omega)^2 - w^2$ in Eq.\ (\ref{stab31}) 
there exists its reciprocal $[(Q\omega)^2 - w^2]^{-1}$ in the function 
$\widetilde{R}(w + l\omega)$ which cancels the apparent singularity at $w^2 = 
(Q\omega)^2$. Only if the frequencies $w_0 + l\omega'$ lie outside the 
bandwidth of $\widetilde{G}(w)$, i.e.\ for $\widetilde{G}(w_0 + l\omega') = 
0$, the zeros $w_0$ will be important for the time behaviour of the 
coefficients $C_l(t)$. Because these coefficients are not influenced by the 
cooling system, they cannot become unstable by an interaction via the 
feedback loop and therefore need not to be considered for our purpose. 

Predictions about the stability of the collective beam motion in a cooling 
system can only be obtained from the frequencies $w_l$ which satisfy the 
second condition (\ref{stab35}), and thus depend on the gain 
$\widetilde{G}(w)$ of the cooling system. The {\sl critical gain} 
$\widetilde{G}_{crit}(w)$ is defined as that value at which the corresponding 
frequencies $w_l$ describe the onset of unstable collective beam motion, and 
therefore determines the stability boundaries of the system. There is a 
continuous functional dependence of the gain $\widetilde{G}(w)$ from the roots 
$w_l$ so that with $\gamma_l = {\sf Im}\,w_l$: 
\[ \gamma_l \longrightarrow 0 \quad \Longleftrightarrow \quad 
   \widetilde{G}(w_l) \longrightarrow \widetilde{G}_{crit}(w_l). \]
To determine $\widetilde{G}_{crit}(w_l)$, we first assume an initially stable 
solution $w_l$ of Eq.\ (\ref{stab35}) with $\gamma_l > 0$, and then we let the 
imaginary part $\gamma_l$ tend to zero, yielding the defining equation for 
the critical gain:
\[
\lim_{\gamma_l \to 0^+} \left[ 1 - \widetilde{G}(w_l + l\omega')\,
\widetilde{R}(w_l + l\omega') \right] = 0, \]
and with the definition $\Omega_l = {\sf Re}\,w_l + l\omega'$ it follows
\begin{equation} \eqlabel{stab36}
1 =  \widetilde{G}_{crit}(\Omega_l)\;
\lim_{\gamma_l \to 0^+}\!\! \widetilde{R}(\Omega_l + i\gamma_l).
\end{equation}
The evaluation of the limit is performed in Appendix \ref{Gcripart1}, leading 
to the stability criterion (\ref{stab41}), 
\begin{eqnarray} \eqlabel{stab41a}
\frac{Q}{\kappa N |\widetilde{G}_{crit}(\Omega_l)|}
& = & i\pi \sum_{(m)} \frac{1}{|m + Q|}\; 
      f\li( \frac{\Omega_l}{m + Q} \re)\, \frac{\Omega_l}{m + Q}\,
      e^{i\bar{\Phi}_m^\delta(\Omega_l)} +                         \nonumber\\
&   & \sum_{(m)}\, {\cal P\!\!\!\!} \int\limits_{-\infty}^\infty
      \dx{\omega} f(\omega)\, \omega\, 
      \frac{e^{i\bar{\Phi}_m^{\cal P}(\Omega_l, \omega)}}
      {(m + Q)\omega + \Omega_l}.
\end{eqnarray}
with the distribution function $f(\omega)$ being redefined by Eq.\ 
(\ref{stab36a}). The phases $\bar{\Phi}_m^\delta(\Omega_l)$ and 
$\bar{\Phi}_m^{\cal P}(\Omega_l, \omega)$ are given by the relations 
(\ref{stab42}) and (\ref{stab43}) respectively.

When applied to cooling systems, Eq.\ (\ref{stab41a}) can be further reduced, 
since
\begin{itemize} \label{coolsys}
\item %
Cooling systems are usually set up such that the phase $\psi(w)$ of the 
feedback gain can be described by a pure delay $\tau$, so we can write
\[ \widetilde{G}(w) = |\widetilde{G}(w)| e^{-iw\tau}. \]
\item %
The signal delay $\tau$ in the electronic components is adjusted to match the 
transit times of the particles between pickup and kicker. As a result of their 
distribution in revolution frequencies the particles have different transit 
times so that this coincidence cannot be accomplished for all particles 
simultaneously. The delay is generally adapted to the center frequency 
$\omega_0$ of the distribution, thus accommodating the most particles. In this 
case we obtain 
\[ \tau = t_0^\DK = \theta^\DK/\omega_0. \]
\item %
The pickup detects the transverse displacements of the particles whereas the 
kicker corrects the angle corresponding to the measured errors, requiring 
the proper betatron phase advance $\varphi^\DK$ between pickup and kicker. 
Ideally, the azimuthal distance $\theta^\DK$ from pickup to kicker is chosen 
such that the betatron phase advance between them is just 
\[ \varphi^\DK = \frac{\pi}{2} + 2\pi n, \qquad n = 0, 1, 2, \ldots \]
\item %
Assuming only small real frequency shifts due to feedback interaction, the 
transverse frequencies can be written as
\[ {\sf Re}\,w_l \approx Q \omega' \qquad\mbox{and}\qquad 
\Omega_l \approx (l + Q)\omega'.
\]
From Eq.\ (\ref{stab31}) it can be seen that $\omega' > 0$ and hence 
$\sigma_{\!\Omega_l} = \sgn{l}$.\footnote{The quantities $\sigma_{\!\Omega_l}$,
$\sgn{l}$ and $\sigma_{\!\omega}$ are defined in Appendix \ref{Gcripart1}.}
\end {itemize}
With these assumptions, Eq.\ (\ref{stab41a}) can be further evaluated (see 
App. \ref{Gcripart1}), and we obtain the following stability criteria, 
\begin{equation} \eqlabel{stab46}
\frac{Q}{\kappa N |\widetilde{G}_{crit}(\Omega_l)|} =  
\pi \sum_{(m)} \frac{\cos\Phi_m^\delta(\Omega_l)}{|m + Q|}
\frac{|\Omega_l|}{|m + Q|}\; f\li( \frac{\Omega_l}{m + Q} \re) +
\sum_{(m)}\, {\cal P\!\!\!\!} \int\limits_{-\infty}^\infty \dx{\omega} 
f(\omega)\, |\omega|\, \frac{\sin\Phi_m^{\cal P}(\Omega_l, \omega)}
{(m + Q)\omega + \Omega_l}
\end{equation}
and
\begin{equation} \eqlabel{stab47}
\pi \sum_{(m)} \frac{\sin\Phi_m^\delta(\Omega_l)}{|m + Q|}
\frac{|\Omega_l|}{|m + Q|}\; f\li( \frac{\Omega_l}{m + Q} \re) =
\sum_{(m)}\, {\cal P\!\!\!\!} \int\limits_{-\infty}^\infty \dx{\omega} 
f(\omega)\, |\omega|\, \frac{\cos\Phi_m^{\cal P}(\Omega_l, \omega)}
{(m + Q)\omega + \Omega_l}
\end{equation}
where the phases are given by 
\begin{equation} \eqlabel{stab44}
\Phi_m^\delta(\Omega_l) \approx 
\sgn{l}\, \Big[ |l + Q| - |m + Q| \Big] \theta^\DK
+ (l + Q)\theta^\DK \frac{\delta\omega'}{\omega_0}
\end{equation}
and
\begin{equation} \eqlabel{stab45}
\Phi_m^{\cal P}(\Omega_l, \omega) \approx 
\Big[ (l + Q) + \sigma_{\!\omega}\, (m + Q) \Big] \frac{\varphi^\DK}{Q}
+ (l + Q)\frac{\Delta\varphi}{Q}\Delta'(\omega)
+ (l + Q) \theta^\DK \frac{\delta\omega'}{\omega_0}.
\end{equation}
These equations which allow for overlapping frequency bands in the beam 
spectrum establish the stability boundaries of the expansion coefficients 
$\widetilde{C}_l(w)$ over the entire frequency range. Eq.\ (\ref{stab46}) 
defines the maximum stable gain of the cooling system: if $|\widetilde{G}(w)|$ 
exceeds the critical value $|\widetilde{G}_{crit}(w)|$ the beam will become 
unstable owing to the feedback interaction. On the other hand, Eq.\ 
(\ref{stab47}) determines the real frequencies $\Omega_l = {\sf Re}\,w_l + 
l\omega'$ at which the beam supports the propagation of collective modes due 
to the critical gain of the cooling system. 

\subsubsection{The Special Case of Non-Overlapping Bands}

The results obtained above include also the special case of non-overlapping 
frequency bands. In this case particles can only interact via frequencies 
lying within the same band, and thus the summation over the bands reduces to 
the single term $m = l$. \\
Using $\Omega_l = (l + Q)\omega'$ and $\omega' > 0$, the Eqs.\ (\ref{stab46}) 
and (\ref{stab47}) now reads 
\begin{equation} \eqlabel{stab48}
\frac{Q}{\kappa N |\widetilde{G}_{crit}(\Omega_l)|} = 
\pi \frac{\cos\Phi_l^\delta(\Omega_l)}{|l + Q|}\; \omega'\, f(\omega') + 
\frac{1}{l + Q}\; {\cal P\!\!\!\!} \int\limits_{-\infty}^\infty \dx{\omega} 
f(\omega)\, |\omega|\, \frac{\sin\Phi_l^{\cal P}(\Omega_l, \omega)}
{\omega + \omega'}
\end{equation}
and
\begin{equation} \eqlabel{stab49}
\pi \frac{\sin\Phi_l^\delta(\Omega_l)}{|l + Q|} \omega'\; f(\omega') =
\frac{1}{l + Q}\; {\cal P\!\!\!\!} \int\limits_{-\infty}^\infty \dx{\omega} 
f(\omega)\, |\omega|\, \frac{\cos\Phi_l^{\cal P}(\Omega_l, \omega)}
{\omega + \omega'}.
\end{equation}
The phases (\ref{stab44}) and (\ref{stab45}) become 
\[
\Phi_l^\delta(\Omega_l) =  \sgn{l}\, \Big[ |l + Q| - |l + Q| \Big]\theta^\DK
+ (l + Q)\theta^\DK\frac{\delta\omega'}{\omega_0}
= (l + Q)\theta^\DK\frac{\delta\omega'}{\omega_0}
\]
and
\[
\Phi_l^{\cal P}(\Omega_l, \omega) =  
\Big[ (l + Q) + \sigma_{\!\omega}\, (l + Q) \Big]\frac{\varphi^\DK}{Q}
+ (l + Q)\frac{\Delta\varphi}{Q}\Delta'(\omega) + 
(l + Q)\theta^\DK\frac{\delta\omega'}{\omega_0}
\] 
respectively. Significant contributions from the principal value integrals in 
the Eqs.\ (\ref{stab48}) and (\ref{stab49}) only arise in the proximity of the 
pole, i.e.\ for $\omega \approx -\omega'$. Since $\omega' > 0$ (see Eq.\ 
(\ref{stab31})) and thus $\sigma_{\!\omega} = -1$, we obtain 
\[
\Phi_l^{\cal P}(\Omega_l, \omega) =  
(l + Q)\frac{\Delta\varphi}{Q}\Delta'(\omega) +
(l + Q)\theta^\DK\frac{\delta\omega'}{\omega_0}.
\]
These results are consistent with the findings in \cite{wei} which have been 
obtained by considering only non-overlapping frequency bands. The more general 
theory (including band overlap) gives us additional information about the 
phases $\Phi_l^\delta(\Omega_l)$ and $\Phi_l^{\cal P}(\Omega_l, \omega)$ which 
in \cite{wei} have been assumed to be zero.

\section{Comparison with the Feedback Theory}
\plabel{stabpart13}

Finally, we will show that the result (\ref{stab31}) for the mode expansion 
coefficients is consistent with the predictions of the multi-bunch feedback 
theory. For that purpose we assume a mono-energetic beam, and thus the same 
revolution frequency $\omega_0$ for all particles, which later will permit to 
identify the beam particles with bunches. Then the distribution function in 
revolution frequencies reads $f_0(\omega) = \delta(\omega - \omega_0)$, and 
Eq.\ (\ref{stab31}) reduces to 
\[
\widetilde{C}_l(w) = \bar{\kappa} \omega_0^2 N\,
     \frac{e^{-iw\tau^\DK}}{(Q\omega_0)^2 - w^2}\,
     \frac{\widetilde{g}(w + l\omega_0) e^{i(w + l\omega) t^\DK}}
     {1 - \widetilde{G}(w + l\omega_0) \widetilde{R}(w + l\omega_0)}
\]
where $\tau^\DK = \varphi^\DK/Q\omega_0$ and $t^\DK = \theta^\DK/\omega_0$ are 
now the same for all particles. Furthermore, Eq.\ (\ref{stab25a}) becomes 
\[
\widetilde{R}(w) = \kappa \omega_0^2 N e^{iwt^\DK} \sum_{(m)} 
\frac{e^{-i(w + m\omega_0)\tau^\DK}}{(Q\omega_0)^2 - (w + m\omega_0)^2}.
\]
Defining the $\omega_0$-periodic function
\[ 
\widehat{R}(w) = \sum_{(m)} \frac{e^{-i(w + m\omega_0)\tau^\DK}}
                  {(Q\omega_0)^2 - (w + m\omega_0)^2},
\]
and writing the feedback gain and the external mode excitation as 
\begin{equation} \eqlabel{stab31a}
\widetilde{G}_C(w) = \widetilde{G}(w) e^{iwt^\DK}
\end{equation}
and
\[ \frac{1}{N}\, \widetilde{F}_l(w) = 
   \widetilde{g}(w + l\omega_0) e^{i(w + l\omega_0)t^\DK}
\]
respectively, we obtain 
\begin{equation} \eqlabel{stab32}
\widetilde{C}_l(w) =  \frac{e^{-iw\tau^\DK}}{(Q\omega_0)^2 - w^2}\, 
\frac{\bar{\kappa} \omega_0^2\, \widetilde{F}_l(w)}
{1 - \Upsilon \widetilde{G}_C(w + l\omega_0) \widehat{R}(w)} 
\end{equation} 
where $\Upsilon = \kappa \omega_0^2 N$. 
The corresponding result from the multi-bunch feedback theory for the $r$-th 
multi-bunch mode is given by \cite{ko1}
\begin{equation} \eqlabel{stab33}
\widetilde{C}_r(w) =  \sum_{(m)} \frac{e^{-i(w + m\omega_0)\tau^\DK}}
{(Q\omega_0)^2 - (w + m\omega_0)^2}\,
\frac{\bar{\kappa} \omega_0^2\, \widetilde{F}_r(w)}
{1 - \Upsilon \widetilde{G}_{C\!N}(w + r\omega_0) \widehat{R}(w)}
\end{equation}
with the definition
\[ \widetilde{G}_{C\!N}(w) = \sum_{(l)} 
\widetilde{G}(w + lN\omega_0) e^{i(w + lN\omega_0)t^\DK}, \]
in which $N$ denotes the number of bunches. $\widetilde{G}_{C\!N}(w)$ is a 
periodic function with period $N\omega_0$.

We now discuss the differences between the Eqs.\ (\ref{stab32}) and 
(\ref{stab33}). To that end we identify the bunches in the feedback theory with
the individual particles of the unbunched beam.

The period $N\omega_0 = 2\pi N f_0$ of $\widetilde{G}_{C\!N}(w)$ equals 
(except for a factor $2\pi$) the bunch frequency $f_B$, i.e.\ the frequency 
with which succeeding bunches appear at fixed locations. If $T_B$ denotes the 
time difference between two adjacent bunches and $T_0$ the revolution time, it 
immediately follows that $f_B = 1/T_B = 1/(T_0/N) = N f_0$. The limit of the 
continuous unbunched beam in which the distance of adjacent particles becomes 
zero (see Sect.\ \ref{stabpart7}) corresponds in the bunched beam to $T_B \to 
0$ or $f_B \to \infty$ so that $\widetilde{G}_{C\!N}(w)$ is no longer 
periodic. In this limit $\widetilde{G}_{C\!N}(w)$ turns into the function 
$\widetilde{G}_C(w)$ which has been defined for the unbunched beam 
(see Eq.\ (\ref{stab31a})).

The second difference between the Eqs.\ (\ref{stab32}) and (\ref{stab33})
shows up in the first term on the right hand side of each equation. The 
expression for the multi-bunch modes contains an additional summation over all 
revolution harmonics $m\omega_0$ which arises from the discrete time structure 
of the signals and expresses the {\sl periodic sampling} of the force at the 
kicker by the bunches (see Sect.\ \ref{basepart5}). The continuous beam, on 
the other hand, is found at the kicker at any time and therefore does {\sl not 
sample} the force at discrete times, and hence the summation over the 
revolution harmonics does not occur in Eq.\ (\ref{stab33}). In this respect 
the continuous beam corresponds to a bunched beam subjected to a force which 
is distributed over the whole ring. Since in this case the force would act 
permanently on the bunches, the summation in Eq.\ (\ref{stab33}) would 
likewise disappear.

If, according to these comments, we replace in Eq.\ (\ref{stab33})
\[ \widetilde{G}_{C\!N}(w) \longrightarrow \widetilde{G}_{C}(w)\]
and disregard the sampling of the force so that
\[ \sum_{(m)} \frac{e^{-i(w + m\omega_0)\tau^\DK}}
{(Q\omega_0)^2 - (w + m\omega_0)^2} \longrightarrow 
\frac{e^{-iw\tau^\DK}}{(Q\omega_0)^2 - w^2},
\]
the result of the multi-bunch feedback theory will yield expression 
(\ref{stab32}) which has been obtained for the unbunched beam. 

\chapter{Stochastic Cooling of Unbunched Beams} \plabel{coolpart1}
\vspace*{-2.5mm}
\section{Overview} \plabel{coolpart1a}

In this chapter we will discuss transverse stochastic cooling of an unbunched 
beam in case of a linear cooling interaction. The basic concepts of stochastic 
cooling are presented on a qualitative level, omitting the detailed 
derivations of the fundamental equations for the most part. Only the 
calculations which are typical of the specific cooling process considered 
here, are explicitly shown in the Appendices \ref{fdkopart1} and 
\ref{parapart1}.

Compared with the instabilities studied in the previous chapter stochastic 
cooling can be regarded as a slow process. Hence it is convenient to 
investigate the cooling in the time domain since then averaging over fast 
changing variables can easily be performed. Here, we will describe the cooling 
process by means of a Fokker-Planck equation for the phase-space density in 
the transverse action variable. Unlike the sample picture of stochastic 
cooling \cite{moe} where the properties of the cooling process have to be 
introduced explicitly on the basis of empirical arguments, these effects 
follow automatically in the Fokker-Planck approach due to the more rigorous 
mathematical treatment of the particle dynamics. In the results we will 
recognize the {\sl mixing} which measures the decay of inter-particle 
correlations as well as the {\sl optimal betatron phase advance} of the 
particles between the pickup and kicker. Furthermore the reason will become 
obvious why stochastic cooling systems take a {\sl shortcut through the ring}. 
The results of the Fokker-Planck equation allow a profound insight into the 
physical origin of these effects and can also provide quantitative predictions 
for the parameters which determine the performance of the cooling system. This 
will be discussed in detail in the Sections \ref{coolpart9} and 
\ref{coolpart10}.

The Fokker-Planck equation represents a statistical description of the cooling 
process, assuming independent particles with random phases. In Section 
\ref{coolpart8} this assumption will be motivated by means of physical 
arguments. Including the collective motion of the particles into this 
formulation poses considerable mathematical difficulties, and hence further 
assumptions have to be made. The cooling model underlying the Fokker-Planck 
equation neglects the collective effects (see Sect.\ \ref{coolpart7}), and 
therefore can only be applied after the beam stability has explicitly been 
proven. The parameters for which the Fokker-Planck equation predicts the most 
efficient cooling operation are compared to the stability criteria of the 
coherent beam modes in order to verify that these predictions are compatible 
with the stability boundaries of the beam. Initially, we will review the 
fundamental physical principles of stochastic cooling.

\section{The Phase-Space Fluctuations}  \plabel{coolpart2}

The beam model used in the previous chapter is capable of describing 
collective effects for which not the single particle behaviour, but the 
common motion of all particles is important. The model considers the 
phase-space volume occupied by the beam as a whole, thus providing only 
information about macroscopic beam quantities averaged over all particles. 
The macroscopic state of the beam is completely characterized by a smooth mean 
phase-space density which no longer contains information about individual 
particles.

The beam cooling aims at increasing the phase-space density by concentrating 
the beam particles in a phase-space volume as small as possible. To manipulate 
the individual particle motions, the cooling interaction needs information 
about the {\sl internal} phase-space structure which to a certain degree 
requires the knowledge of the phase-space coordinates of single particles. 
Stochastic cooling finally depends on how fast and how precisely details of 
the phase space can be resolved in order to obtain the necessary information. 

\begin{figure}[ht]
\parbox[b]{9cm}{\input{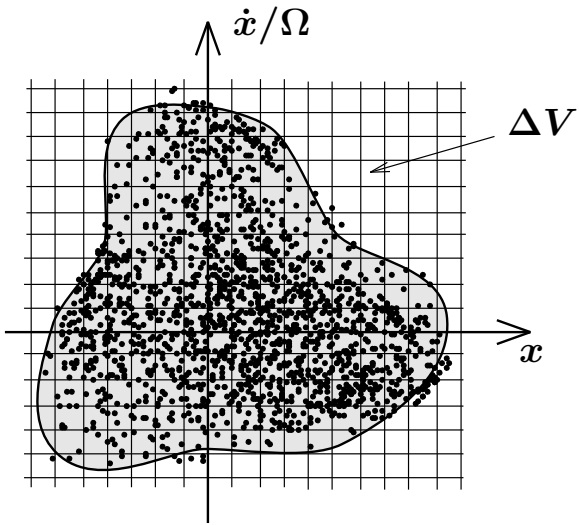}}
\raisebox{10mm}{\parbox[b]{7cm}{\caption{\label{phafluc}\sl Fluctuations of 
the phase-space density due to the discreteness of the beam particles.}}}
\end{figure}

The graininess of the beam, i.e.\ the discrete nature and finite number of 
its particles, makes these information accessible to a pickup. We will 
illustrate this fact with a beam of $N$ independent particles represented by 
$N$ individual points in the phase space (see Fig.\ \ref{phafluc}). To define 
a density, we subdivide the phase space into small volumes in which the 
phase-space coordinates do not change noticeably, and identify the local 
phase-space density with the number of particles in such a subvolume $\Delta 
V$. The density defined in this way is not a continuous function, but 
fluctuates from subvolume to subvolume according to the number of particles 
each volume contains. The smooth mean phase-space density then is the ensemble 
average over all possible particle configurations in the subvolumes. 

The actual phase-space density ${\bf F}({\bf x}, {\bf \dot{x}}, t)$ of the 
beam depends on the instantaneous phase-space coordinates of the particles and 
fluctuates around the smooth mean density $f({\bf x}, {\bf \dot{x}}, t)$. 
Writing these fluctuations as
\[
\delta\! f({\bf x}, {\bf \dot{x}}, t) = {\bf F}({\bf x}, {\bf \dot{x}}, t) -
\langle {\bf F}({\bf x}, {\bf \dot{x}}, t) \rangle = 
{\bf F}({\bf x}, {\bf \dot{x}}, t) - f({\bf x}, {\bf \dot{x}}, t),
\]
we can describe the macroscopic and microscopic beam properties separately by 
means of the corresponding distribution functions, $f({\bf x}, {\bf 
\dot{x}}, t)$ and $\delta\! f({\bf x}, {\bf \dot{x}}, t)$. Since only the 
fluctuations $\delta\! f({\bf x}, {\bf \dot{x}}, t)$ contain the instantaneous 
phase-space coordinates of individual particles, solely they can provide the 
wanted information about the internal structure of the phase space.

\section{The Fluctuation Spectrum} \plabel{coolpart3}

The fluctuations in the phase-space density are observable in the measured 
beam signal $S(t)$ where they lead to fast variations $\delta\! S(t)$ around 
the mean signal. The phase-space fluctuations are determined by the individual 
particle motions, and thus depend on the initial conditions of the particles. 
Apart from very special cases, the initial state of the beam is unknown, 
allowing only a statistical description of the fluctuations which turns them 
into stochastic quantities.

The time behaviour of the fluctuations follows from the 
autocorrelation-function,
\[ C_{\delta\! S}(t, t') := \langle \delta\! S(t)\, \delta\! S(t') \rangle, \]
which measures the correlation time $\tau_{corr}$ of the signal fluctuations. 
The correlation time is given by the time interval $t' - t$ within which the 
autocorrelation-function of the measured signals has non-zero values. In other 
words, the fluctuations will be statistically independent from each other if 
they are separated in time by more than the correlation time $\tau_{corr}$: 
\[ 
\langle \delta\! S(t)\,\delta\! S(t') \rangle \approx 0 
\qquad\mbox{for}\quad \tau_{corr} <  |t' - t|.
\]
For unbunched beams, the autocorrelation-function only depends on the 
difference $\tau = t' - t$, and the Fourier transformation yields the power 
spectrum of the fluctuations which often is referred to as Schottky spectrum, 
\[
P_{\delta\! S}(\omega) = \frac{1}{2\pi} \int\limits_{-\infty}^\infty\! 
\dx{\tau} e^{-i\omega\tau}\, \langle \delta\! S(t)\,\delta\! S(t + \tau) 
\rangle.
\]
At this place, we omit a detailed description of the Schottky spectra, and 
summarize only the properties important for the following discussion. General 
information about the Schottky beam-spectra can be found e.g.\ in \cite{ch1}.

The Schottky spectrum mirrors the revolution frequency distribution of the 
particles so that the results of Section \ref{basepart6} also apply to the 
Schottky spectrum. One obtains a spectrum of bands in which the width of the 
bands increases at higher frequencies, resulting in a band overlap beyond a 
certain frequency. This corresponds to a finite correlation time in the order 
of the reciprocal width of the bands, and thus decreases at higher 
frequencies. We will see later that a finite correlation time, and with it a 
non-zero width of the frequency distribution, are essential for the stochastic 
cooling operation. 

\section{Particle Dynamics in the Cooling System} \plabel{coolpart4}

To describe the stochastic beam cooling, we again start at the single particle 
equation of motion, but this time do not discard the discreteness of the 
particles. Excluding any external excitation, it follows from Eqs.\ 
(\ref{stab12}) and (\ref{stab17}) for the particle $j$
\begin{equation} \eqlabel{cool1}
\frac{d^2}{d\qt{j}^2}\, x_j(\qt{j}) + \Omega_j^2\, x_j(\qt{j}) = 
\Omega_j^2 \beta_k^{3/2} \int \dx{w} e^{iw\qt{j}}\,\widetilde{F}_j(w)
\end{equation}
where
\begin{equation} \eqlabel{cool2}
\widetilde{F}_j(w) = 
     e^{-iw\tau_j^\DK} \;\frac{\omega_j}{\Omega_j} \frac{2\pi R}{\beta_k}\,
     \sum_{(m)} \widetilde{G}(w + m\omega_j) \widetilde{S}(w + m\omega_j)
     e^{i(w + m\omega_j)t_j^\DK} e^{i(w + m\omega_j)t_j^0}.
\end{equation}
Using Eq.\ (\ref{stab6}), the Fourier transform of the signal at the pickup 
reads 
\begin{equation} \eqlabel{cool3}
\widetilde{S}(w) = \frac{\sqrt{\beta_P}}{2\pi} \sum_{j'=1}^N \sum_{(l)} 
       \widetilde{x}_{j'}(w - l\omega_{j'}) e^{-iwt_{j'}^0}.
\end{equation}
We now regard the particle $j$ as a test-particle treated separately from the 
rest of the beam which consists of all the remaining particles $j' \not= j$. 
For this purpose, we split the force on the right-hand side of Eq.\ 
({\ref{cool1}) into a part arising from the test-particle, and the 
contributions of all other particles. In Eq.\ (\ref{cool3}) only the term for 
$j' = j$ enters into the self-interaction, and together with Eq.\ 
(\ref{cool2}) we find 
\begin{equation} \eqlabel{cool4}
F_j^S(\qt{j}) = 
\kappa \omega_j^2 \int \dx{w} e^{iw\qt{j}} e^{-iw\tau_j^\DK} 
\sum_{(m)} \widetilde{G}(w + m\omega_j) e^{i(w + m\omega_j)t_j^\DK}
\sum_{(l)} \widetilde{x}_j(w + l\omega_j)
\end{equation}
where $\kappa = RQ\sqrt{\beta_P\beta_K}$. The contributions of the other 
particles are given by
\begin{equation} \eqlabel{cool5}
F_j^R(\qt{j}) = 
\kappa \omega_j^2 \int \dx{w} e^{iw\qt{j}} e^{-iw\tau_j^\DK} \sum_{(m)} 
\widetilde{G}(w + m\omega_j)\, \widetilde{S}_j(w + m\omega_j)
e^{i(w + m\omega_j)t_j^\DK} e^{i(w + m\omega_j)t_j^0}
\end{equation}
with
\[
\widetilde{S}_j(w) := \sum_{j' \not= j} \sum_{(l)} 
\widetilde{x}_{j'}(w + l\omega_{j'}) e^{-iwt_{j'}^0}.
\]
Writing the cooling force in this way, it becomes more apparent that the 
self-interaction of the test-particle does not depend on the time $t_j^0 = 
\theta_j^0/\omega_j$ of the first passage of the test-particle at the pickup. 
The self-interaction has a fixed phase with respect to the test-particle, and 
thus gives a coherent contribution to the cooling interaction over the entire 
frequency range. On the other hand, the signal $\widetilde{S}_j(w)$ entering 
into the force $F_j^R(\qt{j})$ includes the initial conditions of the 
particles, and shows fast variations in time, as described above. Because this 
fluctuations occur around the smooth macroscopic beam signal, it appears 
obvious to divide the force $F_j^R(\qt{j})$ further into a collective and a 
fluctuation part:
\[ F_j^R(\qt{j}) = \langle F_j^R(\qt{j}) \rangle + \delta\! F_j^R(\qt{j}). \]
The averaging process $\langle \cdot \rangle$ is performed over the phases 
and azimuths of the rest beam particles with $j' \not= j$. The collective 
force $\langle F_j^R(\qt{j}) \rangle$ results from the common motion of the 
particles and is completely determined by the smooth mean phase-space density. 
The fluctuation force $\delta\! F_j^R(\qt{j})$ can be regarded as a purely 
statistical quantity, providing a stochastic part to the interaction. 

The equation of motion of the test-particle describes a stochastic 
differential equation which we write as
\begin{equation} \eqlabel{cool6}
\frac{d^2}{d\qt{j}^2}\, x_j(\qt{j}) + \Omega_j^2\, x_j(\qt{j}) = 
F_j^S(\qt{j}) + \langle F_j^R(\qt{j}) \rangle + \delta\! F_j^R(\qt{j}).
\end{equation}
The stochastic force $\delta\! F_j^R(\qt{j})$ in the interaction renders 
determined predictions of the particle motions impossible so that only the 
statistical properties of the kinetic quantities, obtained by averaging over 
the probability densities of the particles in the phase space, are relevant to 
the cooling description. Since the probability densities of the particles and 
the phase-space density of the beam are connected \cite{kam}, the time 
evolution of the latter fully characterizes the dynamics of the cooling 
process.

\section{The Time Evolution of the Phase-Space Density} \plabel{coolpart6}
\subsection{The Model of the Cooling Interaction} \plabel{coolpart7}

Describing the cooling process in terms of the smooth mean phase-space density 
allows, in principle, the calculation of all statistical moments of the 
dynamic beam quantities. With some assumptions about the forces on the 
right-hand side of Eq.\ (\ref{cool6}), one can derive a Fokker-Planck equation 
for the time evolution of the phase-space density.\footnote{At the beginning, 
the Fokker-Planck equation determines the probability density of each particle 
in the phase space. For statistically independent particles the probability 
densities of all particles are equal and can be identified with the mean 
phase-space density of the beam \cite{kam}.} Before using this equation to 
make predictions about the cooling, we will discuss the consequences of the 
necessary assumptions in order to facilitate a correct physical interpretation 
of the results. The Fokker-Planck description of stochastic cooling follows 
only in the next sections. In particular, the derivation of the Fokker-Planck 
equation is based on the following assumptions:
\begin{itemize}
\item[(1)]%
The correlation time $\tau_{corr}$ of the signal fluctuations is finite. 
Contributions from the stochastic force $\delta\! F_j^R(\qt{j})$ which are 
separated by time differences $\Delta t > \tau_{corr}$, then, do not have a 
definite phase relation with respect to each other and add up incoherently. 
After this time interval, they can be considered as statistically independent. 
\item[(2)]%
The oscillation amplitudes of the particles change noticeably only after a 
time long compared with the correlation time of the fluctuations. Within a 
time interval $\tau_{corr}$ the amplitudes can be regarded almost as constant.
\item[(3)]%
Neglecting all collective particle effects, only the case $\langle 
F_j^R(\qt{j}) \rangle = 0$ is investigated.
\end{itemize}
According to Section \ref{coolpart3}, the requirement (1) is met by a finite 
width of the signal frequency distribution corresponding to a band spectrum 
of the fluctuations which is always present in an unbunched beam, owing to 
the energy spread of the particles. Likewise, stochastic cooling systems 
comply with the assumption (2) of {\sl slow} cooling since typical cooling 
times span a range from a few seconds to many hours always larger than the 
reciprocal width of the relevant frequency bands \cite{moe}. On the other 
hand, the point (3) cannot be justified so easily; it can entail important 
consequences so that its meaning has to be considered in detail. We have 
already seen in Chapter \ref{stabpart1} that a non-zero collective force 
$\langle  F_j^R(\qt{j}) \rangle$ can lead to an unstable beam motion. In this 
situation, the collective force becomes the dominant term in the interaction, 
and obviously the assumption $\langle F_j^R(\qt{j}) \rangle = 0$ makes no 
sense. The verification of point (3) need to be based on the beam stability 
boundaries derived in Section \ref{stabpart12}: if the parameters of the 
cooling system only vary within these boundaries, all coherent modes of the 
beam will be damped, and thus will not contribute to the interaction. 

Neglecting the collective force in Eq.\ (\ref{cool6}), therefore, necessitates 
an {\sl explicit} proof that the predictions for the cooling performance are 
{\sl consistent} with the beam stability criteria of Section \ref{stabpart12}. 
A Fokker-Planck equation which relies on this assumption disregards any 
collective effect of the particles, and provides only purely statistical 
results. Hence such a description could predict that on average each beam 
particles is damped, although the collective beam motion is unstable. The 
Fokker-Planck equation describes the dynamics in a cooling system only partly 
and must be supplemented with the stability analysis elaborated in Chapter 
\ref{stabpart1}.

On the other hand, the damping of the coherent modes required by a stable beam 
motion introduces correlations among the particles which destroy the 
statistical independence of the particles. Now, the signal fluctuations arise 
around a macroscopic beam signal which is modulated by the collective 
interactions through the feedback loop so that the fluctuations become 
dynamically coupled with the collective beam motion. Since including these 
correlations into the mathematical description turns out to be very difficult, 
one simplifies the problem by assuming a small impact of this effect which 
hardly affects the time evolution of the phase-space density \cite{ch2}. The 
particles are again considered uncorrelated. On the other hand, the damping of 
all coherent beam modes eliminates the collective force in the interaction, 
and enables us to set $\langle F_j^R(\qt{j}) \rangle = 0$ in Eq.\ 
(\ref{cool6}). The interaction with the rest beam then reduces to the purely 
stochastic force $\delta\! F_j^R(\qt{j})$.

Having manifested the underlying assumptions and limits of our cooling model, 
we now derive the time evolution of the phase-space density to predict the 
maximum attainable cooling rate. The parameters of the cooling system which 
give the best cooling results will be verified with respect to their 
compatibility with the beam stability in order to decide how far these values 
are physically reasonable.

\subsection{The Fokker-Planck Equation} \plabel{coolpart8}

An appropriate set of variables for the description of transverse stochastic 
cooling is given by the action and angle variables $(I, \varphi)$. Here, the 
action $I$ is connected to the betatron amplitude $A$ by the relation $I = 
A^2/2$ and $\varphi$ is just the betatron phase. The time evolution of the 
corresponding distribution function $\bar{\rho}(I, \varphi, t)$ completely 
characterizes the cooling process, and obeys a two-dimensional Fokker-Planck 
equation. Since only the long-term behaviour of the action $I$ is relevant for 
the cooling description, it is suitable to average the equation over the fast 
varying angle variable $\varphi$, resulting in the Fokker-Planck equation for 
the distribution function $\rho(I, t)$ \cite{we2}
\begin{equation} \label{cool7} 
\frac{\partial}{\partial t} \rho(I, t) = 
- \frac{\partial}{\partial I} \left\{ F(I) \rho(I, t) - \frac{1}{2} D(I)\, 
\frac{\partial}{\partial I} \rho(I, t) \right\}.
\end{equation}
The drift coefficient $F(I)$ and diffusion coefficient $D(I)$ are evaluated 
from the relations
\begin{equation} \label{cool8}
F(I) = \Big\langle \Big\langle \frac{\Delta I^F}{\Delta T} 
\Big\rangle \Big\rangle_{\theta^0, \varphi^0}
\qquad\mbox{and}\qquad
D(I) = \Big\langle \Big\langle \frac{\Delta I^D \Delta I^D}{\Delta T} 
\Big\rangle \Big\rangle_{\theta^0, \varphi^0}
\end{equation}
where $\Delta I^F$ denotes the change of the action during the time interval 
$\Delta T$ which arises from the coherent self-interaction of the particles, 
and $\Delta I^D$ is the corresponding value originating in the incoherent 
fluctuation signal of the other beam particles. The time interval $\Delta T$ 
must be chosen long compared with the correlation time of the fluctuations, 
but still much shorter than the time in which the density $\rho(I, t)$ changes 
appreciably due to the cooling. The brackets $\langle\langle\ \rangle\rangle$ 
in Eq.\ (\ref{cool8}) indicate averaging processes over the initial azimuths 
$\theta^0$ and betatron phases $\varphi^0$ of the beam particles. From the 
finite correlation time of the fluctuations follows that after a first delay 
$\Delta t > \tau_{corr}$ the initial state of the beam at the time $t = 0$ 
becomes insignificant. The starting-time and the initial conditions get an 
arbitrary meaning in so far as the system does not store any information about 
them, and the dynamics at times $t > \tau_{corr}$ no longer depends on the 
initial state of the beam. Hence not the initial conditions at the time $t = 
0$ are important, but the corresponding values of the beam at the beginning of 
the averaging process. Because each averaging extents over a time interval 
$\Delta T > \tau_{corr}$, the so defined {\sl initial conditions} can be 
interpreted as stochastic quantities. In other words, the averaging process is 
the same after each time interval $\Delta T$, and can always be performed over 
statistically independent phases and azimuths. By reason of the additional 
requirement $\Delta T < \tau_\rho$ the change of the phase-space density 
$\rho(I, t)$ during the averaging is negligible so that we consider the 
density $\rho(I, t)$ to remain constant over that period. Fig.\ \ref{tscales} 
illustrates the relations between the different time intervals involved in the 
averaging processes.
\begin{figure}[ht]
\input{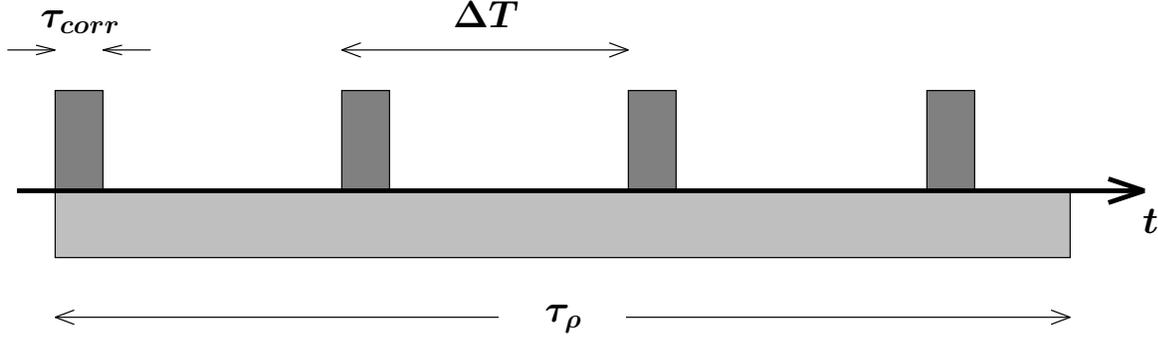}
\caption{\label{tscales}\sl Illustration of the different time scales of the 
averaging processes. The period $\Delta T$ over which the averaging is 
performed is large compared to the correlation time $\tau_{corr}$ of the 
fluctuations, but still short relative to the time $\tau_\rho$ in which the 
distribution changes noticeably.}
\end{figure}

With the preceding remarks we can explicitly calculate the drift and 
diffusion coefficients of the test-particle $j$. The elementary, but lengthy 
derivations can be looked up in Appendix \ref{fdkopart1}, and yield the 
relations (\ref{cool9}) and (\ref{cool10}). The results are discussed in the 
Sections \ref{coolpart9} and \ref{coolpart10}, revealing all the well-known 
properties of stochastic cooling in a quite natural way. 

\subsection{The Drift Coefficient} \plabel{coolpart9} 

According to Eq.\ (\ref{fdko4a}), we write the drift coefficient of the 
particle $j$ as
\begin{equation} \eqlabel{cool9}
F_j(I_j) = \bar{F}_j\, I_j \qquad\mbox{with}\qquad 
\bar{F_j}  =  -\frac{\kappa\omega_j}{Q} \sum_{(m)}\; 
|\widetilde{G}[(m + Q)\omega_j]|\; \sin\Phi_m^j 
\end{equation}
and the phase
\[
\Phi_m^j = (m + Q)\theta^\DK \frac{\delta\omega_j}{\omega_0} + \varphi^\DK.
\]
The drift coefficient describes the intrinsic damping of the action. It 
depends linearly on the gain $|\widetilde{G}(w)|$ of the cooling system, 
however weighted by a sine-function. The phase $\Phi_m^j$ already includes the 
time delay $\tau$ of the correction signals in the cooling electronics. Here, 
it has been assumed that this time delay is adjusted to the travel time of 
the synchronous particles (with the revolution frequency $\omega_0$) from the 
pickup to the kicker, i.e.\ $\tau = \theta^\DK/\omega_0$. These particles 
experience the maximum cooling effect if $\sin\Phi_m^j = 1$, which in 
consequence of $\delta \omega_j = 0$ is just the case for 
\[ \Phi_m^j = \frac{\pi}{2} + 2\pi n \quad\mbox{with}\quad 
n = 0, 1, 2, \ldots \]
This {\sl optimum betatron phase advance} simply manifests the fact that the 
pickup measures the displacements of the particles, but the kicker corrects 
the offsets by applying the corresponding angles. 

On the other hand, the particles with $\delta \omega_j \not= 0$ whose 
revolution frequencies vary from the nominal frequency $\omega_0$ have 
different travel times from the pickup to the kicker, and so do not obtain 
their corrections with the optimum phase. The additional phase shift, often 
called {\sl unwanted mixing}, is taken into account by the first term in the 
phase $\Phi_m^j$. It follows from Eq.\ (\ref{cool6}) that the unwanted mixing 
can be diminished for all particles simultaneously by reducing the azimuthal 
distance $\theta^\DK$ between the pickup and kicker. For that reason 
stochastic cooling systems take a {\sl shortcut through the storage ring}. 

\subsection{The Diffusion Coefficient} \plabel{coolpart10}

Neglecting the electronic noise in the cooling system, the diffusion 
coefficient of particle $j$ reads (see Eq.\ (\ref{fdko10})) 
\[ D_j(I_j) = \bar{D}_j \langle I \rangle I_j \]
with
\begin{equation} \eqlabel{cool10}
\bar{D_j} =  \frac{2\pi\kappa^2\omega_j^2N}{Q^2}\, \sum_{(m)}\,
\left| \widetilde{G}[(m + Q)\omega_j] \right|^2\, \sum_{(l)}
\frac{1}{|l + Q|}\, f\li( \frac{m + Q}{l + Q}\, \omega_j \re).
\end{equation}
The diffusion coefficient is determined by the stochastic part of the 
interaction which originates in the signal fluctuations at the pickup. Due to 
the statistical nature of this interaction, the diffusion coefficient shows a 
quadratic dependence on the system gain, weighted by the spectral particle 
density at the frequencies at which the test-particle samples the correction 
signals. The diffusion increases the mean oscillation amplitudes of the 
particles, and so counteracts the cooling, hence it is also referred to as 
{\sl heating} of the beam. In order to reduce the undesired diffusion without 
lowering the gain which at the same time would decrease the cooling, the 
cooling system has to operate in a frequency range where the frequency bands 
just begin to overlap, and the spectral particle density reaches its minimum 
value. We will discuss this requirement in detail. Defining in Eq.\ 
(\ref{cool10}) the effective frequency distribution $\widehat{f}(\Omega)$ by 
\begin{equation} \label{cool10b}
\widehat{f}(\Omega) =  
\sum_{(l)} \frac{1}{|l + Q|}\, f\li( \frac{\Omega}{l + Q} \re),
\end{equation}
the spectral particle density can be expressed as $dN/d\Omega =N\widehat{f}
(\Omega)$. Owing to the internal summation over all frequency bands, particles 
from different bands can contribute to the value of $\widehat{f}$ at a given 
frequency $\Omega$, and thus the definition (\ref{cool10b}) includes the case 
of overlapping frequency bands (see Sect.\ \ref{basepart6}). For frequencies 
$\Omega$ at which the bands do not yet overlap only the band at the 
corresponding harmonic $m + Q \sim \Omega / \omega_0$ contributes in the 
summation in Eq.\ (\ref{cool10b}). For these harmonics the amplitude of 
$\widehat{f}(\Omega)$ decreases as $1/|m + Q|$ so that higher frequencies 
lead to the wanted reduction of the  spectral particle density, as long as the 
frequency bands do not overlap. In the case of overlapping bands the 
contributions of the individual bands  add up to a nearly constant value, and 
a further increase of the frequency would not leave any profit 
\cite{sac, bi2}. 

We will visualize this behaviour in the time domain. For that purpose we 
consider the case in which neighboring bands just touch so that their width 
$\Delta \Omega_l$ is in the order of the revolution frequency: $\Delta 
\Omega_l \sim \omega_0$. Then the correlation time of the fluctuations is 
 $\tau_{corr} \sim T_0$, i.e.\ signals which are sampled with the revolution 
time $T_0$ are statistically independent. The fluctuation spectrum in such a 
frequency interval has the character of uncorrelated, white noise, similar to 
e.g.\ electronic noise. This situation corresponds to the so-called {\sl 
good} or {\sl perfect mixing}. Accordingly, {\sl bad mixing} refers to the 
case of non-overlapping frequency bands in which the width of each band is 
smaller than the revolution frequency, and therefore the correlations last 
over more than one turn. Due to the larger amplitudes of the non-overlapping 
bands the bad mixing results in an enhanced diffusion. Improving the mixing, 
in general, necessitates broader frequency bands because then the particles 
within a single band spread over a larger frequency interval, and thus the 
spectral particle density becomes smaller.

The unwanted mixing in the drift coefficient (\ref{cool9}), however, increases 
at higher harmonics. If the spectral particle density hardly changes in the 
frequency interval relevant to the stochastic cooling (perfect mixing), higher 
frequencies will only enhance the unwanted mixing, and hence will degrade the 
cooling. In case the undesired phase shifts of the particles exceed even the 
value $\pm \pi/2$, the particles are no longer damped but excited. Therefore 
one has to find an acceptable compromise between the two counteracting effects.

The results which we have derived so far are now used to make predictions for 
the cooling process.

\subsection{The Cooling Rate} \plabel{coolpart11}

The cooling rate $\tau_{cool}^{-1}$ at the frequency $\omega_j$ follows from 
an integration of the Fokker-Planck equation (\ref{cool7}) with the drift and 
diffusion coefficients given by Eqs.\ (\ref{cool9}) and (\ref{cool10}). The 
calculation in Appendix \ref{parapart2} yields for the cooling rate the simple 
expression (\ref{para1}), 
\begin{equation} \eqlabel{cool10a}
\tau_{cool}^{-1} = \bar{F} + \frac{1}{2} \bar{D}.
\end{equation}
The relevance of the system bandwidth for the cooling rate becomes apparent 
from Eqs.\ (\ref{cool9}) and (\ref{cool10}). Both equations contain summations 
over all harmonics $(m + Q)\omega_j$, however only harmonics within the 
bandwidth of the cooling system contribute because only then $|\widetilde{G}
[(m + Q)\omega_j]| \not= 0$. If for each of these harmonics the cooling effect 
exceeds the diffusion, there will remain net cooling contributions which all 
enter into the cooling rate (\ref{cool10a}). Because a larger bandwidth covers 
more harmonics, it results in a faster cooling.

The only parameter which permits to control the ratio of drift and diffusion 
coefficients is the gain $\widetilde{G}(w)$ of the cooling system (see Eqs.\ 
(\ref{cool9}) and (\ref{cool10})). Of course we are interested in the value 
of the gain which results in the most efficient cooling operation, yielding 
the maximum cooling rate. According to Section \ref{coolpart7}, an arbitrary 
choice of the gain is only permitted within the limits of beam stability so 
that the predicted optimum value $\widetilde{G}_{opt}(w)$ has to be compared 
with the critical gain $\widetilde{G}_{crit}(w)$ in order to ensure that 
optimum cooling preserves the beam stability. We will satisfy this requirement 
in two steps:
\begin{itemize}
\item%
First, we calculate the optimum gain $\widetilde{G}_{opt}(w)$ for any possible 
frequency.
\item%
For each frequency, we compare the optimum value with the critical gain 
$\widetilde{G}_{crit}(w)$ with regard to the beam stability.
\end{itemize}

\subsubsection{The Optimum Gain}

To calculate the optimum gain we consider the coefficients $\bar{F}$ and 
$\bar{D}$ in the cooling rate (\ref{cool10a}) at an arbitrary, but fixed 
frequency, and interpret them as pure functions of the gain. The magnitude of 
the optimum gain $|\widetilde{G}_{opt}(\Omega_m)|$ at a given frequency 
$\Omega_m = (m + Q) \omega'$ is derived in Appendix \ref{parapart3}, and reads 
\begin{equation} \eqlabel{cool11}
\frac{1}{|\widetilde{G}_{opt}(\Omega_m)|} = \frac{2\kappa \omega' N}{Q}\, 
\pi \sum_{(l)} \frac{1}{|l + Q|}\, f\li( \frac{\Omega_m}{l + Q} \re).
\end{equation}

\subsubsection{Beam Stability under Optimum Cooling} 

Since the coherent beam modes completely describe the collective beam motion 
(see Sect.\ \ref{stabpart12}), they can be used to analyse the stability of 
the beam. To that end we interpret the frequency $\Omega_m$ as a mode 
frequency, and investigate how the critical and the optimum gain compare at 
this frequency. The magnitude of the critical gain $|\widetilde{G}_{crit}
(\Omega_m)|$ is given by (\ref{stab46}), 
\begin{equation} \eqlabel{cool12}
\frac{Q}{\kappa N |\widetilde{G}_{crit}(\Omega_m)|} =  
\pi \sum_{(l)} \frac{\cos\Phi_l^\delta(\Omega_m)}{|l + Q|}
\frac{|\Omega_m|}{|l + Q|}\; f\li( \frac{\Omega_m}{l + Q} \re) +
\sum_{(l)}\, {\cal P\!\!\!\!} \int\limits_{-\infty}^\infty \dx{\omega} 
f(\omega)\, |\omega|\, \frac{\sin\Phi_l^{\cal P}(\Omega_m, \omega)}
{(l + Q)\omega + \Omega_m}.
\end{equation}
The phases $\Phi_l^\delta(\Omega_m)$ and $\Phi_l^{\cal P}(\Omega_m, \omega)$ 
in this expression contain the phase advance of the modes between the pickup 
and kicker so that the precise value of $|\widetilde{G}_{crit}(\Omega_m)|$ 
depends on the tune and the azimuthal distance from the pickup to the kicker. 
General results for the beam stability in a stochastic cooling system are 
difficult to obtain because the behaviour can largely vary depending on the 
actual choice of the parameters. At this place, we hence can only estimate the 
ratio of the critical and optimum gain by making a few reasonable assumptions. 
Supposing a symmetric frequency distribution $f(\omega)$ and a negligible 
shift of the real mode frequencies, $\delta\Omega_m \approx 0$, we expect only 
a small contribution from the principal value integral in Eq.\ (\ref{cool12}) 
which can be neglected against the first term in this equation. Sizing the 
first term from above, we find a conservative criterion for the critical gain. 
Accordingly, we substitute for the cosine-function its maximum value: 
$\cos\Phi_l^\delta(\Omega_m) \to 1$, and obtain for Eq.\ (\ref{cool12}) 
\begin{equation} \eqlabel{cool13}
\frac{Q}{\kappa N |\widetilde{G}_{crit}(\Omega_m)|} \le
\pi \sum_{(l)} \frac{1}{|l + Q|} \frac{|\Omega_m|}{|l + Q|}\; 
f\li( \frac{\Omega_m}{l + Q} \re).
\end{equation}
The overlap condition for two adjacent frequency bands, say $m +Q$ and 
$l + Q = m + Q + \Delta l$ ($\Delta l = 0, \pm 1, \pm 2, \ldots$), is given by
$(m + Q)\Delta \omega \sim \Delta l\, \omega_0$ where $\Delta \omega$ denotes 
the width of the frequency distribution around the center frequency 
$\omega_0$. Hence the ratio of the two harmonic numbers can be written as
\[ 
\frac{l + Q}{m + Q} = 1 + \frac{\Delta l}{m + Q} \sim 
1 + \frac{\Delta \omega}{\omega_0} \sim 1 \pm 10^{-3} 
\]
which we approximate by one. With $|\Omega_m| = |m +Q| \omega'$ Eq.\ 
(\ref{cool13}) then becomes 
\[ 
\frac{1}{|\widetilde{G}_{crit}(\Omega_m)|} \le \frac{\pi\kappa\omega' N}{Q} 
\sum_{(l)} \frac{1}{|l + Q|}\; f\li( \frac{\Omega_m}{l + Q} \re).
\]
A comparison with the optimum gain (\ref{cool11}) immediately yields 
\[ \frac{1}{|\widetilde{G}_{crit}(\Omega_m)|} \le 
\frac{1}{2|\widetilde{G}_{opt}(\Omega_m)|}. \]
Relying on the above approximations, we find that the optimum gain is at most 
half the critical gain so that beam stability is guaranteed over the entire 
frequency range. A similar result has been obtained earlier in \cite{wei} for 
the case of non-overlapping frequency bands. Of course one should realize that 
the result of this section gives only an estimate because it does not take 
into account the contribution of the principal value integral. Precise 
predictions will, in general, require numerical computations, using the actual 
parameters of the cooling system and storage ring.

\chapter*{Conclusion}
\addcontentsline{toc}{chapter}{Conclusion}

The aim of this thesis was a description of stochastic cooling in the 
framework of control theory in order to allow predictions about the beam 
stability in such systems. This description has to treat the pickup and kicker 
of the cooling system as localized objects in the storage ring which impose a 
discrete time structure on both the generated signals and the forces 
experienced by the particles. The dynamics for a sampled interaction differ 
significantly from the motion under a force acting continuously in time, so 
that a careful stability analysis has to include the sampling.

The investigation presented here is based on the theory of multi-bunch 
feedback systems in which the control theory of discrete time signals had been 
adapted to feedback loops specific for accelerators. This special formulation 
takes strictly into account the positions of the pickup and kicker and the 
sampling of the interaction. 

Relying on this formulation, a general stability criterion has been derived 
for an unbunched beam undergoing linear transverse stochastic cooling. The 
result allows for overlapping frequency bands in the beam spectrum and 
therefore goes beyond existing treatments. For the first time, the boundaries 
of beam stability could be predicted over the entire frequency range. 

For the mathematical description of stochastic cooling a Fokker-Planck 
equation has been employed. This purely statistical approach does not include 
the collective motion of the particles and hence beam stability in the cooling 
system must be ensured. Owing to the results of this work, this prerequisite 
now can be verified. The stability criterion which has been obtained defines 
the boundaries within which the stability requirement is satisfied, and only 
there the Fokker-Planck equation produces physically reasonable results.

With a few assumptions a relation has been derived between the critical gain 
beyond which the beam motion becomes unstable and the optimum gain which 
allows the most efficient cooling operation. The comparison shows that a 
sufficient safety margin exists between these values, so it can be concluded 
that transverse stochastic cooling preserves beam stability even in the case 
of overlapping frequency bands. This result provides the Fokker-Planck 
approach to stochastic cooling with a well-founded physical and mathematical 
basis.

\chapter*{Acknowledgements}

I want to express my gratitude to the DESY directorate for giving me the 
possibility to write a thesis in the field of accelerator physics, and for the 
financial support during this time. 

I am much obliged to Prof.\ Dr.\ R.-D. Kohaupt whose qualified guidance and 
great knowledge of the subject made it possible for me to gain the necessary 
insight into the field of accelerator physics, and who has thus decisively 
contributed to the successful outcome of this thesis. His enthusiasm and 
liveliness in our illuminating discussions had always encouraged me and could 
give me fresh motivation for my work. I would also like to thank him for the 
critical study of my thesis. 

I am grateful to Prof.\ Dr.\ P. Schm\"user for carefully reading my 
manuscript, and for his numerous constructive suggestions which have become 
valuable additions to this work. 

Thanks to Dr.\ J. Feikes who never became tired of answering my questions with 
saintly patience, and who could give me many inspirations in our long 
discussions. 

I want to thank Dr.\ J. R. Maidment, Dr.\ T. Sen, Dr.\ N. Walker and S. G. 
Wipf for a careful reading of the manuscript. 

I express my thanks to my colleagues from MPY and MKK for providing a warm and 
close working atmosphere. 

I am deeply indebted to my family for their forbearing and continuous support 
during this time, and for the enormous patience and understanding which they 
had to summon up. 

\def\thefigure{\Alph{chapter}-\arabic{figure}}

\begin{appendix}

\chapter{Properties of the General Fourier Transformation} \plabel{FTE}

Here, we summarize only the properties of the general Fourier transformation 
used in the calculations of this work. Detailed information about the general 
Fourier transformation can be found e.g.\ in \cite{mys}. 

Let $f(t)$ and $g(t)$ be functions satisfying the condition (\ref{FTBed}). 
Their general Fourier transforms defined by (\ref{FTt2w}) are denoted as 
$\widetilde{f}(w)$ and $\widetilde{g}(w)$ respectively. For the relation 
between the original function $f(t)$ and its transform $\widetilde{f}(w)$ we 
use the symbolic notation $\widetilde{f}(w) = {\cal F\,} \li[f(t)\re]$. 
Then the following properties can be derived \cite{mys}: 

\begin{description}
\item[Damping Theorem]
Given a constant $\tau > 0$, then 
\begin{equation} \eqlabel{FTE1}
{\cal F\,} \li[f(t-\tau)\re] = e^{-iw\tau}\widetilde{f}(w).
\end{equation}

\item[Displacement Theorem]
Let $c$ be a complex constant. Then
\begin{equation} \eqlabel{FTE2}
{\cal F\,} \li[e^{ict}f(t)\re] = \widetilde{f}(w-c).
\end{equation}

\item[Differentiation Theorem]
Let $f^{(n)}(t)$ denote the $n$-th derivative of $f(t)$. Then 
\begin{equation} \eqlabel{FTE3}
{\cal F\,} \li[f^{(n)}(t)\re] = (iw)^n\widetilde{f}(w)
  - \frac{(iw)^{n-1}}{2\pi} f(+0) - \frac{(iw)^{n-2}}{2\pi} f'(+0) - \ldots
  - \frac{1}{2\pi} f^{(n-1)}(+0)
\end{equation}
where 
\[ f^{(n)}(+0) = \lim_{t\to+0}f^{(n)}(t). \]
Especially, 
\begin{equation} \eqlabel{FTE4}
{\cal F\,} \li[f''(t)\re] = -w^2\widetilde{f}(w) - \frac{iw}{2\pi} f(+0) - 
                    \frac{1}{2\pi} f'(+0).
\end{equation}

\item[Convolution Theorem]
For the convolution $(f \ast g)$ of $f(t)$ and $g(t)$, defined by
\[ (f \ast g)(t) = \int\limits_0^t\dx{t'} f(t')g(t-t') \quad\mbox{with}\quad
0 \le t < \infty, \]
it follows that 
\begin{equation} \eqlabel{FTE5}
{\cal F\,} \li[(f \ast g)(t)\re] = \widetilde{f}(w) \widetilde{g}(w).
\end{equation}

\end{description}

\chapter{The Harmonic Oscillator in the Formalism of the General Fourier 
Transformation} \plabel{hoxspart1}

\section{The Free Oscillator with Initial Conditions} \plabel{hoxspart2}

Here, we consider a free harmonic oscillator with frequency $\Omega_0$. 
To solve the equation of motion,
\[ \ddot{x}(t) + \Omega_0^2 \,x(t) = 0, \]
with initial conditions $x_0 = x(0)$ and $\dot{x}_0 = \dot{x}(0)$ we 
transform this equation according to Eq.\ (\ref{FTt2w}). Using the relation 
(\ref{FTE4}) this leads to 
\begin{equation} \eqlabel{hoxs1}
 -w^2\, \widetilde{x}(w) + \Omega_0^2\, \widetilde{x}(w) 
 - \frac{iw}{2\pi} x(0) - \frac{1}{2\pi} \dot{x}(0) = 0
\end{equation}
or
\[ (-w^2 + \Omega_0^2)\, \widetilde{x}(w) = \frac{1}{2\pi} (iwx_0 + \dot{x}_0).
\]
Hence follows 
\[
\widetilde{x}(w) = \frac{1}{2\pi} \frac{iwx_0 + \dot{x}_0}{(-w^2 + \Omega_0^2)}
 =  \frac{1}{2\pi} \frac{iwx_0 + \dot{x}_0}{(-w + \Omega_0)(w + \Omega_0)}
\]
so that the inverse transformation reads
\[
x(t) = \int\limits_C\dx{w} \widetilde{x}(w) e^{iwt} = 
    \frac{1}{2\pi} \int\limits_C\dx{w} 
    \frac{iwx_0 + \dot{x}_0}{(-w + \Omega_0)(w + \Omega_0)}\, e^{iwt}.
\]
The integrand has two poles on the real axis, $w_+=+\Omega_0$ and 
$w_-=-\Omega_0$. To evaluate the integral, we close the integral contour 
in the upper $w$-plane, as is shown in Fig.\ \ref{HOAbb1}. The integral 
along the arc $B$ gives no contribution because the integrand tends to zero 
for $|w| \longrightarrow \infty$. From Cauchy's residue theorem follows that 
only the residues of the poles $w_+$ and $w_-$ contribute, yielding 
\begin{equation} \eqlabel{hoxs2}
x(t) = 2\pi i \: {\rm res}_{w_+}\li\{\widetilde{x}(w)\re\}  e^{iw_+t} +
       2\pi i \: {\rm res}_{w_-}\li\{\widetilde{x}(w)\re\}  e^{iw_-t}.
\end{equation} 
\begin{figure}[htb] 
\centerline{\epsfig{figure=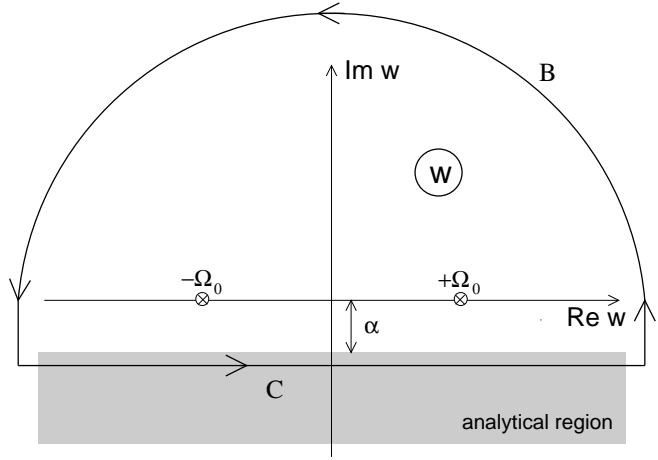,height=6cm}}
\caption{\label{HOAbb1}\sl The poles of the free harmonic oscillator. To apply 
the residue theorem, we have closed the integral contour $C$ in the upper 
$w$-plane.}
\end{figure} 
For the residues we find 
\[ 
{\rm res}_{w_+}\li\{\widetilde{x}(w)\re\}  e^{iw_+t}= 
\lim_{w\to w_+}\left\{(w-w_+)\,\widetilde{x}(w)\right\} e^{iw_+t} = 
-\frac{1}{2\pi} \frac{i\Omega_0 x_0 + \dot{x}_0}{2\Omega_0}\, e^{i\Omega_0t}
\]
and
\[
{\rm res}_{w_-}\li\{\widetilde{x}(w)\re\} e^{iw_-t} = 
\lim_{w\to w_-}\left\{(w-w_-)\,\widetilde{x}(w)\right\} e^{iw_-t} = 
\frac{1}{2\pi} \frac{-i\Omega_0 x_0 + \dot{x}_0}{2\Omega_0}\, e^{-i\Omega_0t}
\]
so that Eq.\ (\ref{hoxs2}) becomes
\begin{eqnarray*}
x(t) & = & \frac{1}{2\Omega_0} \left\{ (\Omega_0 x_0 - i\dot{x}_0) 
           e^{i\Omega_0t} + (\Omega_0 x_0 + i\dot{x}_0) e^{-i\Omega_0t} 
           \right\} \\
     & = & x_0\, \frac{1}{2} (e^{i\Omega_0t} + e^{-i\Omega_0t}) + 
           \frac{\dot{x}_0}{\Omega_0}\, \frac{1}{2i} (e^{i\Omega_0t} - 
            e^{-i\Omega_0t}).
\end{eqnarray*}
Finally, we get
\[ x(t) = x_0 \cos\Omega_0t + \frac{\dot{x}_0}{\Omega_0} \sin\Omega_0t. \]
Of course this is the expected result, but it has been derived without any 
intuitive ansatz. Furthermore the formalism includes the initial conditions
from the beginning of the calculation (see Eq.\ (\ref{hoxs1})).
 
\section{The Oscillator with Feedback Interaction} \plabel{hoxspart3}

We will reconsider the self-interacting oscillator discussed in Section 
\ref{basepart4}. According to Eq.\ (\ref{sys3}), the equation of motion reads
\begin{equation} \eqlabel{hoxs3}
\ddot{x}(t) + \Omega_0^2\, x(t) = \int\limits_0^\infty\dx{t'} G(t-t') x(t') 
 + g(t)
\end{equation}
with the initial conditions $x(0) = \dot{x}(0) = 0$. $\Omega_0$ denotes the 
frequency of the undisturbed oscillator. The impulse response $G(t)$ measures 
the strength of the feedback interaction, and $g(t)$ is the external 
$\delta$-pulse excitation. Using the relations (\ref{FTE4}) and (\ref{FTE5}), 
the general Fourier transformation of Eq.\ (\ref{hoxs3}) yields
\[ 
(-w^2 + \Omega_0^2)\, \widetilde{x}(w) = \widetilde{G}(w)\widetilde{x}(w) 
 + \widetilde{g}(w)
\]
where $\widetilde{g}(w) = A/\,2\pi$ is the transform of the $\delta$-pulse 
$g(t) = A \delta(t)$. From this follows
\begin{equation} \eqlabel{hoxs4}
\widetilde{x}(w) = \frac{1}{2\pi} 
\frac{A}{\left( -w^2 + \Omega_0^2 - \widetilde{G}(w) \right)}.
\end{equation}
According to Eq.\ (\ref{FTw2t}), the inverse transformation is given by 
\begin{equation} \eqlabel{hoxs5}
x(t) = \int\limits_C\dx{w} \widetilde{x}(w) e^{iwt} = 
       \frac{1}{2\pi} \int\limits_C\dx{w}
       \frac{A}{-w^2 + \Omega_0^2 - \widetilde{G}(w)}\, e^{iwt}.
\end{equation}
To proceed in this example, we assume an impedance $\widetilde{G}(w) = -i 2 
\gamma w$ with $|\gamma| < \Omega_0$. Then Eq.\ (\ref{hoxs4}) becomes
\[
\widetilde{x}(w) = \frac{1}{2\pi} \frac{A}{(-w^2 + \Omega_0^2 + i 2\gamma w)}
= -\frac{1}{2\pi} \frac{A}{(w - w_+)(w - w_-)}
\]
where
\[ w_{\pm} = i\gamma \pm \Omega \qquad\mbox{with}\qquad 
   \Omega = \sqrt{\Omega_0^2 - \gamma^2}.
\]
With that, Eq.\ (\ref{hoxs5}) can be written as 
\begin{equation} \eqlabel{hoxs6}
x(t) = -\frac{1}{2\pi} \int\limits_C\dx{w} 
       \frac{A}{(w - w_+)(w - w_-)}\, e^{iwt}.
\end{equation}
\begin{figure}[bt]
\centerline{\epsfig{figure=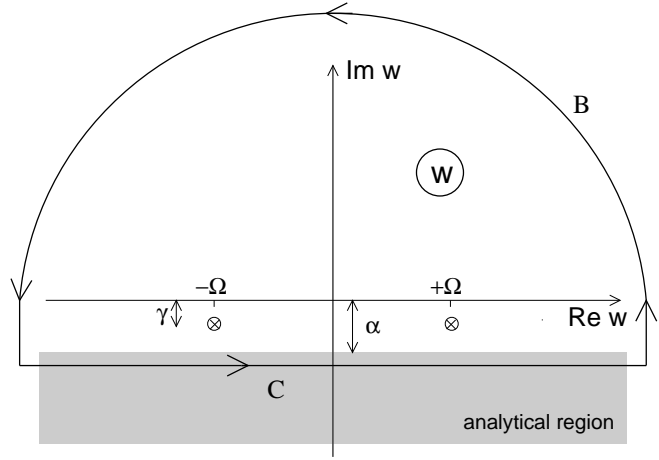,height=6cm}}
\caption{\label{HOAbb2}\sl For $\gamma < 0$ the poles are located in the lower
$w$-plane. The integral contour $C$ has to run below the poles and is 
continued in the upper $w$-plane to a closed contour.}
\end{figure}
The integrand has two poles, $w_+$ and $w_-$, in the upper or lower $w$-plane, 
depending on the sign of $\gamma$. The integral contour $C$ can be chosen as a 
straight line with ${\sf Im}\,w < \gamma$, thus lying in the analytic region 
of $\widetilde{x}(w)$. This is shown in Fig.\ \ref{HOAbb2} for the case 
$\gamma < 0$. The integral (\ref{hoxs6}) is evaluated in the same way as in 
the previous section, i.e.\ the integral contour is closed in the upper 
$w$-plane by the arc $B$ and the residue theorem is applied. This yields
\begin{equation} \eqlabel{hoxs7}
x(t) = 2\pi i \: {\rm res}_{w_+} \li\{ \widetilde{x}(w) \re\} e^{iw_+t} +
       2\pi i \: {\rm res}_{w_-} \li\{ \widetilde{x}(w) \re\} e^{iw_-t}.
\end{equation}
With $w_+ - w_- = 2\Omega$, we obtain for the residues
\[ {\rm res}_{w_+} \li\{ \widetilde{x}(w) \re\} e^{iw_+t} =
   -\frac{1}{2\pi} \frac{A}{(w_+ - w_-)}\, e^{iw_+t} = 
   -\frac{1}{2\pi} \frac{A}{2\Omega}\, e^{iw_+t} \]
and 
\[ {\rm res}_{w_-} \li\{ \widetilde{x}(w) \re\} e^{iw_-t}=
   -\frac{1}{2\pi} \frac{A}{(w_- - w_+)}\, e^{iw_-t} = 
   \frac{1}{2\pi} \frac{A}{2\Omega}\, e^{iw_-t}. \]
Since $iw_{\pm} = -\gamma \pm i\Omega$, Eq.\ (\ref{hoxs7}) becomes
\[ x(t) = -i \frac{A}{2\Omega} \left( e^{iw_+t} - e^{iw_-t} \right) 
        =  \frac{A}{\Omega} e^{-\gamma t}\, \frac{1}{2i} 
           \left( e^{i\Omega t} - e^{-i\Omega t} \right) \]
and is finally written as 
\[ x(t) = \frac{A}{\Omega} e^{-\gamma t} \sin\Omega t 
   \qquad\mbox{with}\qquad \Omega = \sqrt{\Omega_0^2 - \gamma^2}.
\]
Depending on the sign of $\gamma$, we obtain an exponentially increasing 
$(\gamma < 0)$ or decreasing $(\gamma > 0)$ solution, i.e.\ the feedback
interaction via an impedance can lead to an unstable motion of the oscillator. 

\chapter{The Derivation of the Self-consistent Solution for the Unbunched Beam}
\plabel{somopart1}

In this appendix, we will derive the self-consistent solution of the beam 
signal $\widetilde{S}(w)$ at the pickup, and the defining equation of the 
coefficients $\widetilde{C}_l(w)$. The calculation starts with the transformed 
equation of motion (\ref{stab20b}),
\[
(-w^2 + \Omega_j^2)\,\widetilde{x}_j(w) = \Omega_j^2 \beta_K^{3/2} 
e^{-iw\tau_j^\DK} \left( \widehat{F}_j(w) + \widehat{g}_j(w) \right).
\]
Both sides of this equation are multiplied by $e^{-iwt_j^0}$ and divided
by $(\Omega_j^2 - w^2)$, resulting in 
\begin{equation} \eqlabel{stab20a}
\widetilde{x}_j(w) e^{-iwt_j^0} = 
\frac{e^{-iwt_j^0} e^{-iw\tau_j^\DK}}{\Omega_j^2 - w^2}\, 
\Omega_j^2 \beta_K^{3/2} \left( \widehat{F}_j(w) + \widehat{g}_j(w) \right).
\end{equation}
By inserting the decomposition (\ref{modes}) on the left-hand side and 
substituting $w \to w - m\omega_j$, this expression becomes 
\[
\frac{1}{N} \sum_l \widetilde{C}_l(w - m\omega_j) e^{il\theta_j^0} =
\frac{e^{-i(w - m\omega_j)t_j^0} e^{-i(w - m\omega_j)\tau_j^\DK}}
{\Omega_j^2 - (w - m\omega_j)^2}\, \Omega_j^2 \beta_K^{3/2} 
\left( \widehat{F}_j(w - m\omega_j) + \widehat{g}_j(w - m\omega_j) \right).
\]
Now, the periodicity of the functions $\widehat{F}_j(w)$ and 
$\widehat{g}_j(w)$ is used, and both sides of the equation are multiplied by 
$e^{-im\theta_j^0}$, followed by a summation over all particles $j$. Since 
$\theta_j^0 = \omega_j t_j^0$, this yields 
\begin{equation} \eqlabel{stab21}
\frac{1}{N} \sum_j \sum_l \widetilde{C}_l(w - m\omega_j) 
e^{i(l-m)\theta_j^0} = \sum_j \frac{e^{-iwt_j^0} 
e^{-i(w - m\omega_j)\tau_j^\DK}} {\Omega_j^2 - (w - m\omega_j)^2}\,
\Omega_j^2  \beta_K^{3/2} \left( \widehat{F}_j(w) + \widehat{g}_j(w) \right).
\end{equation}
In this expression, we perform the transition to the continuous beam, as 
described in Section \ref{stabpart5}, i.e.\ the summation over the particles 
is replaced by an integration, following the instruction given by Eq.\ 
(\ref{sum2int}). On the left-hand side of Eq.\ (\ref{stab21}), we get 
\[ \frac{1}{N} \sum_j \sum_l \widetilde{C}_l(w - m\omega_j)
   e^{i(l-m)\theta_j^0} \longrightarrow \]
\begin{equation} \eqlabel{stab22}
\sum_{(l)} \int\dx{\omega} f_0(\omega) \widetilde{C}_l(w - m\omega)
\int\!\frac{d\theta^0}{2\pi}\; e^{i(l-m)\theta^0} = 
\int\dx{\omega} f_0(\omega) \widetilde{C}_m(w - m\omega)
\end{equation}
since 
\[ \int\!\frac{d\theta^0}{2\pi}\; e^{i(l-m)\theta^0} = \delta_{lm}. \]
The comments on page \pageref{Qtaudef} permit us to write $\Omega_j = 
Q\omega_j$ and $\tau_j^\DK = \varphi^\DK/Q\omega_j$. Using Eqs.\ 
(\ref{stab17}) and (\ref{stab18}) and the definition $\bar{\kappa} = 2\pi RQ 
\sqrt{\beta_K}$, the first term $T_I(w)$ on the right-hand side of Eq.\ 
(\ref{stab21}) containing the feedback force $\widehat{F}_j(w)$ reads 
\[
T_I(w) = \bar{\kappa} \sum_j \omega_j^2\,
\frac{e^{-i(w - m\omega_j)\varphi^\DK/Q\omega_j}}
{(Q\omega_j)^2 - (w - m\omega_j)^2} 
\sum_{(k)} \widetilde{G}(w + k\omega_j) \widetilde{S}(w + k\omega_j) 
e^{i(w + k\omega_j)\theta^\DK/\omega_j} e^{ik\theta_j^0}.
\]
After the transition to the continuous beam, the integration over the initial
azimuths yields
\[ \int\!\frac{d\theta^0}{2\pi}\; e^{ik\theta^0} = \delta_{k0} \]
so that only the term with $k=0$ contributes to the summation. Hence follows
\begin{equation} \eqlabel{stab23}
T_I(w) \longrightarrow \bar{\kappa} N \int\dx{\omega} f_0(\omega)\, \omega^2\,
      \frac{e^{-i(w - m\omega)\varphi^\DK/Q\omega}}
      {(Q\omega)^2 - (w - m\omega)^2}\, 
      \widetilde{G}(w) \widetilde{S}(w) e^{iw\theta^\DK/\omega}.
\end{equation}
For the second term $T_{I\!I}(w)$ on the right-hand side of Eq.\ 
(\ref{stab21}) coming from the external excitation $\widehat{g}_j(w)$, the 
analogous calculation leads with Eqs.\ (\ref{stab19}) and (\ref{stab20}) to 
\begin{equation} \eqlabel{stab24}
T_{I\!I}(w) \longrightarrow \bar{\kappa} N \int\dx{\omega} f_0(\omega)\,
\omega^2\, \frac{e^{-i(w - m\omega)\varphi^\DK/Q\omega}} 
{(Q\omega)^2 - (w - m\omega)^2}\, \widetilde{g}(w)
e^{iw\theta^\DK/\omega}.
\end{equation}
The combination of the results (\ref{stab22}), (\ref{stab23}) and 
(\ref{stab24}) gives
\[
\int\dx{\omega} f_0(\omega) \widetilde{C}_m(w - m\omega) = 
\bar{\kappa} N \int\dx{\omega} f_0(\omega)\, \omega^2\,
\frac{e^{-i(w - m\omega)\varphi^\DK/Q\omega}}
{(Q\omega)^2 - (w - m\omega)^2} e^{iw\theta^\DK/\omega} \left(
\widetilde{G}(w) \widetilde{S}(w) + \widetilde{g}(w) \right).
\]
Multiplying by $\sqrt{\beta_P}/\, 2\pi$, summing over all values $m$ and using 
Eq.\ (\ref{stab7}),  we can write this equation as 
\begin{equation} \eqlabel{stab25}
\widetilde{S}(w) = \widetilde{R}(w) \widetilde{G}(w) \widetilde{S}(w) 
                 + \widetilde{R}(w) \widetilde{g}(w)
\end{equation}
with the definitions $\kappa = \bar{\kappa}\cdot\sqrt{\beta_P}/\,2\pi$ and
\begin{equation} \eqlabel{stab25b}
\widetilde{R}(w) = \kappa N \int\limits_0^\infty \dx{\omega} f_0(\omega)\,
\omega^2 \sum_{(m)} \frac{e^{-i(w + m\omega)\varphi^\DK/Q\omega}}
{(Q\omega)^2 - (w + m\omega)^2}\, e^{iw\theta^\DK/\omega}.
\end{equation}
Solving Eq.\ (\ref{stab25}) for $\widetilde{S}(w)$, we finally find the 
self-consistent solution to the coherent beam motion at the pickup:
\begin{equation} \eqlabel{stab26a}
\widetilde{S}(w) = \frac{\widetilde{R}(w) \widetilde{g}(w)}
{1 - \widetilde{R}(w) \widetilde{G}(w)}.
\end{equation}
From this result, the expansion coefficients $\widetilde{C}_l(w)$ in Eq.\ 
(\ref{modes}) can be derived. Starting at Eq.\ (\ref{stab20a}), we insert the 
decomposition (\ref{modes}) and obtain 
\[
\frac{1}{N} \sum_m \widetilde{C}_m(w) e^{im\theta_j^0} = 
\frac{e^{-iwt_j^0} e^{-iw\tau_j^\DK}}{(Q\omega_j)^2 - w^2}\, 
(Q\omega_j)^2 \beta_K^{3/2} \left( \widehat{F}_j(w) + \widehat{g}_j(w) \right).
\]
This equation is multiplied by $ e^{-il\theta_j^0}$ and summed over all 
particles $j$, yielding
\begin{equation} \eqlabel{stab27}
\frac{1}{N} \sum_j  \sum_m \widetilde{C}_m(w) e^{i(m-l)\theta_j^0} = 
\sum_j \frac{e^{-iw\tau_j^\DK} e^{-i(w + l\omega_j)t_j^0}}
{(Q\omega_j)^2 - w^2}\, (Q\omega_j)^2 \beta_K^{3/2} 
\left( \widehat{F}_j(w) + \widehat{g}_j(w) \right).
\end{equation}
Now we replace the summation over the particles by an integration, according 
to (\ref{sum2int}). On the left-hand side, the integral over $\theta^0$ 
contributes only if $m = l$, and owing to the normalization $\int\!\! d\omega 
f_0(\omega)= 1$, we thus obtain
\begin{equation} \eqlabel{stab28}
\frac{1}{N} \sum_j  \sum_m \widetilde{C}_m(w) e^{i(m-l)\theta_j^0}
\longrightarrow \widetilde{C}_l(w).
\end{equation}
On the right-hand side of Eq.\ (\ref{stab27}), the terms $T_I(w)$ and 
$T_{I\!I}(w)$ containing the feedback force and the external excitation 
respectively are again treated separately. Using Eqs.\ (\ref{stab17}) and 
(\ref{stab18}), the feedback term reads
\[
T_I(w) = \bar{\kappa} \sum_j \omega_j^2\,
\frac{e^{-iw\varphi^\DK/Q\omega_j}} {(Q\omega_j)^2 - w^2} 
\sum_{(k)} \widetilde{G}(w + k\omega_j) \widetilde{S}(w + k\omega_j) 
e^{i(w + k\omega_j)\theta^\DK/\omega_j} e^{i(k-l)\theta_j^0}
\]
so that the transition to the continuous beam leads to 
\begin{equation} \eqlabel{stab29}
T_I(w) \longrightarrow \bar{\kappa} N \int\dx{\omega} f_0(\omega)\, \omega^2\,
       \frac{e^{-iw\varphi^\DK/Q\omega}}{(Q\omega)^2 - w^2}\, 
       \widetilde{G}(w + l\omega) \widetilde{S}(w + l\omega) 
       e^{i(w + l\omega)\theta^\DK/\omega}.
\end{equation}
Similarly, we obtain for the external excitation
\begin{equation} \eqlabel{stab30}
T_{I\!I}(w) \longrightarrow \bar{\kappa} N \int\dx{\omega} f_0(\omega)\,
       \omega^2\, \frac{e^{-iw\varphi^\DK/Q\omega}}{(Q\omega)^2 - w^2}\, 
       \widetilde{g}(w + l\omega) e^{i(w + l\omega)\theta^\DK/\omega}.
\end{equation}
Combining the Eqs.\ (\ref{stab28}), (\ref{stab29}) and (\ref{stab30}) results 
in 
\[
\widetilde{C}_l(w) = \bar{\kappa} N \int\dx{\omega} f_0(\omega)\, \omega^2\,
     \frac{e^{-iw\varphi^\DK/Q\omega}}{(Q\omega)^2 - w^2}\,
     e^{i(w + l\omega)\theta^\DK/\omega} \left(
     \widetilde{G}(w + l\omega) \widetilde{S}(w + l\omega) +
     \widetilde{g}(w + l\omega) \right). 
\]
Inserting here the self-consistent solution (\ref{stab26a}) reduces the 
expression in the brackets. Writing $w' = w + l\omega$, it can easily be shown
that 
 \[
\widetilde{G}(w') \widetilde{S}(w') + \widetilde{g}(w') =
\widetilde{G}(w')\, \frac{\widetilde{R}(w') \widetilde{g}(w')}
{1 - \widetilde{G}(w') \widetilde{R}(w')} + \widetilde{g}(w') =
\frac{\widetilde{g}(w')}{1 - \widetilde{G}(w') \widetilde{R}(w')}.
\]
From this finally follows 
\begin{equation} \eqlabel{stab31b}
\widetilde{C}_l(w) = \bar{\kappa} N \int\limits_0^\infty \dx{\omega} 
f_0(\omega)\, \omega^2\, \frac{e^{-iw\varphi^\DK/Q\omega}}{(Q\omega)^2 - w^2}\,
\frac{\widetilde{g}(w + l\omega) e^{i(w + l\omega)\theta^\DK/\omega}}
{1 - \widetilde{G}(w + l\omega) \widetilde{R}(w + l\omega)}.
\end{equation}

\chapter{The Derivation of the Beam Stability Criteria}
\plabel{Gcripart1}

Starting at the stability condition (\ref{stab36}), we will present the 
calculation which lead to the stability criteria (\ref{stab46}) and 
(\ref{stab47}). For the evaluation of the limit in Eq.\ (\ref{stab36}), it is 
convenient to define a new distribution function $f(\omega)$ by 
\begin{equation} \eqlabel{stab36a}
f(\omega) = \frac{1}{2} \Big\{ f_0(\omega) + f_0(-\omega) \Big\} 
\qquad\mbox{with}\qquad
\int\limits_{-\infty}^{+\infty} \dx{\omega} f(\omega) = 1 
\quad\mbox{and}\quad f(-\omega) = f(\omega).
\end{equation}
Using this definition, we can write the Eq.\ (\ref{stab25a}) in the following 
form: 
\begin{eqnarray*}
\widetilde{R}(w) 
& = & \frac{\kappa N}{2 Q} \sum_{(\pm)} \sum_{(m)}\, \int\limits_0^\infty 
      \dx{\omega} f_0(\omega)\, \omega\, 
      \frac{\pm e^{-iw\Delta\varphi/Q\omega}}
      {w + (m \pm Q)\omega}\, e^{-im\varphi^\DK/Q} \\
& = & \frac{\kappa N}{Q} \sum_{(m)}\, 
      \int\limits_{-\infty}^\infty \dx{\omega} f(\omega)\, \omega\, 
      \frac{e^{-i\sigma_{\!\omega}w\Delta\varphi/Q\omega}}
      {w + (m + Q)\omega}\, e^{-i\sigma_{\!\omega}m\varphi^\DK/Q}
\end{eqnarray*}
where $\sigma_{\!\omega} = \omega/|\omega|$ indicates the sign of $\omega$, 
and $\Delta\varphi = \varphi^\DK - Q\theta^\DK$. Then $\widetilde{R}(\Omega_l 
+ i\gamma_l)$ can be expressed as 
\begin{eqnarray*}
\widetilde{R}(\Omega_l + i\gamma_l) 
& = & \frac{\kappa N}{Q} \sum_{(m)}\,
      \int\limits_{-\infty}^\infty \dx{\omega} f(\omega)\, \omega\, 
      \frac{e^{-i\sigma_{\!\omega}\Omega_l\Delta\varphi/Q\omega}\, 
      e^{\gamma_l\sigma_{\!\omega}\Delta\varphi/Q\omega}}
      {\Omega_l + (m + Q)\omega + i\gamma_l}\, 
      e^{-i\sigma_{\!\omega}m\varphi^\DK/Q}                            \\
& = & \frac{\kappa N}{Q} \sum_{(m)}
      \frac{1}{m + Q} \int\limits_{-\infty}^\infty \dx{\omega} f(\omega)\, 
      \omega\, \frac{e^{-i\sigma_{\!\omega}\Omega_l\Delta\varphi/Q\omega}\,
      e^{\gamma_l\sigma_{\!\omega}\Delta\varphi/Q\omega}} 
      {\omega + \frac{\Omega_l}{m + Q} + i\sgn{m}
      \frac{\gamma_l}{|m + Q|}}\, e^{-i \sigma_{\!\omega}m\varphi^\DK/Q}.
\end{eqnarray*}
Here, $\sgn{m} = (m + Q)/|m + Q|$ gives the sign of $(m + Q)$. The limit 
$\gamma_l\to 0^+$ of $\widetilde{R}(\Omega_l + i\gamma_l)$ can be computed 
using the identities
\[ \lim_{\epsilon \to 0^+} \frac{1}{x \pm i\epsilon} = 
   {\cal P}(\frac{1}{x}) \mp i\pi \delta(x) \]
where ${\cal P}(\frac{1}{x})$ denotes the principal value. In the present 
case, we obtain 
\begin{equation} \eqlabel{stab37}
\lim_{\gamma_l \to 0^+} \widetilde{R}(\Omega_l + i\gamma_l) = 
\frac{\kappa N}{Q} \left\{ I_{\cal P}(\Omega_l) - i\pi
I_\delta(\Omega_l) \right\}
\end{equation}
with the principal value integral 
\begin{equation} \eqlabel{stab38}
I_{\cal P}(\Omega_l) = \sum_{(m)}\, {\cal P\!\!\!\!} 
\int\limits_{-\infty}^\infty \dx{\omega} f(\omega)\, \omega\, 
\frac{e^{-i\sigma_{\!\omega}\Omega_l\Delta\varphi/Q\omega}}
{(m + Q)\omega + \Omega_l}\; 
e^{-i\sigma_{\!\omega}m\varphi^\DK/Q}, 
\end{equation}
and the integral containing the $\delta$-function
\[
I_\delta(\Omega_l) = \sum_{(m)} \frac{\sgn{m}}{m + Q} 
\int\limits_{-\infty}^\infty \dx{\omega} f(\omega)\, \omega\, 
e^{-i\sigma_{\!\omega}\Omega_l\Delta\varphi/Q\omega}\,
e^{-i\sigma_{\!\omega}m\varphi^\DK/Q}\,
\delta \li( \omega + \frac{\Omega_l}{m + Q} \re).
\]
Contributions from the $\delta$-function only arise at the frequencies $\omega 
= - \Omega_l/(m + Q)$ so that $\sigma_{\!\omega} = - \sigma_{\!\Omega_l} 
\sgn{m}$, depending on the sign $\sigma_{\!\Omega_l}$ of $\Omega_l$. Together 
with $\sgn{m}(m + Q) = |m + Q|$ and $f(-\omega) = f(\omega)$, this leads to 
\begin{equation} \eqlabel{stab40}
I_\delta(\Omega_l) = - \sum_{(m)} \frac{1}{|m + Q|}\; 
     f \li( \frac{\Omega_l}{m + Q} \re)\, \frac{\Omega_l}{m + Q}\,
     e^{i\sigma_{\!\Omega_l}\sgn{m}(m\theta^\DK - \Delta\varphi)}.
\end{equation}
Combining the Eqs.\ (\ref{stab36}), (\ref{stab37}), (\ref{stab38}) and 
(\ref{stab40}), we obtain the defining equation for the critical gain 
$\widetilde{G}_{crit}(w)$: 
\begin{eqnarray*}
1 & = & \widetilde{G}_{crit}(\Omega_l)\, \frac{i\pi\kappa N}{Q} 
        \sum_{(m)} \frac{1}{|m + Q|}\; f\li( \frac{\Omega_l}{m + Q} \re)\,
        \frac{\Omega_l}{m + Q}\,
        e^{i\sigma_{\!\Omega_l}\sgn{m}(m\theta^\DK - \Delta\varphi)} + \\
  &   & \widetilde{G}_{crit}(\Omega_l)\, \frac{\kappa N}{Q}
        \sum_{(m)}\, {\cal P\!\!\!\!}
        \int\limits_{-\infty}^\infty \dx{\omega} f(\omega)\, \omega\, 
        \frac{e^{-i\sigma_{\!\omega}\Omega_l\Delta\varphi/Q\omega}}
        {(m + Q)\omega + \Omega_l}\; e^{-i\sigma_{\!\omega}m\varphi^\DK/Q}.
\end{eqnarray*}
Expressing the critical gain in terms of amplitude and phase, i.e.\ 
$\widetilde{G}_{crit}(w) = |\widetilde{G}_{crit}(w)|e^{i\psi(w)}$, the above 
equation can be written as
\begin{eqnarray} \eqlabel{stab41}
\frac{Q}{\kappa N |\widetilde{G}_{crit}(\Omega_l)|}
& = & i\pi \sum_{(m)} \frac{1}{|m + Q|}\; 
      f\li( \frac{\Omega_l}{m + Q} \re)\, \frac{\Omega_l}{m + Q}\,
      e^{i\bar{\Phi}_m^\delta(\Omega_l)} +                         \nonumber\\
&   & \sum_{(m)}\, {\cal P\!\!\!\!} \int\limits_{-\infty}^\infty
      \dx{\omega} f(\omega)\, \omega\, 
      \frac{e^{i\bar{\Phi}_m^{\cal P}(\Omega_l, \omega)}}
      {(m + Q)\omega + \Omega_l}
\end{eqnarray}
where the phases are 
\begin{equation} \eqlabel{stab42}
\bar{\Phi}_m^\delta(\Omega_l) = \psi(\Omega_l) + \sigma_{\!\Omega_l}\sgn{m}\, 
(m\theta^\DK - \Delta\varphi)
\end{equation}
and
\begin{equation} \eqlabel{stab43}
\bar{\Phi}_m^{\cal P}(\Omega_l, \omega) =
  \psi(\Omega_l) - \sigma_{\!\omega}\, \Omega_l\Delta\varphi/Q\omega 
  - \sigma_{\!\omega}\, m\varphi^\DK/Q.
\end{equation}
The assumptions about stochastic cooling systems (see page \pageref{coolsys}) 
allow the further evaluation of the phases $\bar{\Phi}_m^\delta(\Omega_l)$ and 
$\bar{\Phi}_m^{\cal P}(\Omega_l, \omega)$. With $\omega' = \omega_0 + 
\delta\omega'$ and $\Delta\varphi = \varphi^\DK - Q\theta^\DK$, it follows for 
Eq.\ (\ref{stab42}) that 
\begin{eqnarray*}
\bar{\Phi}_m^\delta(\Omega_l) 
& \approx & -(l + Q)\theta^\DK\frac{\omega'}{\omega_0} + \sgn{l}\sgn{m}\,
            (m\theta^\DK - \varphi^\DK + Q\theta^\DK)                     \\
& \approx & -\sgn{l}\sgn{m}\, \varphi^\DK 
            - \Big[ (l + Q) - \sgn{l}\sgn{m}\, (m + Q) \Big]\theta^\DK
            - (l + Q)\theta^\DK\frac{\delta\omega'}{\omega_0}             \\   
& \approx & -\sgn{l}\sgn{m}\, \varphi^\DK 
            -\sgn{l}\, \Big[ |l + Q| - |m + Q| \Big]\theta^\DK
            - (l + Q)\theta^\DK\frac{\delta\omega'}{\omega_0}.
\end{eqnarray*}
Defining the relative frequency difference $\Delta'(\omega) = (\omega' - 
|\omega|) / |\omega|$ of a frequency $\omega$ from the frequency $\omega'$, 
Eq.\ (\ref{stab43}) becomes 
\begin{eqnarray*}
\bar{\Phi}_m^{\cal P}(\Omega_l, \omega)
& \approx & -(l + Q)\theta^\DK\frac{\omega'}{\omega_0}
            - \sigma_{\!\omega}\, (l + Q)\frac{\Delta\varphi}{Q}
              \frac{\omega'}{\omega}
            - \sigma_{\!\omega}\, m\frac{\varphi^\DK}{Q}                    \\
& \approx & -(l + Q)\theta^\DK(1 + \frac{\delta\omega'}{\omega_0})
            - (l + Q)\frac{\Delta\varphi}{Q}(1 + \Delta'(\omega))
            - \sigma_{\!\omega}\,  (m + Q)\frac{\varphi^\DK}{Q}
            + \sigma_{\!\omega}\, \varphi^\DK                               \\
& \approx & \sigma_{\!\omega}\, \varphi^\DK
            - \Big[ (l + Q) + \sigma_{\!\omega}\, (m + Q) \Big]
              \frac{\varphi^\DK}{Q}
            - (l + Q)\frac{\Delta\varphi}{Q}\Delta'(\omega)
            - (l + Q)\theta^\DK\frac{\delta\omega'}{\omega_0}.
\end{eqnarray*}
It is convenient to write the phases as $\bar{\Phi}_m^\delta = -\sgn{l}\sgn{m} 
\varphi^\DK - \Phi_m^\delta$ and $\bar{\Phi}_m^{\cal P} = \sigma_{\!\omega} 
\varphi^\DK - \Phi_m^{\cal P}$ because then the betatron phase advance 
$\varphi^\DK$ can be considered explicitly. According to the remarks on page 
\pageref{coolsys}, we set $\varphi^\DK = \pi/2 + 2\pi n, n = 0, 1, \dots$, and 
obtain for the real and imaginary parts of Eq.\ (\ref{stab41}) 
\begin{equation} \eqlabel{stab46a}
\frac{Q}{\kappa N |\widetilde{G}_{crit}(\Omega_l)|} =  
\pi \sum_{(m)} \frac{\cos\Phi_m^\delta(\Omega_l)}{|m + Q|}
\frac{|\Omega_l|}{|m + Q|}\; f\li( \frac{\Omega_l}{m + Q} \re) +
\sum_{(m)}\, {\cal P\!\!\!\!} \int\limits_{-\infty}^\infty \dx{\omega} 
f(\omega)\, |\omega|\, \frac{\sin\Phi_m^{\cal P}(\Omega_l, \omega)}
{(m + Q)\omega + \Omega_l}
\end{equation}
and
\begin{equation} \eqlabel{stab47a}
\pi \sum_{(m)} \frac{\sin\Phi_m^\delta(\Omega_l)}{|m + Q|}
\frac{|\Omega_l|}{|m + Q|}\; f\li( \frac{\Omega_l}{m + Q} \re) =
\sum_{(m)}\, {\cal P\!\!\!\!} \int\limits_{-\infty}^\infty \dx{\omega} 
f(\omega)\, |\omega|\, \frac{\cos\Phi_m^{\cal P}(\Omega_l, \omega)}
{(m + Q)\omega + \Omega_l}.
\end{equation}
These equations are the basic relations for the stability analysis carried out 
in this work.

\chapter{The Calculation of the Drift and Diffusion Coefficients} 
\plabel{fdkopart1}

In this section, we will derive the drift and diffusion coefficients of the 
Fokker-Planck equation (\ref{cool7}). To the order of approximation, the 
evaluation of the expressions in (\ref{cool8}) uses the unperturbed zero-order 
particle trajectories. In this case the transverse motion can be written 
as\footnote{A dot on top of a variable symbolizes the derivative with respect 
to the quasi-time $\bar{\tau}$.}
\[ 
x(\bar{\tau}) = \,A \cos(\Omega\bar{\tau} + \varphi^0) \qquad\mbox{and}\qquad
\dot{x}(\bar{\tau}) = - A\Omega\, \sin(\Omega\bar{\tau} + \varphi^0)
\]
where $A$, $\Omega = Q \omega$ and $\varphi^0$ are the amplitude, frequency 
and initial phase of the betatron oscillation respectively. The action 
$I(\bar{\tau})$ of a particle along its unperturbed trajectory is given by
\[
I(\bar{\tau}) = \frac{1}{2} \left\{ x^2(\bar{\tau}) + 
\frac{1}{\Omega^2}\, \dot{x}^2(\bar{\tau}) \right\}
\]
from which immediately follows that 
\[
\dot{I}(\bar{\tau}) = \frac{1}{\Omega^2}\, \dot{x}(\bar{\tau}) \left\{
\ddot{x}(\bar{\tau}) + \Omega^2 x(\bar{\tau}) \right\} =
\frac{1}{\Omega^2}\, \dot{x}(\bar{\tau}) K(\bar{\tau}) 
\]
where $K(\bar{\tau})$ denotes the force acting on the particle. The change 
$\Delta I$ of the action after a time $\Delta T$ can be obtained by integrating
$\dot{I}(\bar{\tau})$ over this time interval which then allows a calculation
of the drift and diffusion coefficients. Again it should be pointed out that 
this approximation implicitly presumes a stable beam motion, and therefore
the expressions for the drift and diffusion coefficients are only valid within 
the boundaries of beam stability. The consequences arising from this are 
discussed in the introduction and in Section \ref{coolpart7}.

\subsubsection{The Drift Coefficient}

The drift coefficient of the particle $j$ is given by
\begin{equation} \eqlabel{fdko1}
F_j(I_j) = \Big\langle \Big\langle \frac{\Delta I_j^F}{\Delta T} 
\Big\rangle \Big\rangle_{\varphi_j^0}.
\end{equation}
The change $\Delta I_j^F$ of the action experienced by the particle in a time 
interval $\Delta T$ due to its self-interaction can be evaluated from
\begin{equation} \eqlabel{fdko2}
\Delta I_j^F = \frac{1}{\Omega^2} \int\limits_0^{\Delta T} \dx{\qt{j}}
\dot{x}(\qt{j}) F_j^S(\qt{j}).
\end{equation}
According to Eq.\ (\ref{cool4}), the self-force $F_j^S(\qt{j})$ reads
\begin{equation} \eqlabel{fdko3}
F_j^S(\qt{j}) = 
\kappa \omega_j^2 \int \dx{w} e^{iw\qt{j}} e^{-iw\tau_j^\DK} \widehat{G}_j(w)
\sum_{(l)} \widetilde{x}_j(w + l\omega_j)
\end{equation}
with the periodic function $\widehat{G}_j(w)$ defined by
\[
\widehat{G}_j(w) = \sum_{(m)} \widetilde{G}(w + m\omega_j)\,
e^{i(w + m\omega_j)t_j^\DK}.
\]
The transform of the unperturbed motion has already been derived in Section
\ref{hoxspart2}, yielding 
\begin{equation} \eqlabel{fdko3a}
\widetilde{x}_j(w) = \frac{1}{2\pi} 
\frac{iwx_j^0 + \dot{x}^0_j}{\Omega_j^2 - w^2} \qquad\mbox{with}\qquad
x_j^0 = A_j\, \cos\varphi_j^0 \quad\mbox{and}\quad 
\dot{x}^0_j = -A_j\Omega_j\, \sin\varphi_j^0.
\end{equation}
The integral over $w$ in Eq.\ (\ref{fdko3}) only contributes at the poles 
of the function $\widetilde{x}_j(w)$ (see Sect.\ \ref{hoxspart2}), and the 
elementary calculation leads to 
\[
F_j^S(\qt{j}) = 
\frac{\kappa \omega_j^2}{2} A_j  \sum_\pm \sum_{(l)} 
\widehat{G}_j(\pm \Omega_j)\, e^{-i(l\omega_j \pm \Omega_j)\tau_j^\DK}
e^{\pm i\varphi_j^0} e^{i(l\omega_j \pm \Omega_j)\qt{j}}.
\]
Inserting this expression into Eq.\ (\ref{fdko2}) and recalling $I_j = 
A_j^2/2$, we obtain 
\[
\Delta I_j^F = \frac{\kappa\omega_j}{2iQ}\, I_j \sum_{\pm'} \sum_\pm \sum_{(l)}
(\mp')\, \widehat{G}_j(\pm \Omega_j)\, e^{-i(l\omega_j \pm \Omega_j)\tau_j^\DK}
e^{\pm i\varphi_j^0} e^{\pm' i\varphi_j^0} \int\limits_0^{\Delta T} \dx{\qt{j}}
e^{i(l\omega_j \pm \Omega_j \pm' \Omega_j)\qt{j}}.
\]
The time integration can easily be carried out, resulting in 
\[
\int\limits_0^{\Delta T} \dx{\qt{j}} 
e^{i(l\omega_j \pm \Omega_j \pm' \Omega_j)\qt{j}} = 
\frac{\sin(l\omega_j \pm \Omega_j \pm' \Omega_j)\Delta T/2}
{(l\omega_j \pm \Omega_j \pm' \Omega_j)/2}\;
e^{i(l\omega_j \pm \Omega_j \pm' \Omega_j)\Delta T/2}.
\]
Since we assumed that $\Delta T \gg T_0$, a significant contribution from this 
expression arises only if $l\omega_j \pm \Omega_j \pm' \Omega_j \approx 0$, 
thus requiring $l = 0$ {\sl and} $\pm \Omega_j = \mp' \Omega_j$ at the same 
time. Hence follows  
\[
\frac{\sin(l\omega_j \pm \Omega_j \pm' \Omega_j)\Delta T/2}
{(l\omega_j \pm \Omega_j \pm' \Omega_j)/2}\;
e^{i(l\omega_j \pm \Omega_j \pm' \Omega_j)\Delta T/2} \longrightarrow \Delta T
\qquad\mbox{for}\qquad 
l\omega_j \pm \Omega_j \pm' \Omega_j \longrightarrow 0.
\]
Since $\varphi^\DK = \Omega_j\tau_j^\DK$ and $\theta^\DK = \omega_j t_j^\DK$, 
the drift coefficient (\ref{fdko1}) then reads 
\[ F_j(I_j) = \bar{F_j}\, I_j \]
with 
\begin{equation} \eqlabel{fdko4}
\bar{F_j} = \frac{\kappa\omega_j}{2iQ} \sum_\pm \sum_{(m)} (\pm)\,
\widetilde{G}(m\omega_j \pm \Omega_j)\, e^{i(m \pm Q)\theta^\DK}
e^{\mp i\varphi^\DK}.
\end{equation}
According to the comments about stochastic cooling sytems on page 
\pageref{coolsys}, we assume 
\[
\widetilde{G}(\omega) = |\widetilde{G}(\omega)|\, e^{-i\omega\tau}
\qquad\mbox{with}\qquad \tau = \theta^\DK/\omega_0.
\]
Writing the frequency difference of the particle $j$ from the nominal 
frequency $\omega_0$ by $\delta\omega_j = \omega_j - \omega_0$ and using the 
relation $\widetilde{G}(-\omega) = \widetilde{G}^*(\omega)$ (see Sect.\ 
\ref{basepart3}), we obtain for Eq.\ (\ref{fdko4}) 
\begin{equation} \eqlabel{fdko4a}
\bar{F_j} = -\frac{\kappa\omega_j}{Q} \sum_{(m)}\;
\left| \widetilde{G}[(m + Q)\omega_j] \right|\; \sin\Phi_m^j
\end{equation}
with the phase
\[
\Phi_m^j = (m + Q)\theta^\DK \frac{\delta\omega_j}{\omega_0} + \varphi^\DK.
\]

\subsubsection{The Diffusion Coefficient}

The diffusion coefficient of the particle $j$ in Eq.\ (\ref{cool8}) is derived 
from 
\begin{equation} \eqlabel{fdko5}
D_j(I_j) = \Big\langle \Big\langle \frac{\Delta I_j^D \Delta I_j^D}{\Delta T} 
\Big\rangle \Big\rangle_{\varphi_j^0}.
\end{equation}
Here, $\Delta I_j^D$ describes the change of the action resulting from the
interaction with the rest beam during the time $\Delta T$, given by
\begin{equation} \eqlabel{fdko6}
\Delta I_j^D = \frac{1}{\Omega^2} \int\limits_0^{\Delta T} \dx{\qt{j}}
\dot{x}(\qt{j}) F_j^R(\qt{j}).
\end{equation}
According to Eq.\ (\ref{cool5}), the expression for $F_j^R(\qt{j})$ reads 
\[
F_j^R(\qt{j}) = 
\kappa \omega_j^2 \int \dx{w} e^{iw\qt{j}} e^{-iw\tau_j^\DK} \sum_{(m)} 
\widetilde{G}(w + m\omega_j)\, \widetilde{S}_j(w + m\omega_j)
e^{i(w + m\omega_j)t_j^\DK} e^{i(w + m\omega_j)t_j^0}
\]
with
\[
\widetilde{S}_j(w) = \sum_{j' \not= j} \sum_{(l)} 
\widetilde{x}_{j'}(w + l\omega_{j'}) e^{-iwt_{j'}^0}.
\]
Again $\widetilde{x}_{j'}(w)$ denotes the transform of the unperturbed motion 
given by (\ref{fdko3a}). As in the case of the drift coefficient, only the 
poles of $\widetilde{x}_{j'}(w)$ contribute in the $w$-integration of the 
previous equation. Performing this integration and inserting the result into 
Eq.\ (\ref{fdko6}) followed by the remaining integration over the time 
$\qt{j}$, we obtain the expression
\begin{eqnarray} \eqlabel{fdko7}
\Delta I_j^D 
& = & \frac{\kappa\omega_j}{2iQ}\, A_j \sum_{j'\not=j} \sum_{(m)} \sum_{(l)}
      \sum_\pm \sum_{\pm'} (\pm)\, e^{\pm i\varphi_j^0}
      \widetilde{G}(-l\omega_{j'} \pm' \Omega_{j'})
      e^{-i(l\omega_{j'} \mp' \Omega_{j'})t_j^\DK}               \nonumber \\
&   & \frac{\sin(m\omega_j \pm \Omega_j - l\omega_{j'} \pm' \Omega_{j'})
      \Delta T/2}{(m\omega_j \pm \Omega_j - l\omega_{j'} \pm' \Omega_{j'})/2}\,
      e^{i(m\omega_j \pm \Omega_j - l\omega_{j'} \pm' \Omega_{j'})\Delta T/2}
                                                                 \nonumber \\
&   & e^{-i(m\omega_j - l\omega_{j'} \pm' \Omega_{j'})\tau_j^\DK}
      e^{-i(l\omega_{j'} \mp' \Omega_{j'})t_j^0}\, 
      e^{i(l\omega_{j'} \mp' \Omega_{j'})t_{j'}^0}\, \frac{A_{j'}}{2}\, 
      e^{\pm' i\varphi_{j'}^0}.
\end{eqnarray}
For the calculation of the diffusion coefficient (\ref{fdko5}), two 
expressions (\ref{fdko7}) are multiplied having independent summation indices, 
say $\{j', m, l, \pm, \pm'\}$ and $\{j'', m', l', [\!\pm\!], \pm''\}$. This 
product is averaged over the betatron phases and azimuths of the particles at 
the beginning of the time integration. According to the arguments in Section 
\ref{coolpart8}, these variables can be considered as statistically 
independent. Averaging over the betatron phase $\varphi_j^0$ of the 
test-particle, we encounter expressions of the form 
\[
\Big\langle e^{\pm i\varphi_j^0} e^{[\!\pm\!] i\varphi_j^0} 
\Big\rangle_{\varphi_j^0} = \delta_{\pm [\!\mp\!]} 
\]
giving contributions only if the phases cancel each other. The second 
averaging process involves both the betatron phases and azimuths of the other 
beam particles. For uncorrelated, equally distributed betatron phases, it 
follows that 
\[
\Big\langle e^{\pm' i\varphi_{j'}^0} e^{\pm'' i\varphi_{j''}^0} 
\Big\rangle_{\varphi_{j'}^0, \varphi_{j''}^0} = \delta_{\pm' \mp''}\,
\delta_{j'j''}.
\] 
With that, the remaining averaging over the azimuths yields 
\[
\Big\langle e^{il\theta_{j'}^0} e^{il'\theta_{j'}^0} 
\Big\rangle_{\theta_{j'}^0} = \delta_{l, -l'}
\]
so that after all we obtain 
\begin{eqnarray} \eqlabel{fdko8}
D_j(I_j)
& = & \frac{1}{\Delta T} \frac{\kappa^2\omega_j^2}{4Q^2}\, I_j 
      \sum_{j'\not=j} \sum_{(m)} \sum_{(m')} \sum_{(l)} \sum_\pm \sum_{\pm'}\,
      \left| \widetilde{G}(l\omega_{j'} \mp' \Omega_{j'}) \right|^2 I_{j'}  
                                                                   \nonumber \\
&   & \frac{\sin(m\omega_j \pm \Omega_j - l\omega_{j'} \pm' \Omega_{j'})
      \Delta T/2}{(m\omega_j \pm \Omega_j - l\omega_{j'} \pm' \Omega_{j'})/2}
      \,e^{im\omega_j\Delta T/2} e^{-im\omega_j\tau_j^\DK}         \nonumber \\
&   & \frac{\sin(m'\omega_j \mp \Omega_j + l\omega_{j'} \mp' \Omega_{j'})
      \Delta T/2}{(m'\omega_j \mp \Omega_j + l\omega_{j'} \mp' \Omega_{j'})/2}
      \,e^{im'\omega_j\Delta T/2} e^{-im'\omega_j\tau_j^\DK}.
\end{eqnarray}
The value of $D_j(I_j)$ will differ significantly from zero only if the 
frequency differences in this expression obey the conditions \\
\noindent
\settowidth{\breite}{and}
\parbox{\breite}{and}
\addtolength{\breite}{5mm}
\hspace{-\breite}
\parbox{\textwidth}{
\begin{eqnarray*}
& -\, 1/\Delta T < m\omega_j \pm \Omega_j - l\omega_{j'} \pm' \Omega_{j'} <  
 1/\Delta T & \\
& -\, 1/\Delta T <  m'\omega_j \mp \Omega_j + l\omega_{j'} \mp' \Omega_{j'} <
 1/\Delta T.&
\end{eqnarray*}}
Adding both equations yields
\[ -\, 2/\Delta T <  (m + m')\omega_j < 2/\Delta T. \]
According to Section \ref{coolpart8}, the time interval $\Delta T$ is much 
larger than the correlation time $\tau_{corr}$, and thus $\Delta T \gg T_0$ so 
that both requirements can be satisfied together only by $m' = -m$. 

We now replace in Eq.\ (\ref{fdko8}) the summation over the particles $j'$ by 
an integration over the corresponding distribution function $\bar{f}(\omega, 
I)$. Provided the frequency $\omega$ and the action $I$ are independent 
variables, this distribution function factorizes and can be written as 
$\bar{f}(\omega, I) = f(\omega) \rho(I)$. The normalized distribution 
functions obey the conditions\footnote{The distribution function $f(\omega)$ 
is defined by Eq.\ (\ref{stab36a}) on page \pageref{stab36a}.}
\[
\int\limits_0^\infty \dx{I} \rho(I) = 1, \qquad\qquad
\int\limits_{-\infty}^\infty \dx{\omega} f(\omega) = 1 
\quad\mbox{and}\quad  f(-\omega) = f(\omega).
\]
Then the integration over the action can be carried out immediately, yielding
\[ \int\limits_0^\infty \dx{I} I \rho(I) = \langle I \rangle \]
where it has been presumed that the distribution $\rho(I)$ does not change 
over the time interval $\Delta T$. This assumption is justified since $\Delta 
T$ is small compared to the time in which $\rho(I)$ alters noticeably due to 
the cooling, i.e.\ $\Delta T \ll \tau_{cool}$. Performing the integration over 
the action, we are hence permitted to use a time-independent, constant 
distribution $\rho(I)$, determined by its value at the beginning of the time 
interval $\Delta T$.

Writing the betatron frequency as $\Omega = Q\omega$ and using the relations
$\widetilde{G}^*(\omega) = \widetilde{G}(-\omega)$ and $f(-\omega) = 
f(\omega)$, we can derive for Eq.\ (\ref{fdko8}) the expression 
\[ D_j(I_j) = \bar{D}_j \langle I \rangle I_j  \] \\[-10mm]
with 
\begin{equation} \eqlabel{fdko9}
\bar{D_j} = \frac{\kappa^2\omega_j^2}{Q^2} \sum_{(m)} \sum_{(l)}
            \int\limits_{-\infty}^\infty \dx{\omega} f(\omega)\, 
            \left| \widetilde{G}[(l + Q)\omega] \right|^2\,
            \frac{\sin^2[(m + Q)\omega_j - (l + Q)\omega] \Delta T/2}
            {[(m + Q)\omega_j - (l + Q)\omega]^2 \Delta T/4}.
\end{equation}
We next consider the integral over the frequency $\omega$ in detail. 
Generally, 
\[
\frac{\sin^2(\omega - \omega_0)\Delta T}{(\omega - \omega_0)^2 \Delta T}
\longrightarrow \pi \delta(\omega - \omega_0) \qquad\mbox{for}\qquad
\Delta T \longrightarrow \infty
\]
so that for sufficiently large $\Delta T$ and smooth functions $g(\omega)$ we 
can write
\[
g(\omega)\, \frac{\sin^2(\omega - \omega_0)\Delta T}
{(\omega - \omega_0)^2 \Delta T} \approx
g(\omega_0)\, \frac{\sin^2(\omega - \omega_0)\Delta T}
{(\omega - \omega_0)^2 \Delta T}.
\]
which allows in Eq.\ (\ref{fdko9}) the substitutions 
\[
\left| \widetilde{G}[(l + Q)\omega] \right|^2 \longrightarrow
\left| \widetilde{G}[(m + Q)\omega_j] \right|^2 \qquad\mbox{and}\qquad
f(\omega) \longrightarrow f\li( \frac{m + Q}{l + Q}\, \omega_j \re).
\]
The remaining integral can be evaluated in a closed form, yielding \cite{bro}
\begin{eqnarray*}
\int\limits_{-\infty}^\infty \dx{\omega} 
\frac{\sin^2[(m + Q)\omega_j - (l + Q)\omega] \Delta T/2}
{[(m + Q)\omega_j - (l + Q)\omega]^2 \Delta T/4}
& = & \frac{8}{(l + Q)^2\Delta T} \int\limits_0^\infty \dx{\omega}
      \frac{\sin^2 (l + Q)\omega\Delta T/2}{\omega^2}         \\
      = \frac{8}{(l + Q)^2\Delta T}\; \frac{\pi}{2}\; 
      \frac{|l + Q|\Delta T}{2} 
& = & \frac{2\pi}{|l + Q|}.
\end{eqnarray*}
With that, Eq.\ (\ref{fdko9}) finally becomes  
\begin{equation} \eqlabel{fdko10}
\bar{D_j} = \frac{2\pi\kappa^2\omega_j^2N}{Q^2}\, \sum_{(m)}\,
            \left| \widetilde{G}[(m + Q)\omega_j] \right|^2\,
            \sum_{(l)} \frac{1}{|l + Q|}\,
            f\li( \frac{m + Q}{l + Q}\, \omega_j \re).
\end{equation}
This expression allows for overlapping frequency bands in the beam spectrum. 
In the special case that the bands are still separated, Eq.\ (\ref{fdko10}) 
reduces because then the summations will only contribute for $l = m$, 
resulting in 
\[
\bar{D_j} = 
\frac{2\pi\kappa^2\omega_j^2N}{Q^2}\, f(\omega_j) \sum_{(m)}\,
\frac{\left| \widetilde{G}[(m + Q)\omega_j] \right|^2}{|m + Q|}.
\]

\chapter{Parameters of the Cooling System} \plabel{parapart1}
\section{The Calculation of the Cooling Rate} \plabel{parapart2}

For linear transverse stochastic cooling, we can deduce an expression for the 
cooling rate $\tau_{cool}^{-1}$ by integrating the Fokker-Planck equation.
Here, the cooling rate is defined by 
\[ 
\tau_{cool}^{-1} = \frac{1}{\langle I \rangle}\frac{d}{dt} \langle I \rangle.
\]
Assuming a distribution function $\rho(I, t)$ normalized to unity which 
vanishes beyond a maximum value $I_{max}$, we stipulate the boundary 
conditions 
\[ \int\limits_0^\infty \dx{I} \rho(I, t) = 1 \qquad\mbox{and}\qquad
\rho(I, t) \equiv 0 \quad\mbox{for}\quad I > I_{max}. \]
Starting with the Fokker-Planck equation (\ref{cool7}), 
\[ 
\frac{\partial}{\partial t} \rho(I, t) = 
- \frac{\partial}{\partial I} \left\{ \bar{F} I \rho(I, t) - \frac{1}{2} 
\bar{D} \langle I \rangle I\, \frac{\partial}{\partial I} \rho(I, t) \right\},
\]
we operate with $\int\!dI\,I$ on both sides of this equation, and obtain
\[
\frac{d}{dt} \langle I \rangle \; = \; 
\int\limits_0^\infty \dx{I} I\, \frac{\partial}{\partial t} \rho(I, t) \; = \;
- \int\limits_0^\infty \dx{I} I\, \frac{\partial}{\partial I} \left\{ 
\bar{F} I \rho(I, t) - \frac{1}{2} \bar{D} \langle I \rangle I\, 
\frac{\partial}{\partial I} \rho(I, t) \right\}.
\]
Taking into account the boundary conditions, subsequent partial integrations 
yield
\[
\frac{d}{dt} \langle I \rangle \; = \;
\bar{F} \int\limits_0^\infty \dx{I} I \rho(I, t) +  \frac{1}{2} \bar{D} 
\langle I \rangle \int\limits_0^\infty \dx{I} \rho(I, t) \; = \;
\left\{ \bar{F} + \frac{1}{2} \bar{D} \right\} \langle I \rangle.
\]
Hence the cooling rate follows as 
\begin{equation} \eqlabel{para1}
\tau_{cool}^{-1} = \bar{F} + \frac{1}{2} \bar{D}.
\end{equation}

\section{The Derivation of the Optimum Gain} \plabel{parapart3}

In this section, we will calculate the optimum gain which provides the most 
effective cooling operation and thus results in the maximum cooling rate. 
To that end, we investigate the gain $\widetilde{G}(\Omega)$ at a given 
frequency $\Omega_n = (n + Q)\omega'$ and henceforth consider the cooling rate 
as a function of the amplitude $G_{n} = |\widetilde{G}(\Omega_n)|$. The 
coefficients (\ref{fdko4a}) and (\ref{fdko10}) entering into the cooling rate 
(\ref{para1}) therefore contain only the contribution from the frequency band 
investigated, i.e.\ $m = n$. To obtain an upper limit for cooling rate, we 
further assume that all particles experience the maximum cooling interaction, 
permitting us to set $\sin\Phi_n = 1$ in Eq.\ (\ref{fdko4a}). Then the 
coefficients can be written as
\[ \bar{F_n} = -\frac{\kappa\omega'}{Q}\, |\widetilde{G}(\Omega_n)| \]
and
\[
\bar{D_n} =  \frac{2\pi\kappa^2\omega'^2N}{Q^2}\, 
\left| \widetilde{G}(\Omega_n) \right|^2\, 
\sum_{(l)} \frac{1}{|l + Q|}\, f\li( \frac{\Omega_n}{l + Q} \re).
\]
The amplitude $|\widetilde{G}_{opt}(\Omega_n)|$ of the optimum gain resulting 
in the maximum cooling rate $\tau_n^{-1}$ at the frequency $\Omega_n$ can be 
inferred from 
\[ \frac{d\tau_n^{-1}}{dG_{n}} = 
\frac{d}{dG_{n}} \left( \bar{F_n} + \frac{1}{2} \bar{D_n} \right) = 0
\qquad\mbox{for}\qquad G_{n} = |\widetilde{G}_{opt}(\Omega_n)|. \] 
The elementary calculation yields 
\begin{equation} \eqlabel{para2}
\frac{1}{|\widetilde{G}_{opt}(\Omega_n)|} = \frac{2\pi\kappa \omega' N}{Q}\, 
\sum_{(l)} \frac{1}{|l + Q|}\, f\li( \frac{\Omega_n}{l + Q} \re).
\end{equation}

\end{appendix}

\end{document}